\documentclass[aps,pre,showpacs,amsmath,amssymb,amsfonts,lengthcheck,twocolumn,longbibliography,superscriptaddress]{revtex4-2}
\usepackage{graphicx}
\usepackage{subfigure}
\usepackage{amsthm}
\usepackage{amsmath}
\usepackage{verbatim}
\usepackage{dcolumn}
\usepackage{bm}
\usepackage{color}
\usepackage[colorlinks=true,citecolor=blue,linkcolor=blue,urlcolor=blue]{hyperref}%
\usepackage{xcolor}
\usepackage{dsfont}
\usepackage{longtable}
\newcommand{\bra}[1]{\left\langle #1\right|}
\newcommand{\ket}[1]{\left|#1\right\rangle}
\newcommand{\braket}[2]{\left\langle #1|#2\right\rangle}

\newcommand{\tr}[1]{\mathrm{tr}\left\{#1\right\}}

\newcommand{\la}{\left\langle}
\newcommand{\ra}{\right\rangle}
\newcommand{\pd}{\partial}

\newcommand{\bla}{bla\\bla\\bla\\bla\\bla}

\begin{document}

\title{Thermodynamics of Statistical Anyons}
\date{\today}
\author{Nathan M. Myers}
\email{myersn1@umbc.edu}
\affiliation{Department of Physics, University of Maryland Baltimore County, Baltimore, MD 21250, USA}
\author{Sebastian Deffner}
\email{deffner@umbc.edu}
\affiliation {Department of Physics, University of Maryland Baltimore County, Baltimore, MD 21250, USA}
\affiliation {Instituto de F\'{i}sica `Gleb Wataghin', Universidade Estadual de Campinas, 13083-859, Campinas, S\~{a}o Paulo, Brazil}

\begin{abstract}
In low-dimensional systems, indistinguishable particles can display statistics that interpolate between bosons and fermions. Signatures of these “anyons” have been detected in two-dimensional quasiparticle excitations of the fractional quantum Hall effect, however experimental access to these quasiparticles remains limited. As an alternative to these ``topological anyons," we propose ``statistical anyons" realized through a statistical mixture of particles with bosonic and fermionic symmetry. We show that the framework of statistical anyons is equivalent to the generalized exclusion statistics (GES) pioneered by Haldane, significantly broadening the range of systems to which GES apply. We develop the full thermodynamic characterizations of these statistical anyons, including both equilibrium and nonequilibrium behavior. To develop a complete picture, we compare the performance of quantum heat engines with working mediums of statistical anyons and traditional topological anyons, demonstrating the effects of the anyonic phase in both local equilibrium and fully nonequilibrium regimes. In addition, methods of optimizing engine performance through shortcuts to adiabaticity are investigated, using both linear response and fast forward techniques.
\end{abstract}

\maketitle

\section{Introduction} 

A unique aspect of quantum particles is that they may be truly \textit{identical} - that is to say there exists no method or feature by which we would be able to distinguish one from another. The linear nature of quantum mechanics accounts for this in the form of a superposition state for the system of all possible permutations of individual particle states \cite{Griffiths}. Such a composition is not unique, with two possible  solutions distinguished by a phase of $\pm 1$ picked up under particle exchange. The symmetric (+1) solution defines a class of fundamental particles commonly called bosons, while the antisymmetric (-1) solution defines a class commonly called fermions. 

This symmetrization requirement has a profound physical consequence in the form of \textit{exchange forces}. A simple evaluation of the separation distance between two identical particles (see for example \cite{Griffiths}) shows that bosons will tend to bunch together, while fermions will tend to be found farther apart. In fact, for fermions the probability of two particles being found in the exact same state is identically zero, which is the origin of the familiar Pauli exclusion principle. 

The field of quantum statistics was significantly expanded when it was discovered by Leinaas and Myrheim that the differences in topology in one- and two-dimensional systems allows for the existence of a continuum of fractional statistics solutions, represented by a general phase factor of $e^{i \pi \nu}$ \cite{Leinaas},     
\begin{equation}
\Psi(\mathbf{r}_1,\mathbf{r}_2) = e^{i \pi \nu} \Psi(\mathbf{r}_2,\mathbf{r}_1).
\end{equation}
For $\nu = 2n$, where $n = 0,1,2,...$, the bosonic case is recovered, and for $\nu = 2n+1$ we obtain the fermionic case. Shortly thereafter Wilczek proposed a realization of these ``any"-ons in the form of a two-dimensional quasiparticle made up of a charged particle orbiting a magnetic flux tube \cite{Wilczek}. The name \textit{anyons} signifies that interchange of particles (accomplished by successive half-rotations of each quasiparticle around the other) can produce an arbitrary phase between that of fermions and bosons as a consequence of the Aharonov-Bohm effect \cite{Wilczek}.  

The study of anyons was brought into the forefront when it was proven by Arovas, Schrieffer, and Wilczek that quasiparticles entering the fractional quantum Hall effect posses not only fractional charge, but also obey fractional statistics \cite{Arovas}. Interest in anyons received another boost when it was found that non-abelian anyons \cite{Moore1991} could be used as key components in the development of a fault-tolerant quantum computer. For non-abelian anyons, exchange (``braiding") does not just introduce a complex phase, but acts as a unitary transformation on the state. In this manner, combinations of braids can act as quantum logic gates \cite{Kitaev2003}. As small local perturbations do not change the braiding properties of the anyons, this method of quantum computation is very robust against noise \cite{Kitaev2003}.  Experimental access to non-abelian anyons remains a challenge, however indirect evidence of non-abelian anyonic states has been found \cite{Clarke2013, Kasahara2018, Banerjee2018}. See Ref. \cite{Nayak2008} for a more detailed review on topological quantum computation and the role of anyons. While non-abelian anyons are necessary for implementing topological quantum computation, understanding the properties of abelian anyons can provide important insight into the general behavior of anyons that may be useful in controlling their non-abelian counterparts.        

Discussion of anyons in one-dimension requires some additional subtlety, as exchange by rotation is no longer valid. In this context, Haldane introduced a dimension-independent approach to anyonic statistics, generalized exclusion statistics (GES), based on a generalization of the Pauli exclusion principle \cite{Haldane}. In Haldane's approach, particle nature is quantified by a statistical interaction that determines the degree to which two identical particles can occupy the same state \cite{Haldane}. Notably, this approach classifies particle type solely on degree of repulsion and as such identifies ``hardcore-bosons" and fermions as equivalent, despite differences in low-energy behavior \cite{Haldane}. While closely linked to Wilczek's formulation of anyons, Haldane's GES anyons are distinctly different, as their behavior is based on generalizing the exclusion principle, rather than the exchange behavior \cite{Murthy1994}.  

Haldane's work was expanded upon by Wu, who derived the statistical distribution for an ideal anyon gas \cite{Wu}. This approach continued to receive attention when it was shown by Yang and Yang that a 1D system of interacting bosons could be treated as non-interacting 1D anyonic system obeying Haldane's generalized exclusion statistics \cite{Yang1969, Yang1970}. Similar mappings to interacting anyonic systems have been demonstrated for more complex potentials \cite{Kundu1999}.  

While the theoretical study of anyons is well-established, it is only in recent years that experimental techniques have advanced to the point that detection of anyons is feasible. For the case of fractional quantum Hall state anyons, a popular detection method is based on implementing quasiparticle interferometers whose interference effects depend directly on the anyonic phase \cite{Chamon1997, Ji2003, Law2006, Bonderson2006}. Numerous experimental attempts have followed this approach \cite{Camino2005, Ofek2010, McClure2012, Willett2013, Nakamura2019, Willett2019} but distinguishing signatures of the anyonic phase from interference effects arising from other factors, such as Coulomb blockading and Aharonov-Bohm interference, has proven elusive until this year \cite{Bartolomei2020}.

The difficulty in accessing fractional quantum Hall state anyons has led to alternative methods of experimentally studying their properties. A promising route that we will focus on in this work is motivated by the celebrated Hong-Ou-Mandel (HOM) effect first observed in photonic interferometry \cite{Hong1987}.

In section \ref{sec:2} we establish an implementation of anyons from a statistical mixture of boson and fermion pairs, which we refer to as ``statistical anyons." We show that statistical anyons, despite different construction, behave equivalently to Haldane's generalized exclusion principle anyons. In section \ref{sec:3} we examine the equilibrium thermodynamic properties of statistical and abelian topological anyons, such as those proposed by Leinaas, Myrheim, and Wilczek, including entropy, heat capacity, and free energy. In section \ref{sec:4} we compare the performance of an endoreversible quantum Otto engine with a working medium of statistical anyons to one with a working medium of topological anyons. In section \ref{sec:5}, we extend our analysis of the statistical anyon engine to the fully nonequilibrium regime. In section \ref{sec:6} we explore the role of anyonic statistics in the implementation of \textit{shortcuts to adiabaticity}, which can be used to enhance thermal machine performance by increasing power without loss of efficiency. In section \ref{sec:7} we conclude with a perspective on future research directions.    

\section{``Statistical" Anyons}
\label{sec:2}
In the typical HOM effect a pair of entangled photons are incident symmetrically on a two-port 50/50 beamsplitter. Assuming all other degrees of freedom of the photons are identical (such as frequency and polarization), the initial bosonic spatial state is described by a symmetric superposition of each input,
\begin{equation}
\ket{\psi_{\mathrm{i}}^{\mathrm{B}}} = \frac{1}{\sqrt{2}}\Big(\ket{a}_1\ket{b}_2+\ket{b}_1\ket{a}_2\Big).
\end{equation}    
The operation of the beamsplitter evolves state $\ket{a} \rightarrow \frac{1}{\sqrt{2}}(\ket{c}+i\ket{d})$ and state $\ket{b} \rightarrow \frac{1}{\sqrt{2}}(i\ket{c}+\ket{d})$ with the imaginary component denoting the phase shift of $\pi$ picked up upon reflection. Carrying out this evolution on each input state while keeping track of the particle indices we find that the only states to survive are those in which both photons exit the same beamsplitter port,
\begin{equation}
\ket{\psi_{\mathrm{f}}^{\mathrm{B}}} = \frac{i}{\sqrt{2}}\Big(\ket{c}_1\ket{c}_2+\ket{d}_1\ket{d}_2\Big).
\end{equation}
Physically, this is a manifestation of the effective attraction between bosons (typically this is referred to as ``boson bunching") \cite{Hong1987}. 

The HOM effect can be extended to cases in which the other degrees of freedom of the photons are not identical \cite{Zeilinger1998}. Consider a case in which the photons are prepared in a Bell state basis in polarization. The four possible Bell pairs are, 
\begin{equation}
\begin{split}
\ket{\Phi_A} = \frac{1}{\sqrt{2}}\Big( \ket{00}+\ket{11}\Big), \\
\ket{\Phi_B} = \frac{1}{\sqrt{2}}\Big( \ket{00}-\ket{11}\Big), \\
\ket{\Phi_C} = \frac{1}{\sqrt{2}}\Big( \ket{01}+\ket{10}\Big), \\
\ket{\Phi_D} = \frac{1}{\sqrt{2}}\Big( \ket{01}-\ket{10}\Big),
\end{split}
\end{equation}
where here $\ket{0}$ and $\ket{1}$ represent orthogonal polarization states. We can use this additional degree of freedom to encode fermionic, and ultimately anyonic, statistics into the behavior of the photons \cite{Rarity1990, Markus1996, Wang2007, Matthews2013}. In this manner the photons can be thought of as a ``quantum substrate" on which we will construct our desired statistics. 

\begin{figure}
	\subfigure[]{
		\includegraphics[width=.25\textwidth]{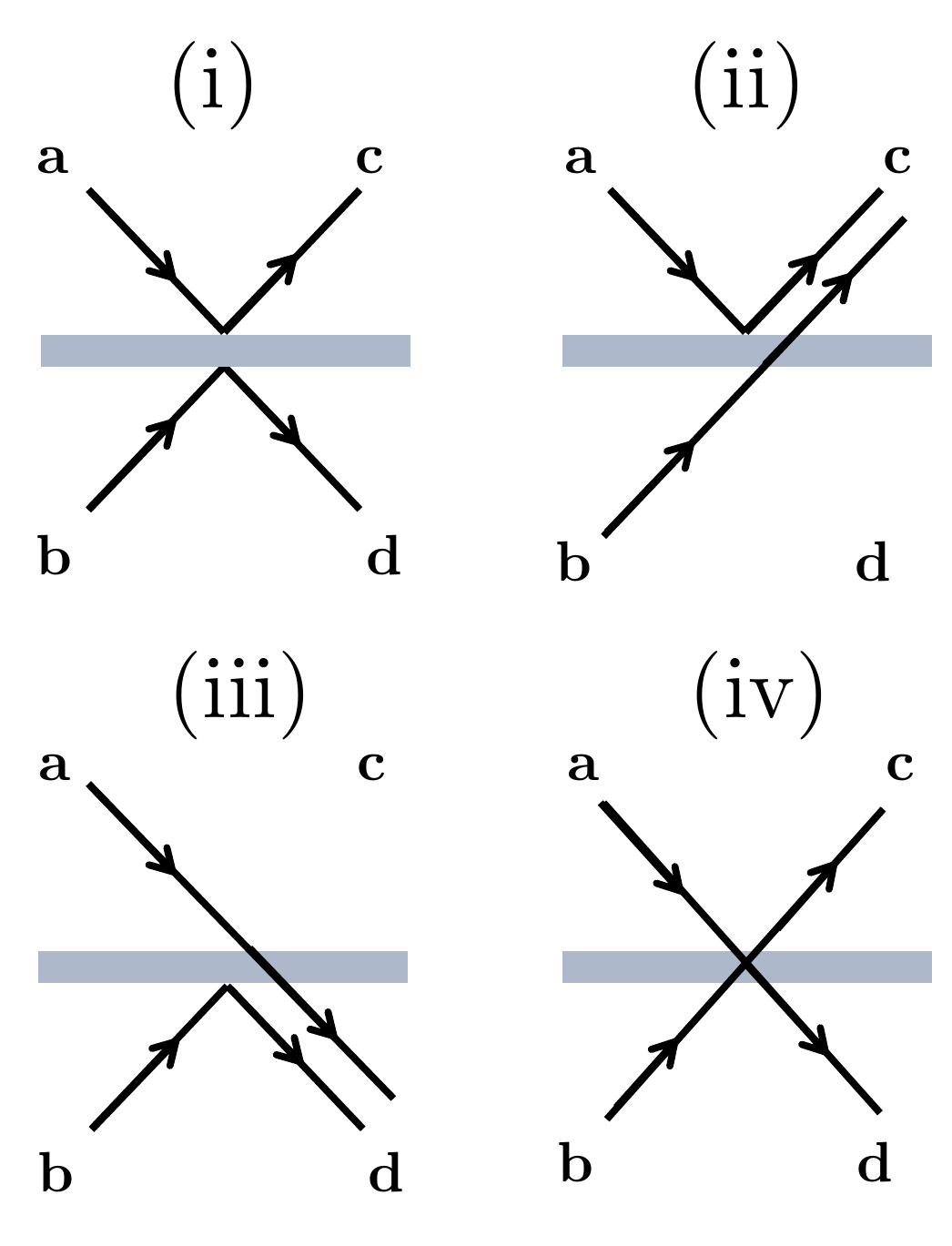}
	}
	\hspace{5mm}
	\subfigure[]{
		\includegraphics[width=.13\textwidth]{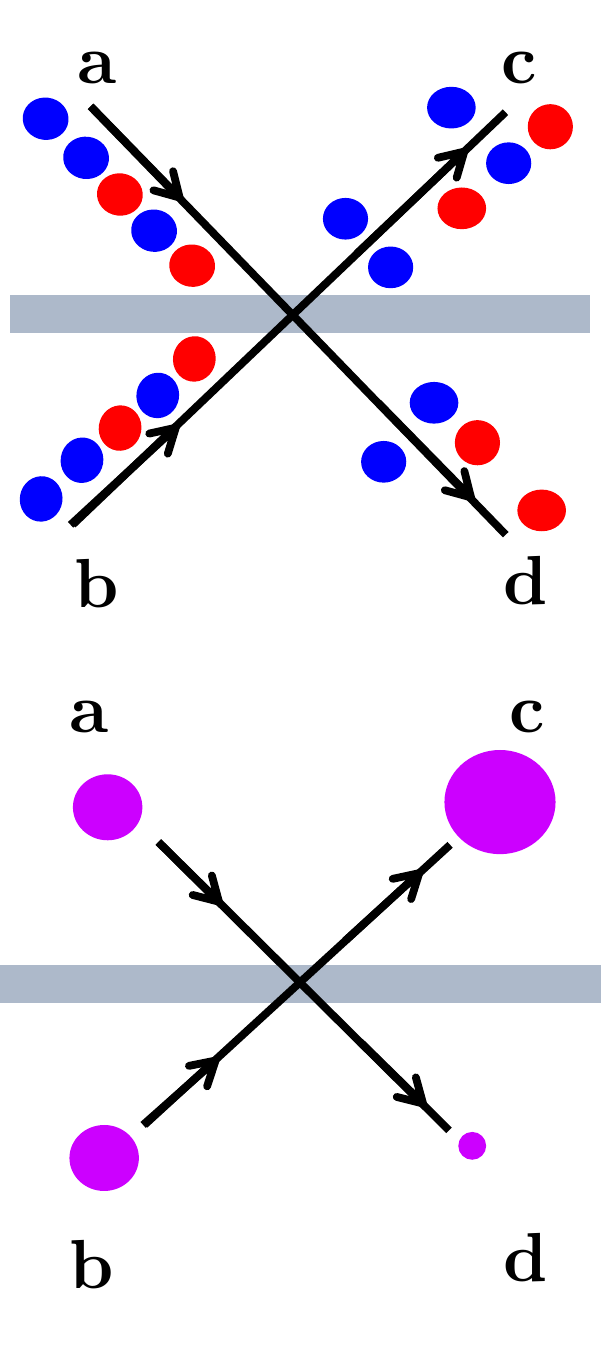}
	}
	\caption{\label{fig:HOMa} (a) All possible photon transmission and reflection combinations. For symmetric (bosonic) states possibilities (i) and (iv) interfere deconstructively while for antisymmetric (fermionic) states they interfere constructively. (b) Representation of statistical anyons for $N=5$. Top figure: N particle pairs are incident, with blue particles representing pairs with symmetric entanglement and red pairs with antisymmetric entanglement. Bottom figure: The equivalent statistical anyonic representation for each particle pair, with particle size representing the probability density of finding a particle at that output port.}
\end{figure}

We note that $\ket{\Phi_A}$, $\ket{\Phi_B}$, and $\ket{\Phi_C}$ are symmetric under exchange, while $\ket{\Phi_D}$ is antisymmetric. As photons are bosons, their overall wave function must still be symmetric. This can occur in one of two ways: (i) a symmetric spatial state paired with one of the three symmetric polarization states,
\begin{equation}
\ket{\Psi_{\mathrm{i}}^{\mathrm{B}}} = \frac{1}{\sqrt{2}}\Big(\ket{a}_1\ket{b}_2+\ket{b}_1\ket{a}_2\Big)\otimes \ket{\Phi_j}, 
\end{equation}
where $j = \{A,B,C\}$, or (ii) an antisymmetric spatial state paired with the antisymmetric polarization state,
\begin{equation}
\label{spatialandpol}
\ket{\Psi_{\mathrm{i}}^{\mathrm{F}}} = \frac{1}{\sqrt{2}}\Big(\ket{a}_1\ket{b}_2-\ket{b}_1\ket{a}_2\Big)\otimes \ket{\Phi_D}, 
\end{equation}
Assuming it is non-polarizing, the beamsplitter acts only on the spatial portion of the wavefunction. For case (ii), as shown in Eq. \eqref{spatialandpol}, the output spatial state is,
\begin{equation}
\ket{\psi_{\mathrm{f}}^{\mathrm{F}}} = \frac{1}{\sqrt{2}}\Big(\ket{c}_1\ket{d}_2-\ket{d}_1\ket{c}_2\Big),
\end{equation}
guaranteeing that both photons exit opposite ports of the beamsplitter. This is a result of the effective repulsion between fermions, and can be seen as a manifestation of the Pauli exclusion principle. Despite the fact that photons are fundamentally bosons, since the beamsplitter only accesses the antisymmetric portion of the overall state, they behave exactly as fermions. Figure \ref{fig:HOMa}a summarizes the possible outcomes for the two-photon input states. Notably, both bunching and anti-bunching behavior vanishes if the photons become distinguishable (say by increasing the flight time for just one port of the beamsplitter) \cite{Zeilinger1998}.
                    
This effect can be extended to anyonic statistics by introducing a phase that tunes between the symmetric and antisymmetric Bell states. Phase control can be achieved using a rotated polarization beamsplitter \cite{Matthews2013}, adjusting path length \cite{Rarity1990, Markus1996}, or introducing phase plates \cite{Wang2007}, making this method of simulating anyonic behavior very experimentally accessible. Here we propose introducing this anyonic phase through a statistical mixture of symmetrically and antisymmetrically entangled photon pairs. Given $N$ photon pairs incident on the beam splitter, the average behavior of each photon pair can then be represented by an anyonic wave function in which the anyonic phase gives the probability that any two incident particles will exit the same port, $p_{\mathrm{B}}$,
\begin{equation}
\ket{\psi_{\mathrm{i}}^{\mathrm{A}}} = \frac{1}{\sqrt{2}}\Big(\ket{a}_1\ket{b}_2+e^{i \pi \nu(p_{\mathrm{B}}) }\ket{b}_1\ket{a}_2\Big).
\end{equation} 
This behavior is shown pictorial in Fig. \ref{fig:HOMa}b.

\begin{figure*}
	\subfigure[]{
		\includegraphics[width=.3\textwidth]{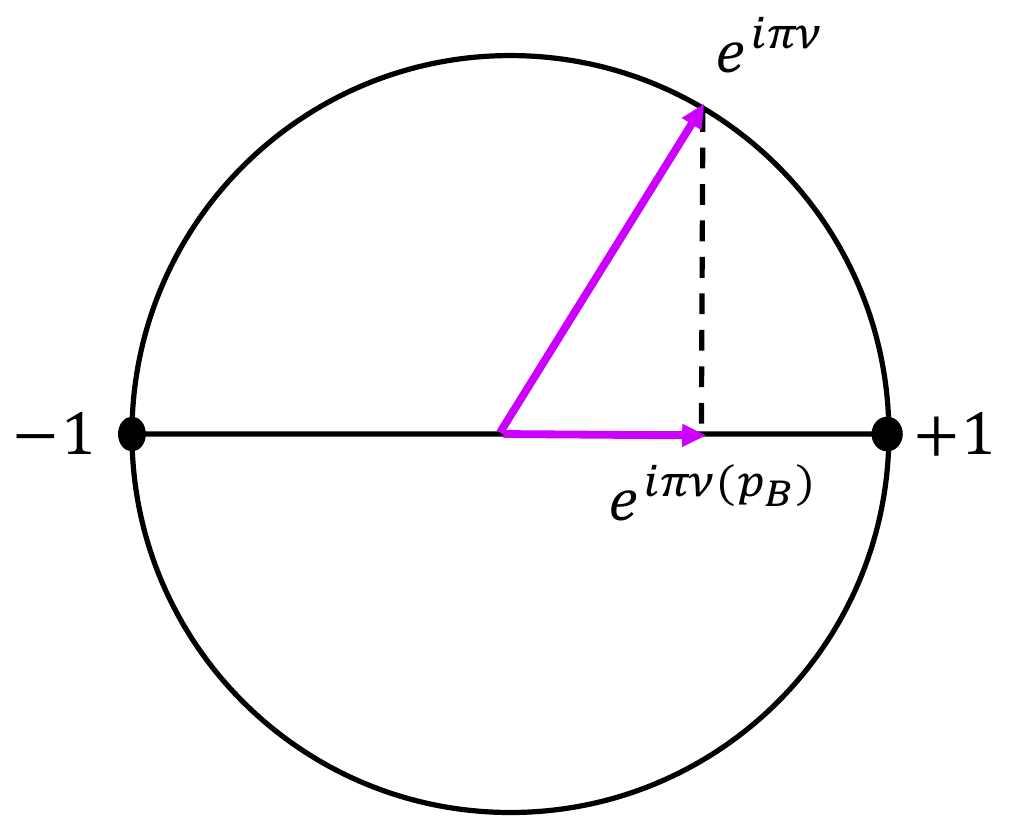}
	}
	\hspace{5mm}
	\subfigure[]{
		\includegraphics[width=.6\textwidth]{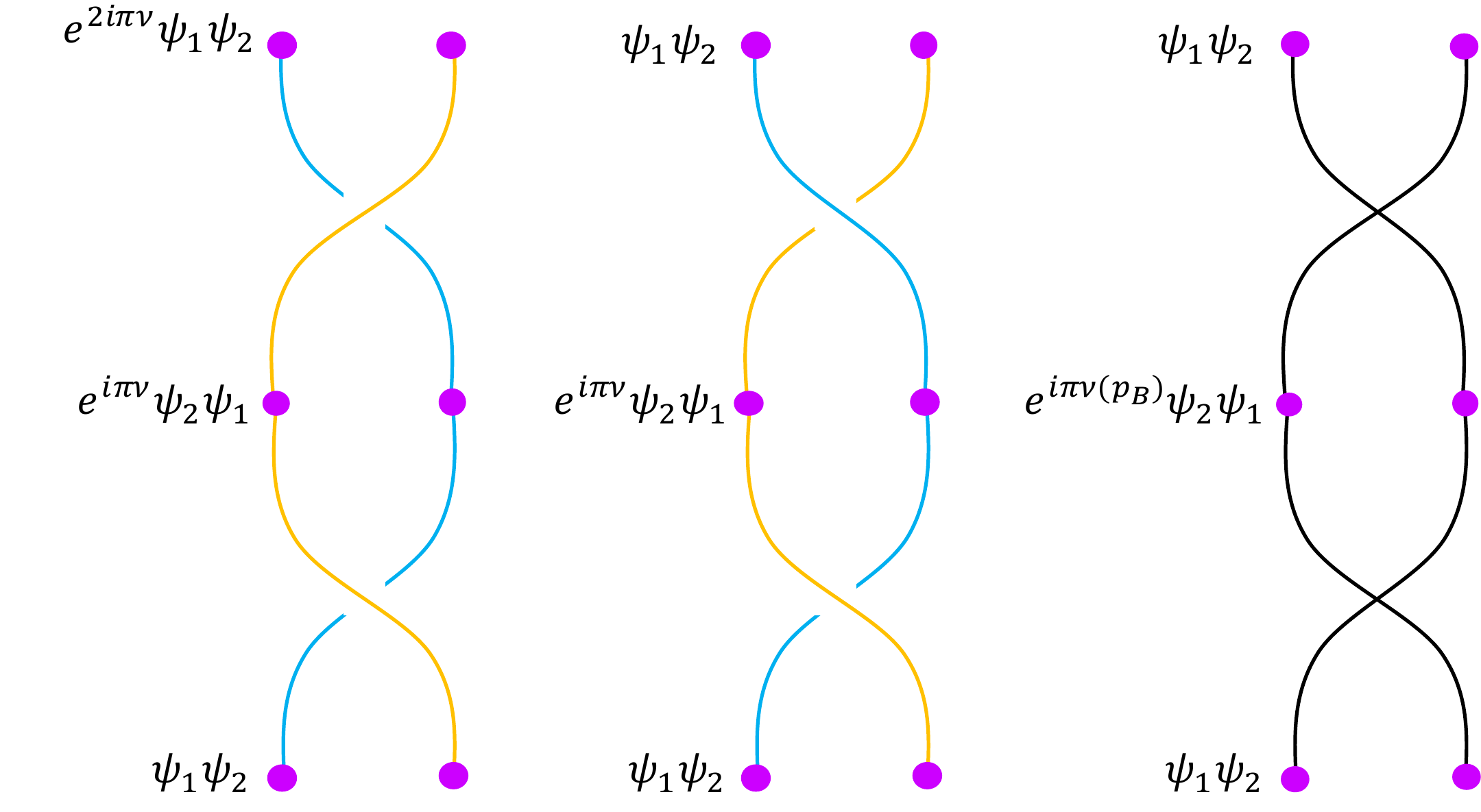}
	}
	\caption{\label{fig:StatVsTop} (a) Comparison between the exchange phases of topological and statistical anyons. For topological anyons the direction of the exchange (clockwise or counterclockwise) determines the sign of the anyonic phase. As such, the space of possible topological anyon exchange phases can be represented by a circle, with bosonic and fermionic symmetry lying on opposite sides. For statistical anyons, the anyonic phase is an average over $N_{\mathrm{B}}$ bosonic phase factors and $N_{\mathrm{F}}$ fermionic phase factors. As such, the space of possible statistical anyon phases is confined to a line between $-1$ and $+1$. (b) Comparison between topological and statistical anyons undergoing two interchanges. For topological anyons two exchanges in the same direction results in the original wavefunction multiplied by an additional phase of $e^{2 i \pi \nu}$ (left diagram), while two exchanges in opposite directions leaves the wavefunction unchanged (middle diagram). For statistical anyons the direction of the exchange is irrelevant, with two exchanges always resulting in an unchanged wavefunction (right diagram).}
\end{figure*}

We refer to this implementation as ``statistical anyons" in contrast to ``topological anyons," that appear in Wilczek's charge and flux tube realization \cite{Wilczek}, or in quasiparticle excitations in the fractional quantum Hall regime \cite{Arovas}. While both formulations pick up an anyonic phase under particle exchange and display behavior interpolating between that of fermions and bosons, in the present statistical anyon framework this is a result of averaging over the behavior of a large number of particles, while for topological anyons it is a property of the quasiparticles themselves. Note that when we refer to topological anyons, we are considering specifically \textit{abelian} particles, whose exchange introduces a phase a $e^{i \pi \nu}$ \cite{Lerda1992}.

From a topological standpoint, there is another important distinction between statistical and topological anyons. It is well established that the configuration space of topological anyon statistics is a representation of the braid group, rather than the permutation group, as is the case for bosonic and fermionic statistics \cite{Leinaas, Lerda1992}. This difference arises as in two dimensions, two repeated exchanges is a topologically distinct operation from doing nothing, while in three dimensions these operations are topologically equivalent \cite{Leinaas, Lerda1992}. This leads to the physical consequence that the \textit{direction} of the exchange is significant for topological anyons, as two exchanges in the same direction will result in a wavefunction different from the original by a phase of $e^{2 i \pi \nu}$. Since statistical anyons are constructed from an average over a mixture of bosons and fermions, whose permutation group statistics ensures that two repeated exchanges will always return the original wavefunction, the anyonic phase of statistical anyons is \textit{independent} of exchange direction. As a result, the space of statistical anyon phases is of a dimension smaller than the space of topological anyon phases. We can represent the two-dimensional space of possible topological anyon phases as a circle, with bosonic and fermionic phases given by diametrically opposite points. The space of possible statistical anyon phases is then represented by the diameter line of the circle. In this geometric picture, the statistical anyon phase corresponding to the topological anyon phase of $e^{i \pi \nu}$ would be the projection of that radial vector onto the diameter line. These differences between topological and statistical anyons are illustrated graphically in Fig. \ref{fig:StatVsTop}. 

We see from this comparison that, as a representation of the braid group, topological anyons show increased complexity in comparison to statistical anyons. This leads to the motivating questions of this work -- How does the thermodynamics of topological and statistical anyons compare? Can the simpler, more experimentally accessible statistical anyons replicate any of the unique properties of topological anyons? Do statistical anyons themselves display intricate thermodynamic behavior that can be exploited?

\subsection{Statistical Anyon Wavefunction} 
       
To better understand the relationship between statistical and topological anyons, let us consider extending the notion of statistical anyons beyond the realm of quantum optics. For a single pair of bosonic particles the spatial wave function is given by,
\begin{equation}
\Psi_{\mathrm{B}}(\mathbf{x}) = \frac{1}{\sqrt{2(1+\delta_{n_1,n_2})}} \left[ \psi_{n_1} (x) \psi_{n_2} (y) + \psi_{n_1} (y) \psi_{n_2} (x) \right],
\end{equation}
where $\psi_{n} (x)$ is the normalized single-particle eigenstate corresponding to quantum number $n$ and $\mathbf{x} = (x,y)$. Note that the normalization coefficient for bosons changes when both particles occupy the same state, as in this case the un-normalized wavefunction becomes identical to the wavefunction of two distinguishable particles with an additional factor of two. Similarly, the wave function of a fermionic particle pair is,
\begin{equation}
\Psi_{\mathrm{F}}(\mathbf{x}) = \frac{1}{\sqrt{2}} \left[ \psi_{n_1} (x) \psi_{n_2} (y) - \psi_{n_1} (y) \psi_{n_2} (x) \right].
\end{equation}   
Here we require no delta function in the normalization, as two fermions will never occupy the same state. The total wave function for $N$ independent particle pairs, of which $N_{\mathrm{B}}$ of the pairs are symmetric under exchange and $N_{\mathrm{F}} = N-N_{\mathrm{B}}$ of the pairs are antisymmetric under exchange is then,
\begin{equation}
\Psi(\mathbf{x}_1,\mathbf{x}_2,...,\mathbf{x}_N) =\displaystyle\prod_{j=1}^{N_{\mathrm{B}}} \Psi_{\mathrm{B}}(\mathbf{x}_j) \displaystyle\prod_{k=N_{\mathrm{B}}+1}^{N}\Psi_{\mathrm{F}}(\mathbf{x}_k).
\end{equation}
This wave function can be equivalently represented in the framework of statistical anyons as,
\begin{equation}
\label{eq:anyonWF}
\Psi(\mathbf{x}_1,\mathbf{x}_2,...,\mathbf{x}_N) =\displaystyle\prod_{j=1}^{N} \Psi_{\mathrm{A}}(\mathbf{x}_j),
\end{equation}
where,
\begin{equation}
\label{eq:anyon}
\begin{split}
\Psi_{\mathrm{A}}(\mathbf{x}_j) = \frac{1}{\sqrt{2(1+\delta_{n_1,n_2})}} [ \psi_{n_1} (x_j) \psi_{n_2} (y_j) \\+ e^{i \pi \nu_j} \psi_{n_1} (y_j) \psi_{n_2} (x_j) ].
\end{split}
\end{equation}
We pause here to note a significant difference between statistical and topological anyons. As we see in Eq. \eqref{eq:anyon}, the statistical anyon wave function can always be constructed from a superposition of the single particle wave functions. This is \textit{not} generally true for the wave function of topological anyons, outside of the bosonic and fermionic limits of the anyonic phase \cite{Lerda1992, Myrheim1999, Khare2005}. 

The product in Eq. \eqref{eq:anyonWF} must give $N_{\mathrm{B}}$ symmetric wave functions and $N_{\mathrm{F}}$ antisymmetric wave functions. This condition is fulfilled if the anyonic phase in Eq. \eqref{eq:anyon} is given by $\nu_j = \Theta\left(j - N_{\mathrm{B}} -1 \right)$, where $\Theta(\cdot)$ is the Heaviside step function, using the convention $\Theta(0) = 1$. In this framework, the anyonic factor in Eq. \eqref{eq:anyon} can be thought of as giving the ``average" phase picked up under exchange for any one of the $N$ pairs of particles. For large $N$ we can express the number of bosonic and fermionic particle pairs as $N_{\mathrm{B}} = N p_{\mathrm{B}}$ and $N_{\mathrm{F}} = N p_{\mathrm{F}} = N(1-p_{\mathrm{B}})$ where $p_{\mathrm{B}}$ ($p_{\mathrm{F}}$) is the probability of a particle pair having bosonic (fermionic) symmetry. 

In the limit of a single two-particle system, the anyonic wave function becomes,              
\begin{equation}
\begin{split}
\Psi_{\mathrm{A}}(\mathbf{x}) = \frac{1}{\sqrt{2(1+\delta_{n_1,n_2})}} \Big[ \psi_{n_1} (x) \psi_{n_2} (y) \\ +e^{i \pi (p_{\mathrm{B}} + 1)} \psi_{n_1} (y) \psi_{n_2} (x) \Big],
\end{split}
\end{equation}
where $p_{\mathrm{B}} = \{0,1\}$, as in reality each individual pair of particles must be either bosons or fermions.

\subsection{Statistical Anyons in Second Quantization}

The statistical anyon framework can also be extended to second quantization, from which we can determine the appropriate commutation relations. For topological anyons, the creation and annihilation operators can be determined from the bosonic or fermionic operators via the Jordan-Wigner transformation, and obey the following commutation relations \cite{Lerda1993},
\begin{align}
 a (\mathbf{x}_{\mathcal{C}})a (\mathbf{y}_{\mathcal{C}}) - e^{-i \pi \nu} a (\mathbf{y}_{\mathcal{C}})a (\mathbf{x}_{\mathcal{C}}) = 0,&  \nonumber \\
 a (\mathbf{x}_{\mathcal{C}})a^{\dagger} (\mathbf{y}_{\mathcal{C}}) - e^{i \pi \nu} a^{\dagger} (\mathbf{y}_{\mathcal{C}})a (\mathbf{x}_{\mathcal{C}}) = 0, &\\
 a^{\dagger} (\mathbf{x}_{\mathcal{C}})a (\mathbf{y}_{\mathcal{C}}) - e^{i \pi \nu} a (\mathbf{y}_{\mathcal{C}})a^{\dagger} (\mathbf{x}_{\mathcal{C}}) = 0. &\nonumber \\
 a^{\dagger} (\mathbf{x}_{\mathcal{C}})a^{\dagger} (\mathbf{y}_{\mathcal{C}}) - e^{-i \pi \nu} a^{\dagger} (\mathbf{y}_{\mathcal{C}})a^{\dagger} (\mathbf{x}_{\mathcal{C}}) = &0, \nonumber
\end{align}
Note that the commutation relations are dependent on the curve $\mathcal{C}$, which indicates the direction of the exchange rotation. As the topological anyon operators are a representation of the braid group, the phase picked up depends on whether or not the exchange was performed clockwise or counterclockwise (as illustrated in Fig. \ref{fig:StatVsTop}). We see that $\nu = 0$ restores the boson commutation relations and $\nu = 1$ restores the fermion anticommutation relations.   

Notably, for particles at the same position the anyonic phase cancels out and the canonical commutation relation reduces to \cite{Lerda1993, Khare2005},
\begin{equation}
\label{eq:CCRtop}
a (\mathbf{x}_{\mathcal{C}})a^{\dagger} (\mathbf{x}_{\mathcal{C}}) \pm a^{\dagger} (\mathbf{x}_{\mathcal{C}})a (\mathbf{x}_{\mathcal{C}}) = 1,
\end{equation}     
with the $(+)$ occurring if the anyonic operators are constructed from transformed fermionic operators, and the $(-)$ if they are constructed from transformed bosonic operators. In general, topological anyons act as ``hard-core" particles for all $\nu$ other than $\nu = 0$, obeying an exclusion principle \cite{Khare2005}. In this sense, the anyonic phase parameter can be thought of as quantifying the degree of repulsion between two identical particles \cite{Khare2005}.  

To determine the commutation relations for statistical anyon creation and annihilation operators, we follow a similar approach to the construction of the statistical anyon wave function in the first quantization. Let us consider $N$ Fock states. Starting each in the vacuum state, we can construct a single occupancy state from the application of the corresponding creation operator. We apply the bosonic creation operator to $N_{\mathrm{B}}$ states and the fermionic creation operator to $N_{\mathrm{F}} = N-N_{\mathrm{B}}$ states,
\begin{equation}
\ket{1}_1\ket{1}_2...\ket{1}_N = \displaystyle\prod_{j=1}^{N_{\mathrm{B}}} b_j^{\dagger} \ket{0}_j \displaystyle\prod_{k=N_{\mathrm{B}}+1}^{N} f_k^{\dagger} \ket{0}_k.
\end{equation}
We can equivalently represent this state in the statistical anyon picture using a statistical anyon creation operator,
\begin{equation}
\ket{1}_1\ket{1}_2...\ket{1}_N = \displaystyle\prod_{j=1}^{N} s_j^{\dagger} \ket{0}_j.
\end{equation}
The operator $s_j^{\dagger}$ must reduce to the bosonic creation operator for $j \leq N_{\mathrm{B}}$ and to the fermionic creation operator for $j > N_{\mathrm{B}}$. We can similarly define a statistical anyon annihilation operator that must reduce to the bosonic and fermionic annihilation operators under the same conditions. With these restrictions, we can construct the statistical anyon commutation relations as follows,
\begin{align}
& s_j^{\dagger} s_j - e^{i \pi \Theta\left(j - N_{\mathrm{B}} -1 \right)} s_j s_j^{\dagger} = 1, \nonumber \\
& s_j s_j^{\dagger} - e^{i \pi \Theta\left(j - N_{\mathrm{B}} -1 \right)} s_j^{\dagger} s_j = 1, \\
& s_j^{\dagger} s_j^{\dagger} - e^{i \pi \Theta\left(j - N_{\mathrm{B}} -1 \right)} s_j^{\dagger} s_j^{\dagger} = 0, \nonumber \\
& s_j s_j - e^{i \pi \Theta\left(j - N_{\mathrm{B}} -1 \right)} s_j s_j = 0. \nonumber
\end{align}
In the limit of a single particle the canonical commutation reduces to,
\begin{equation}
s_j s_j^{\dagger} + e^{i \pi (p_{\mathrm{B}} + 1)} s_j^{\dagger} s_j = 1, 
\end{equation} 
where $p_{\mathrm{B}} = \{0,1\}$, as in reality the particle is either a boson or fermion. This echoes Eq. \eqref{eq:CCRtop}, where the form of the commutation relation depends on whether the anyonic operators are constructed from transformed fermionic or bosonic operators. 

\subsection{Statistical anyons and Generalized Exclusion Statistics}
A notable difference between statistical and topological anyons is the lack of an exclusion principle, except in the fermionic limit. Instead, statistical anyons admit ``partially occupied" states that arise from averaging over the occupancy of all $N$ systems. In this sense, the statistical anyon framework is more akin to Haldane's generalized exclusion statistics \cite{Haldane}. GES is constructed by extending the Pauli exclusion principle through the definition of a parameterized differential relation that quantifies the change in the dimension of the Hilbert space of a discrete-state system upon a change in the particle number \cite{Haldane},
\begin{equation}
\label{eq:GenAnyonRel}
\Delta d_{\mathrm{GES}} = -g \Delta N. 
\end{equation}
For bosons, with infinite possible state occupancy, the dimension is independent of the particle number, thus $g=0$. For fermions, subject to the full exclusion principle, the dimension scales directly with each additional particle, thus $g = 1$. GES statistics anyons have been shown to manifest in confined, interacting gasses, such as the Calogero-Sutherland model gas, consisting of bosons or fermions in a harmonic potential with an inverse square law interaction \cite{Murthy19942, Isakov1996, Meljanac2004, Campo2016}. Notably, this implementation of GES anyons can be directly mapped to topological anyons confined to the lowest Landau level \cite{Wu, Dasnieres1994, Ouvry2018}. Other physical systems shown to host GES anyons include Lieb–Liniger and  hard core Tonks-Girardeau gasses with $\delta$-function potentials \cite{Kundu1999, Batchelor2006, Girardeau2006, Batchelor2007, Calabrese2007, Patu2007, Batchelor2008, Hao2008, Campo2008, Patu2008, Patu2008_2, Santachiara2008} as well as Hubbard chains \cite{Vitoriano2009, Tang2015}, whose anyonic behavior can be imitated in ultracold gasses \cite{Yannouleas2019}.    

In the statistical anyon framework, the change in the Hilbert space dimension will be given by the sum of the dimension change if the particles are bosons and the dimension change if they are fermions, weighted by the respective probabilities, 
\begin{equation}
\Delta d_{\mathrm{SA}} = p_{\mathrm{B}} \Delta d_{\mathrm{B}} + p_{\mathrm{F}} \Delta d_{\mathrm{F}}
\end{equation}
Noting $\Delta d_{\mathrm{B}} = 0$ and $\Delta d_{\mathrm{F}} = - \Delta N$, this simplifies to,   
\begin{equation}
\label{eq:StatAnyonRel}
\Delta d_{\mathrm{SA}} = -p_{\mathrm{F}} \Delta N.
\end{equation}
Comparing Eq. \eqref{eq:StatAnyonRel} to Eq. \eqref{eq:GenAnyonRel} we see that the parameter $g$ in the GES framework is identical to the fermion probability in the statistical anyon framework. This demonstrates that the statistical anyon framework is fully equivalent to GES anyons.

For bookkeeping ease, we conclude this section by comparing the features of topological, statistical, and GES anyons in Table \ref{table:AnyonComp}.
\begin{table*}
\caption{\label{table:AnyonComp} Comparison of the features and behavior of topological, statistical, and GES anyons.}
\begin{ruledtabular}
	\begin{tabular}{lllll}
	\textbf{\,} & \textbf{Dim.} & \textbf{Rep. Group} & \textbf{Exclusion Principle} & \textbf{Origin of Anyonic Behavior}\\
	\, & \, & \, & \, & \, \\
	\textbf{Topological} & 2D & Braid (abelian) & Hard core for all $\nu \neq 0$ & Complex phase introduced by exchange\\
	\, & \, & \, & \, & \, \\
	\textbf{Statistical} & Arbitrary & Permutation & Generalized exclusion & Statistical average of bosonic and fermionic behavior\\
	\, & \, & \, & \, & \, \\
	\textbf{GES} & Arbitrary & Permutation & Generalized exclusion & Generalization of state occupancy arising from \\
	\, & \, & \, & \, & interparticle interactions\\
	\end{tabular}
\end{ruledtabular}
\end{table*} 
 
\section{Equilibrium Thermodynamics of Anyons}
\label{sec:3}
\subsection{1D Statistical Anyons}
The first step in understanding the thermodynamics of anyonic systems is to examine how the thermodynamic quantities such as internal energy, entropy, and heat capacity depend on the anyonic phase. To determine this, we need the proper partition function for our system. Motivated by the manifestation of GES in trapped, interacting gasses, let us consider a gas of two statistical anyons in a 1D harmonic potential. However, in this case we will consider the particles to be \textit{non-interacting}. Trapped boson-fermion mixtures have seen previous study both theoretically \cite{Fang2011, Dehkharghani2017, Decamp2017} and experimentally \cite{Fukuhara2009, Onofrio2016, Sowinski2019}, but this work has focused primarily on effects on the ground state configurations that arise from interactions between the boson and fermion species. In the statistical anyon framework we instead consider the average collective behavior that arises in the ideal gas limit of such a mixture.          

The Hamiltonian for our system of harmonically confined non-interacting statistical anyons reads, 
\begin{equation}
\label{eq:1DHamil}
H = \frac{p_1^2+p_2^2}{2m}+\frac{1}{2} m \omega^2 (x_1^2+x_2^2).
\end{equation}  
In the statistical anyon framework the partition function for a pair of anyons evolving under this Hamiltonian is,
\begin{equation}
\label{eq:partition}
Z_{\mathrm{SA}} = (Z_{\mathrm{B}})^{p_{\mathrm{B}}} (Z_{\mathrm{F}})^{1-p_{\mathrm{B}}},
\end{equation} 
 where $Z_{\mathrm{B}}$ is the partition function for two bosons and $Z_{\mathrm{F}}$ is the partition function for two fermions.  See Appendix \ref{Appendix A} for a full derivation of the partition function. Note that Eq. \eqref{eq:partition} is identical to that of a GES Calogero-Sutherland model gas \cite{Murthy19942}. This is further demonstration of the equivalence of statistical and GES anyons, however, we see that for statistical anyons the anyonic behavior arises purely out of the average properties, rather than from an interaction term in the Hamiltonian.     
\begin{figure*}
	\centering
	\subfigure[]{
		\includegraphics[width=.22\textwidth]{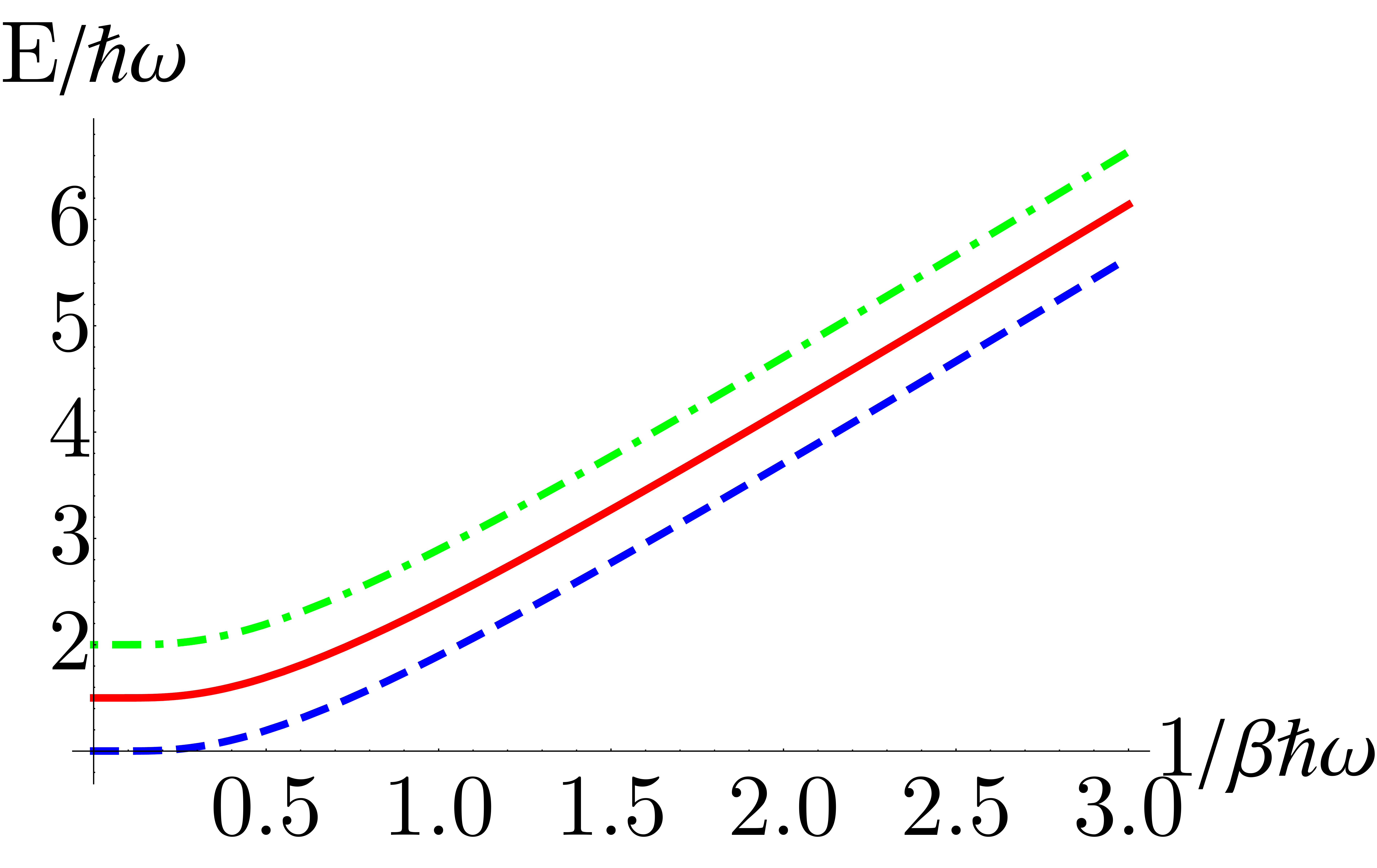}
	}
	\subfigure[]{
		\includegraphics[width=.22\textwidth]{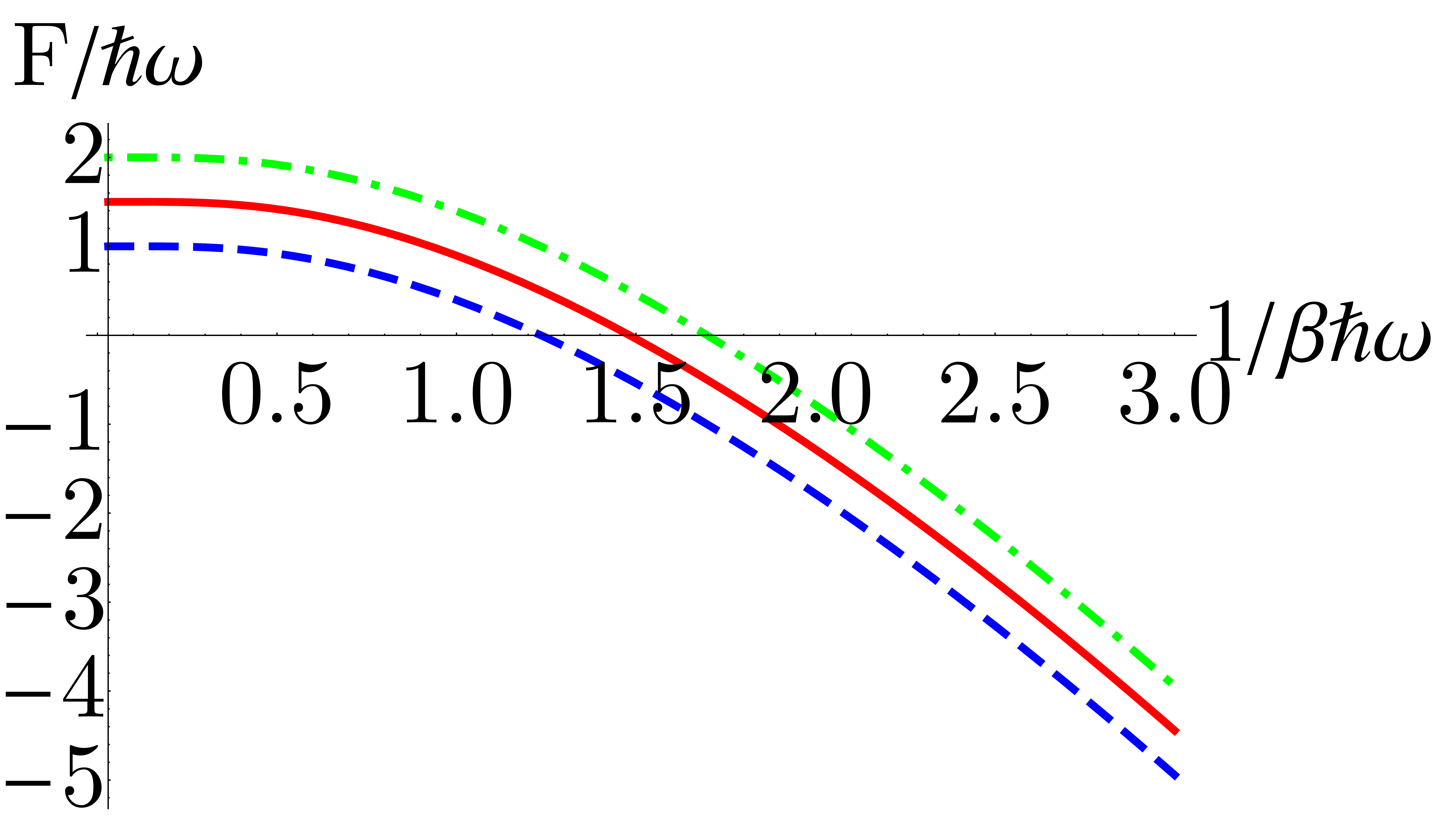}
	}
	\subfigure[]{
		\includegraphics[width=.22\textwidth]{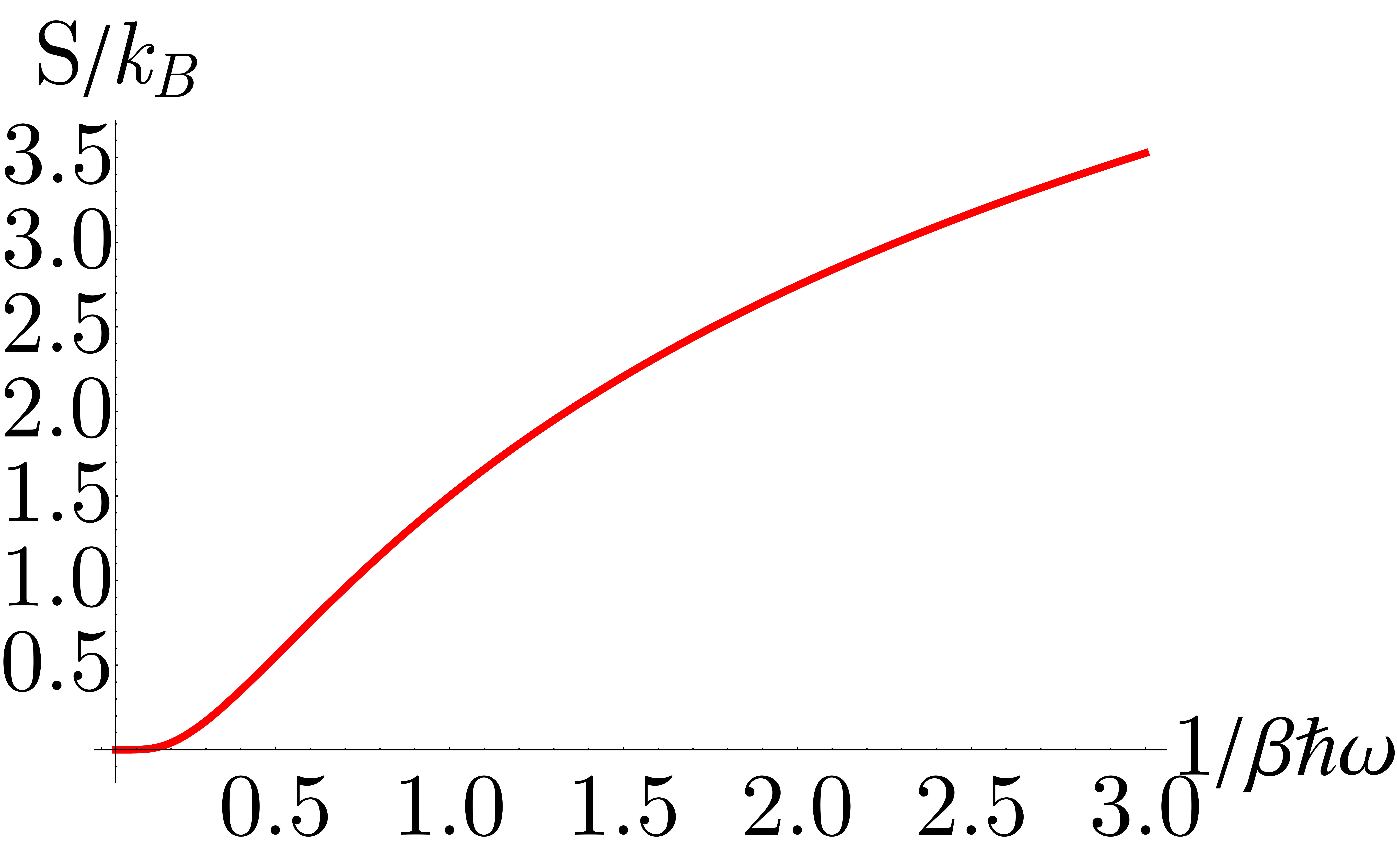}
	}
	\subfigure[]{
		\includegraphics[width=.22\textwidth]{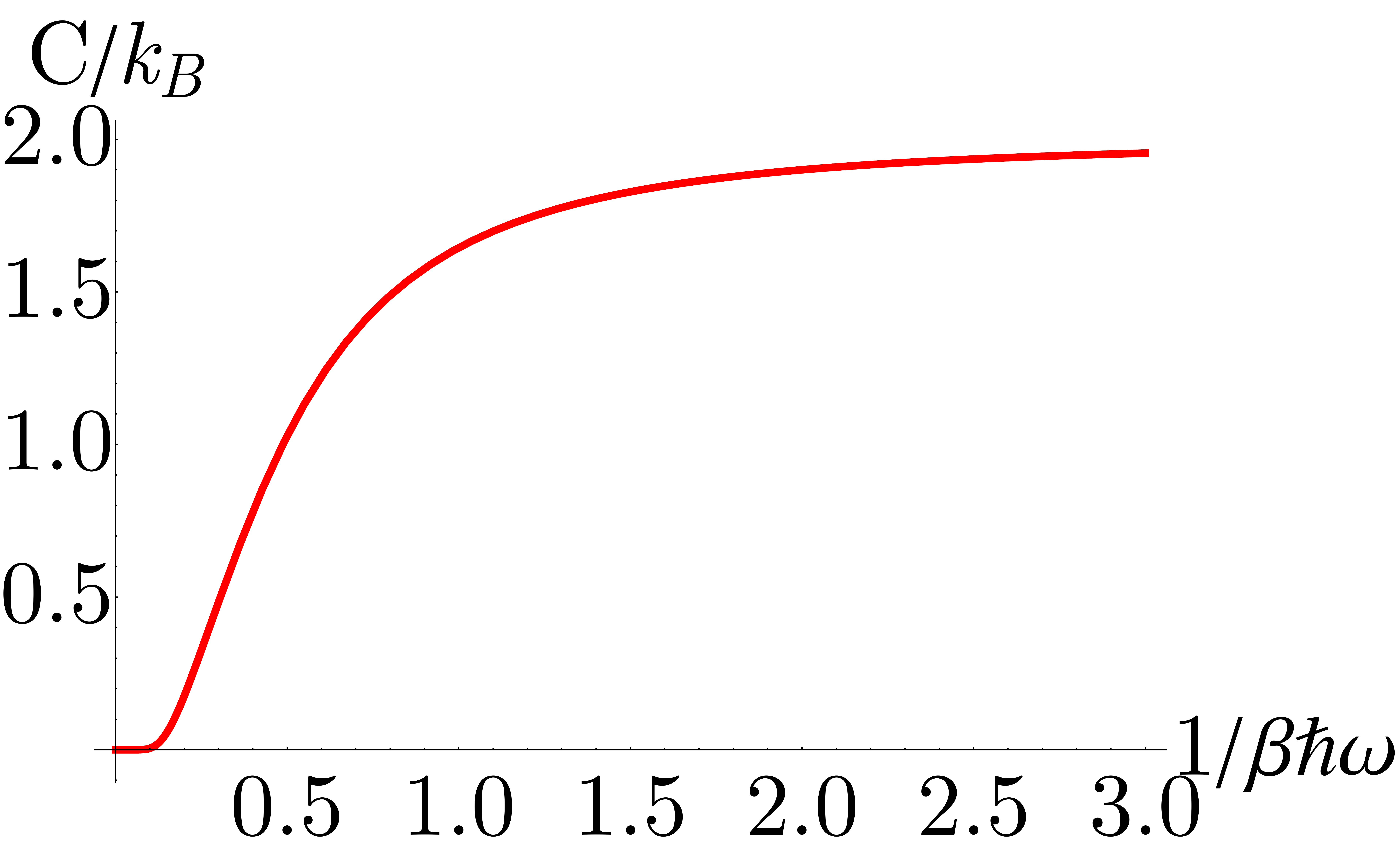}
	}
	\caption{\label{fig:thermo_quant} Equilibrium (a) internal energy, (b) free energy, (c) entropy, and (d) heat capacity for two statistical anyons in one dimension with anyonic phase corresponding to $p_{\mathrm{B}} = 1$ (blue, dashed), $p_{\mathrm{B}} = 1/2$ (red, solid), and $p_{\mathrm{B}} = 1$ (green, dot-dashed).}
\end{figure*} 
 
Using Eq. \eqref{eq:partition} we can determine the internal energy, free energy, entropy, and heat capacity as follows, 
\begin{align}
\label{eq:thermofunc}
& E = - \frac{\partial }{\partial \beta} \ln(Z_{\mathrm{SA}}), \qquad \,
F = - \frac{1}{\beta} \ln(Z_{\mathrm{SA}}), \\
& S = k_{\mathrm{B}} \beta^2 \frac{\partial F}{\partial \beta}, \qquad \quad \,
C = -k_{\mathrm{B}} \beta^2 \frac{\partial E}{\partial \beta}, \nonumber
\end{align}
where $\beta$ is the inverse temperature and $k_{\mathrm{B}}$ is Boltzmann's constant. Plugging Eq. \eqref{eq:partition} into Eq. \eqref{eq:thermofunc} we find, 
\begin{align}
\label{eq:1DThermoFunc}
& E = \frac{1}{2} \hbar \omega \left[ 3 \coth (\beta \hbar \omega)+\text{csch}(\beta \hbar \omega)-2 p_{\mathrm{B}}+1 \right] \nonumber \\
& F = \frac{1}{\beta} \ln\left[\frac{1}{8} \text{csch}^2(\frac{\beta \hbar \omega}{2})-\frac{1}{4}\text{csch}(\beta \hbar \omega)\right] - p_{\mathrm{B}} \hbar \omega \nonumber \\
& S = \frac{1}{2} k_{\mathrm{B}} \beta \hbar \omega \left[3 \coth(\beta \hbar \omega) + \text{csch}(\beta \hbar \omega)+1\right] \\
& \quad \,\,\,\,\, + k_{\mathrm{B}} \ln \left[\frac{1}{8} \text{csch}^2(\frac{\beta \hbar \omega}{2})-\frac{1}{4}\text{csch}(\beta \hbar \omega)\right] \nonumber \\
& C = \frac{1}{2} k_{\mathrm{B}} \beta^2 \hbar^2 \omega^2 \text{csch}^2(\beta \hbar \omega) \left[\cosh(\beta \hbar \omega)+3\right] \nonumber
\end{align} 

Plots of each as a function of temperature are shown in Fig. \ref{fig:thermo_quant}. We see that both the internal energy and free energy are shifted by a constant proportional to $p_{\mathrm{B}}$. Physically this offset arises from the generalized exclusion principle, as outlined in Section \ref{sec:2}. The constant shifts the lowest energy state of the system from the bosonic limit, with both particles in the ground state of the oscillator, to the fermionic limit, with one particle in the ground state and the other in the first excited state. 

We find that the dependence on $p_{\mathrm{B}}$ cancels out exactly in the entropy and heat capacity, leaving them independent of the anyonic phase. This is expected, as in the thermal equilibrium state both fermions and bosons have an equivalent, countably infinite, number of available states. Note that the behavior of the heat capacity is consistent with that of a one-dimensional, two-oscillator Einstein solid \cite{Simon}. However, since topological anyons do not exist in one dimension, in order to properly compare their thermodynamic behavior to that of statistical anyons, we must extend the above analysis to two dimensions.

\subsection{2D Statistical Anyons}

\begin{figure*}
	\centering
	\subfigure[]{
		\includegraphics[width=.22\textwidth]{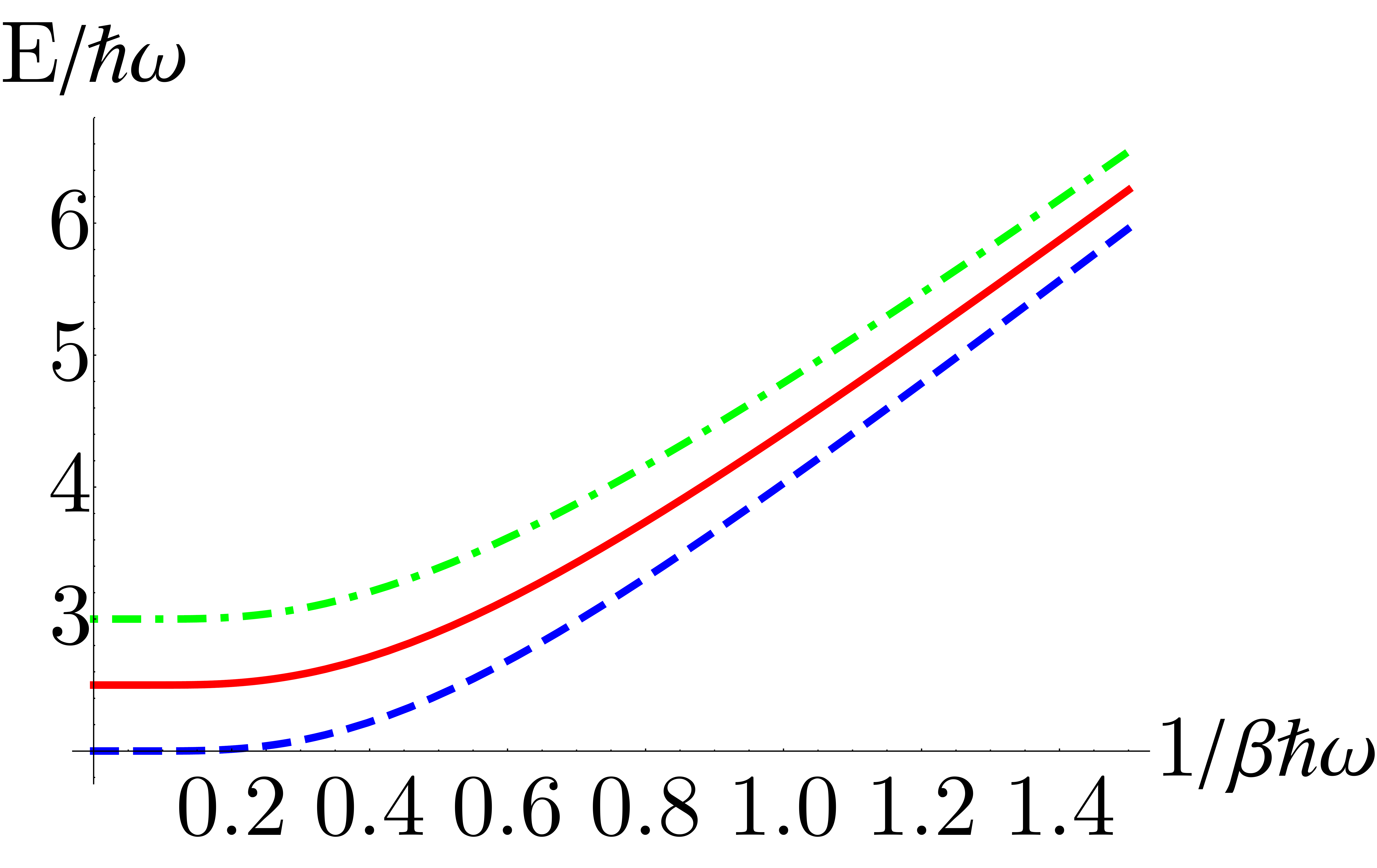}
	}
	\subfigure[]{
		\includegraphics[width=.22\textwidth]{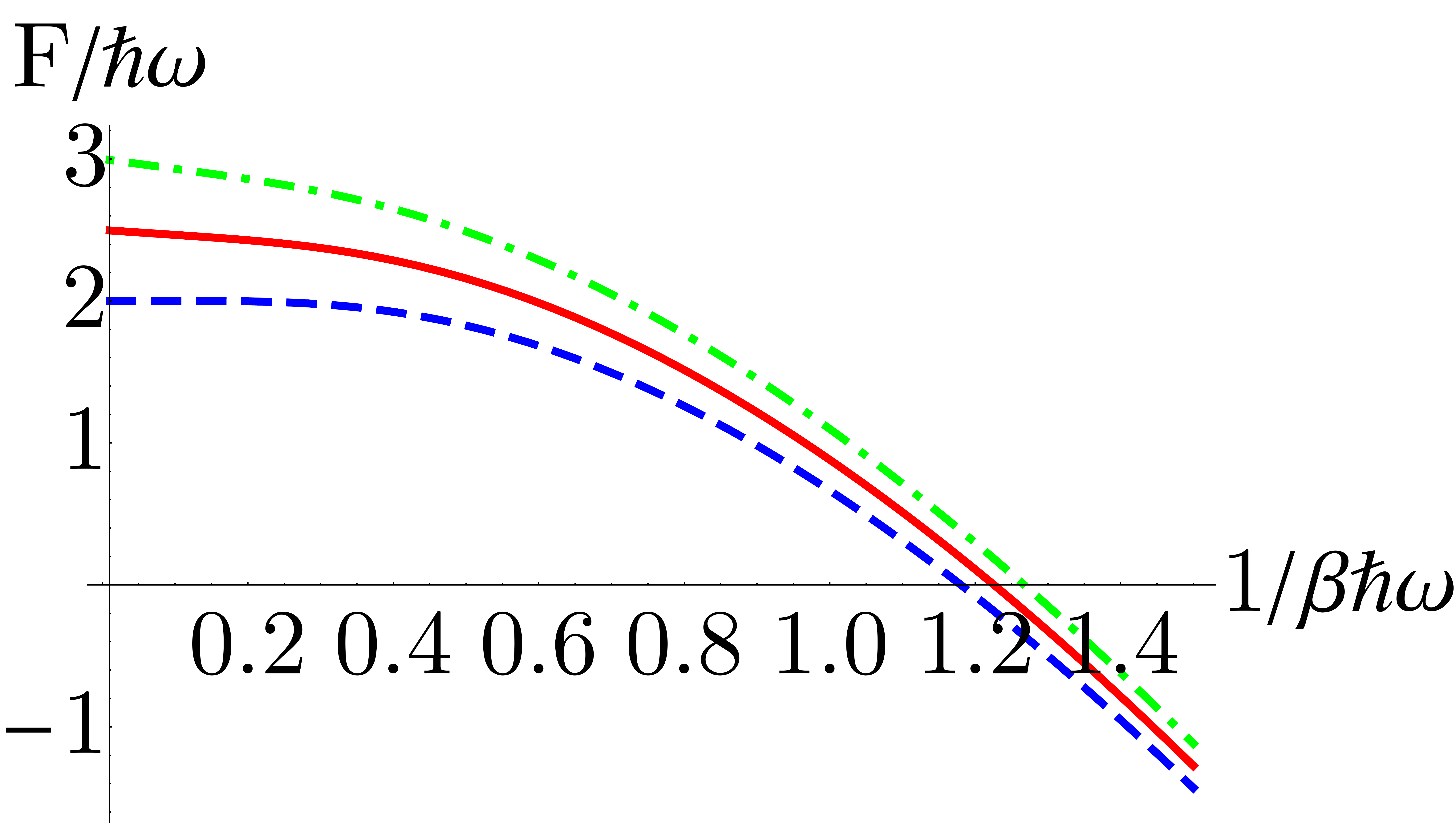}
	}
	\subfigure[]{
		\includegraphics[width=.22\textwidth]{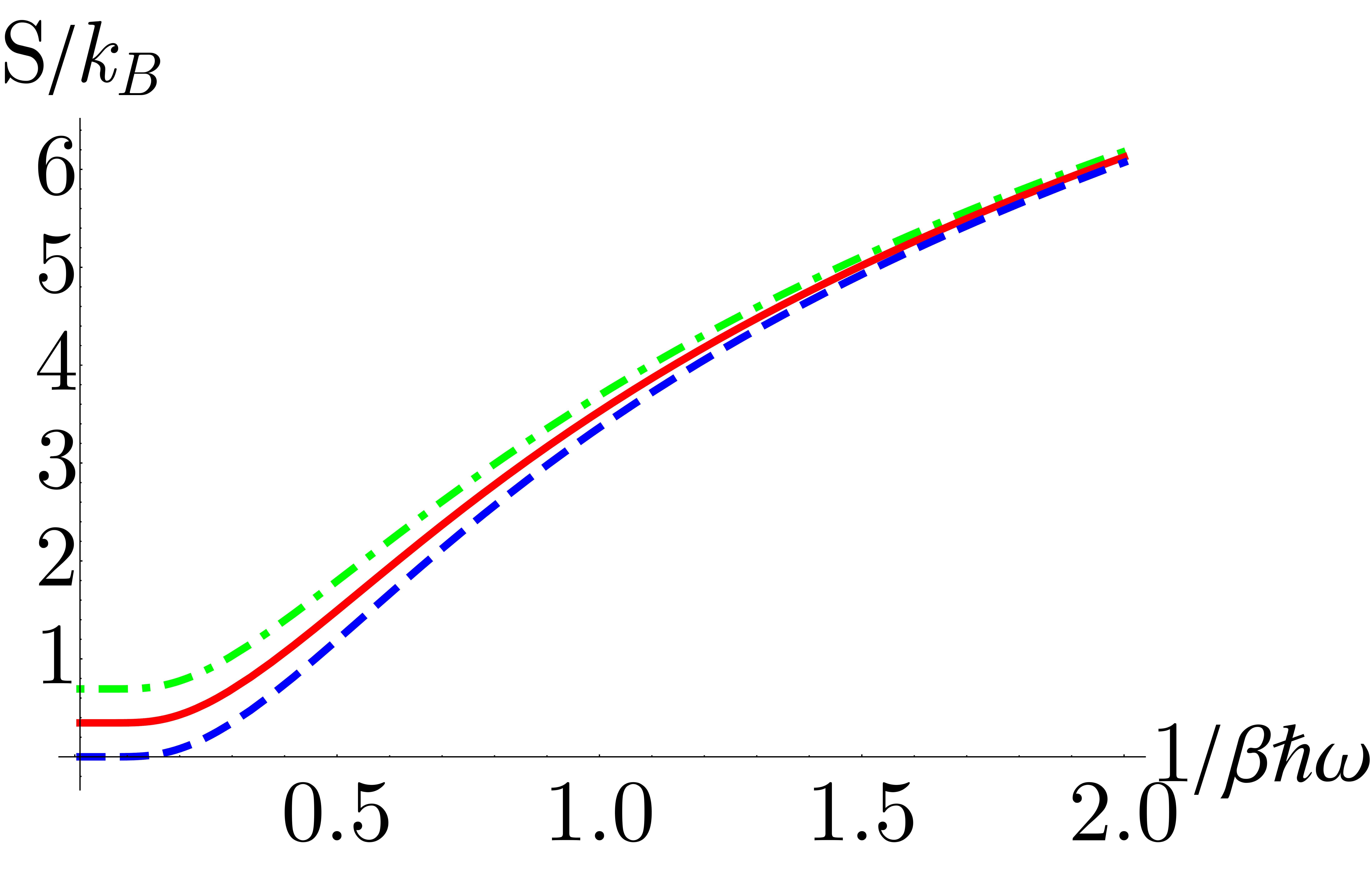}
	}
	\subfigure[]{
		\includegraphics[width=.22\textwidth]{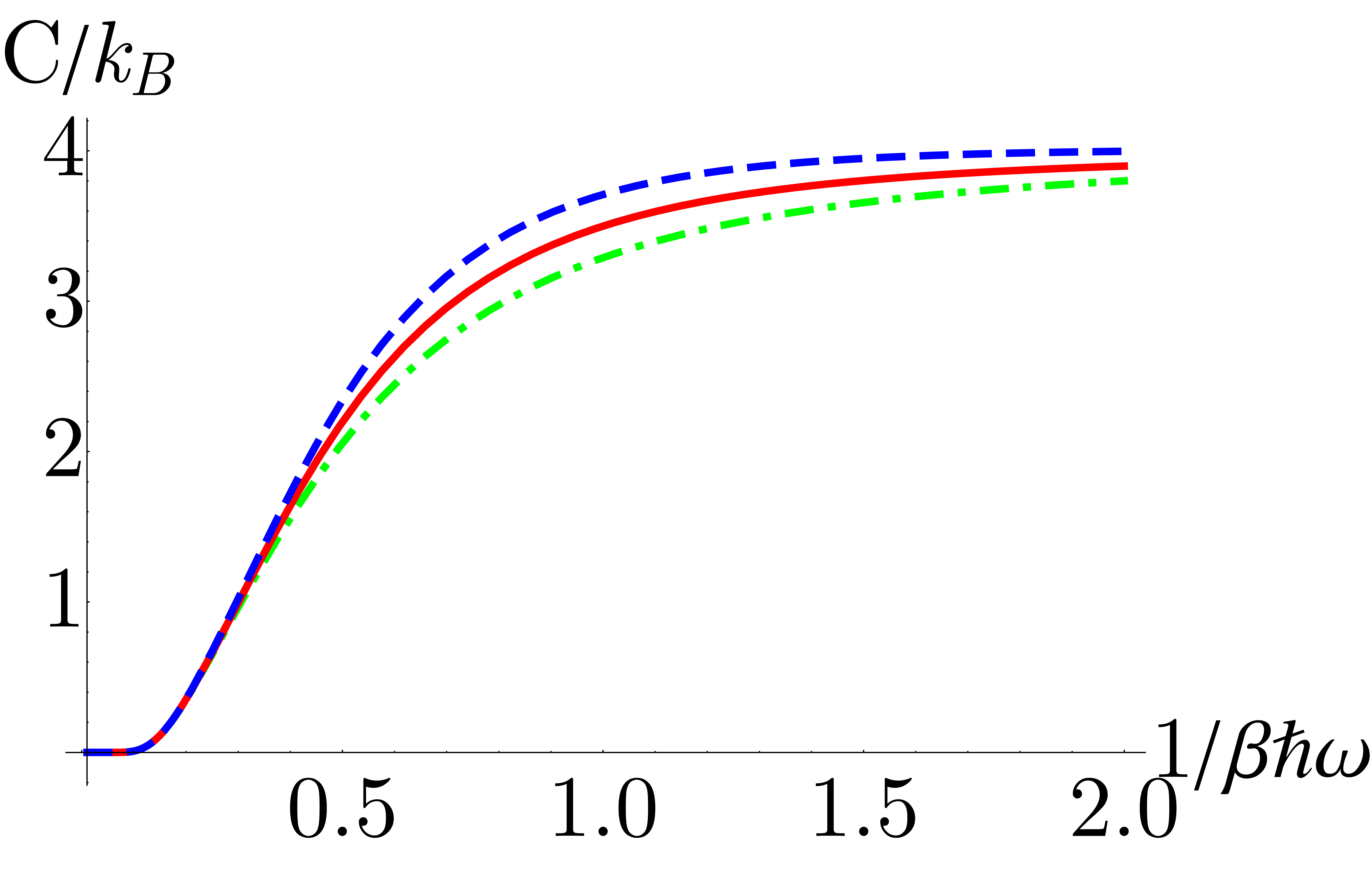}
	}
	\caption{\label{fig:thermo_quant_SA2D} Equilibrium (a) internal energy, (b) free energy, (c) entropy, and (d) heat capacity for two statistical anyons in two dimensions with anyonic phase corresponding to $p_{\mathrm{B}} = 1$ (blue, dashed), $p_{\mathrm{B}} = 1/2$ (red, solid), and $p_{\mathrm{B}} = 1$ (green, dot-dashed).}
\end{figure*}

We can repeat the thermodynamic analysis for two statistical anyons in a two-dimensional harmonic potential. To avoid clutter, we give the full expressions for the internal energy, free energy, entropy, and heat capacity in Appendix \ref{Appendix B}. We plot each as a function of temperature in Fig. \ref{fig:thermo_quant_SA2D}. In contrast to the one-dimensional case, we see that now the entropy and heat capacity do depend on $p_{\mathrm{B}}$. The origin of this difference is clear if we think of the entropy for bosons and fermions in the zero-temperature limit. For bosons, only one configuration is available -- both particles in the ground state in both dimensions. However, the fermion ground state is degenerate. The Pauli exclusion principle requires that one particle must be in the ground state and the other in the first excited state, but the excited state can be in either dimension. This results in a non-zero entropy at $T=0$. The entropy of generic statistical anyons interpolates smoothly between the boson and fermion limits as $p_{\mathrm{B}}$ changes from one to zero.        

\subsection{2D Topological Anyons} 

\begin{figure*}
	\centering
	\subfigure[]{
		\includegraphics[width=.22\textwidth]{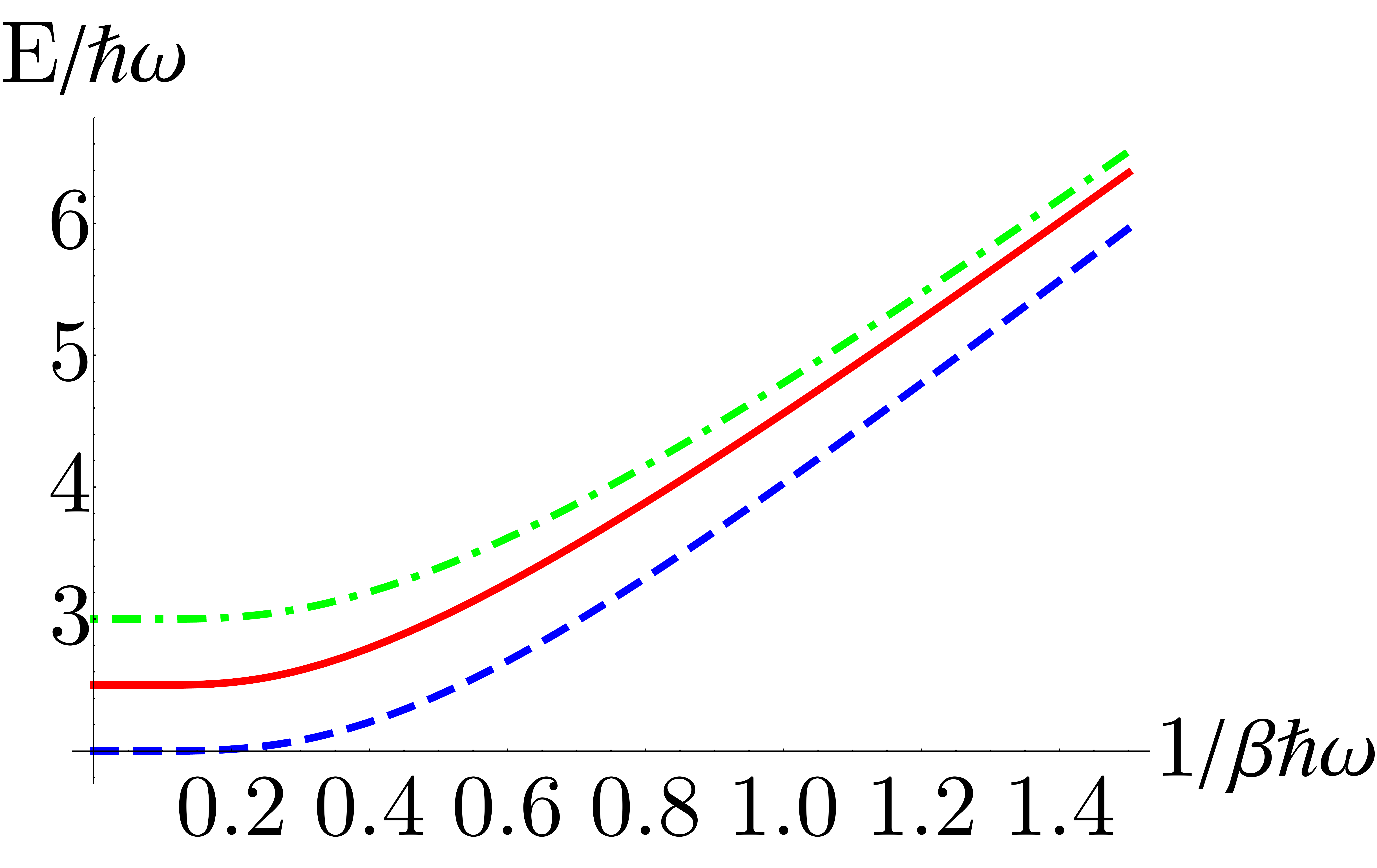}
	}
	\subfigure[]{
		\includegraphics[width=.22\textwidth]{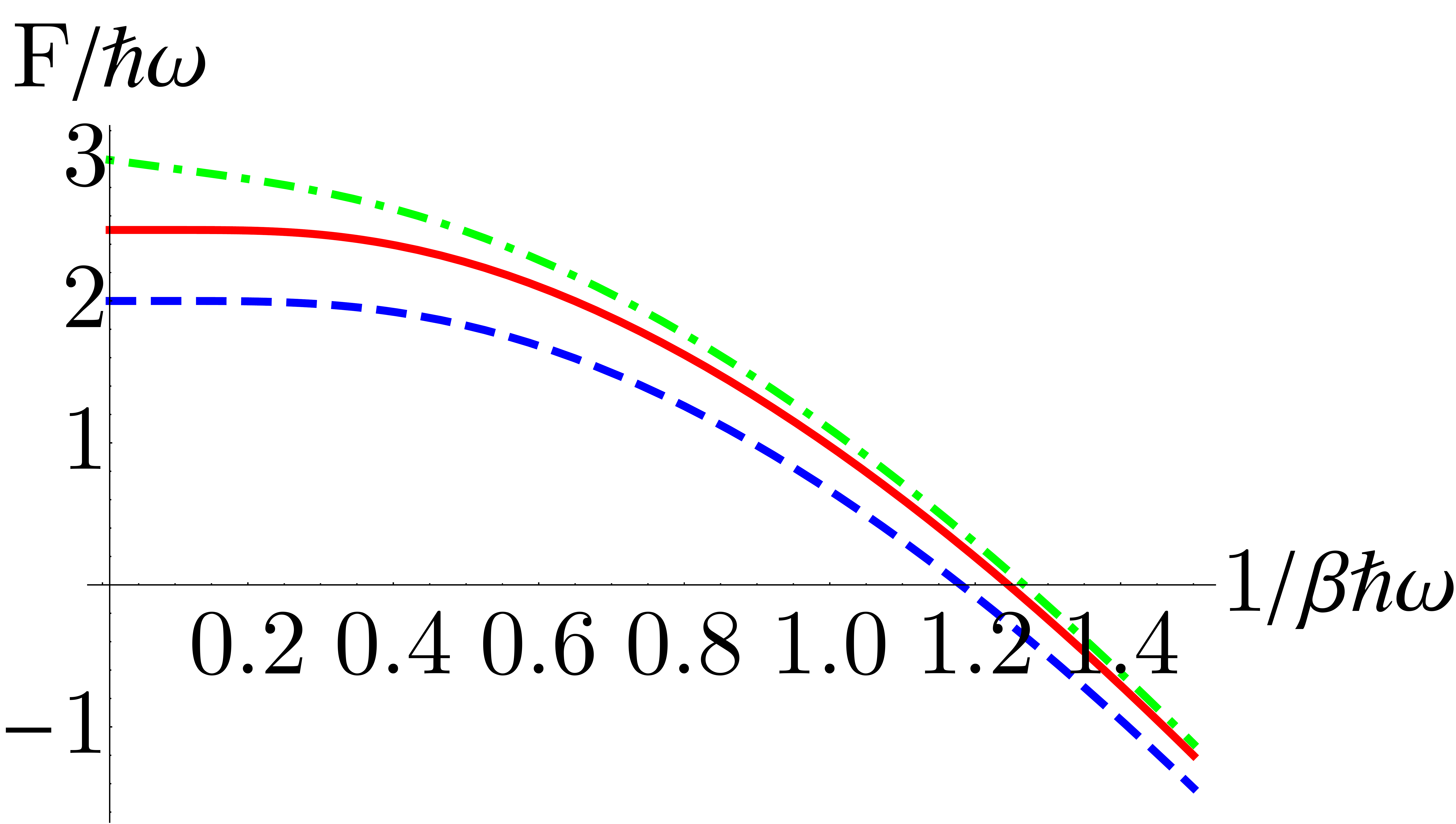}
	}
	\subfigure[]{
		\includegraphics[width=.22\textwidth]{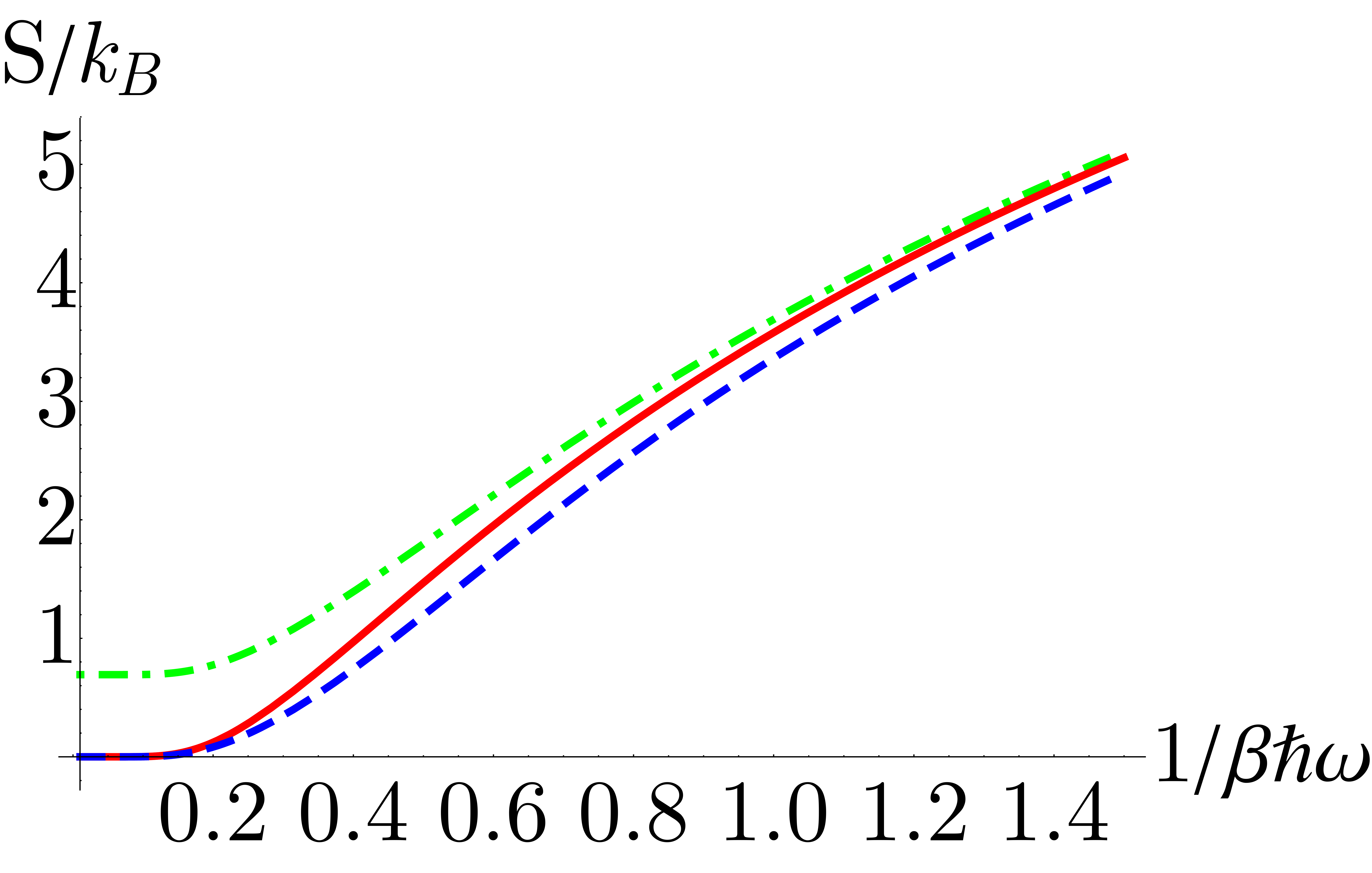}
	}
	\subfigure[]{
		\includegraphics[width=.22\textwidth]{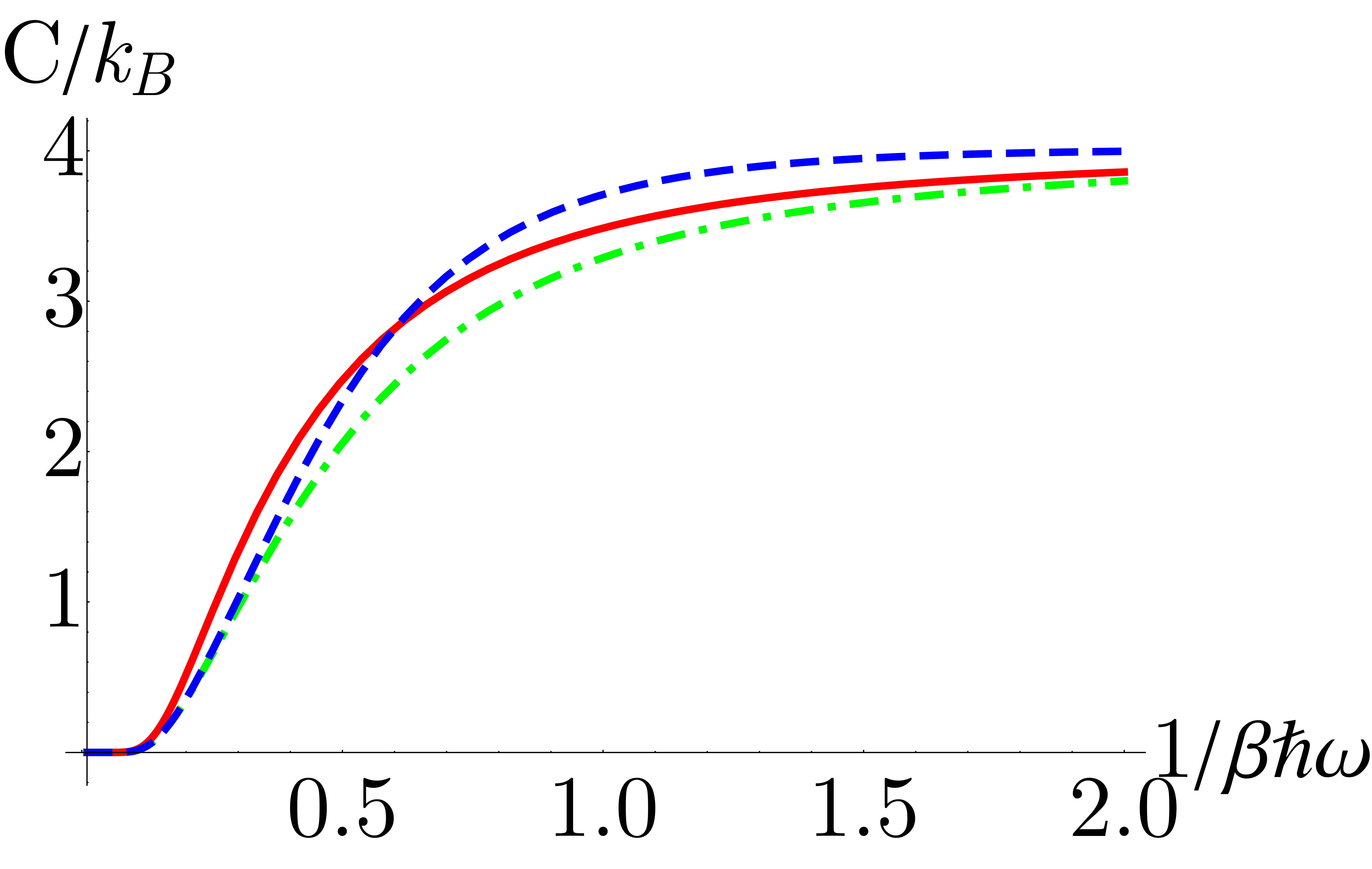}
	}
	\caption{\label{fig:thermo_quant_true} Equilibrium (a) internal energy, (b) free energy, (c) entropy, and (d) heat capacity for two topological anyons with anyonic phase corresponding to $\nu = 0$ (blue, dashed), $\nu = 1/2$ (red, solid), and $\nu = 1$ (green, dot-dashed).}
\end{figure*}
              
The partition function for two topological anyons in a two-dimensional harmonic potential has been previously derived (see for example \cite{Lerda1992, Myrheim1999}). It is given by,
\begin{equation}
\label{eq:partitionTrue}
Z_{\mathrm{TA}}=\frac{e^{-\beta \hbar \omega(2+\nu)}+e^{-\beta \hbar \omega(4-\nu)}}{(1-e^{-\beta \hbar \omega})^2(e^{-2 \beta \hbar \omega})^2},
\end{equation}
where $\nu$ is the anyonic phase. The first thing we note is that, unlike the statistical anyon/GES anyon partition function, the topological anyon partition function cannot be expressed as a function of the boson and fermion partition functions. This arises from the increased complexity of the topological anyon wave function. In the two anyon problem, the the anyonic phase can be considered as a shift in the value of the relative motion angular momentum \cite{Khare2005}. However, accounting for the fact that the phase depends on the \textit{direction} of rotation (a consideration unique to the braid group-based topological anyons) leads to a multi-valued wave function \cite{Myrheim1999}. Accounting for both branches of the wave function results in the two separate anyonic phase-dependent terms seen in the numerator of Eq. \eqref{eq:partitionTrue}.         

Following the same process as for the statistical anyons, we determine the equilibrium internal energy, free energy, entropy, and heat capacity for the two anyon system, with the full expressions given in Appendix \ref{Appendix B}. Plots of each as a function of temperature are given in Fig. \ref{fig:thermo_quant_true}. Here we see qualitatively similar behavior to the two-dimensional statistical anyons, with some notable differences. We see that the entropy of topological anyons converges to zero in the zero-temperature limit, indicating the existence of a unique ground state configuration for all values of $\nu$ except $\nu =1$, corresponding to pure fermions. Another significant discrepancy is seen in the behavior of the heat capacity. For intermediate values of the anyonic phase we see that, at low temperatures, the topological anyon heat capacity is higher than that of both the bosonic and fermionic values.

Comparing Figs. \ref{fig:thermo_quant_SA2D} and \ref{fig:thermo_quant_true} we see that, in each plot, for the statistical anyons the line corresponding to $p_{\mathrm{B}} = 1/2$ remains evenly spaced between the bosonic and fermionic limits, while for topological anyons the line corresponding to $\nu = 1/2$ bends more towards the fermionic behavior. This difference has its origin in the complicated energy spectrum that arises from the topological anyons' ``hard-core" exclusion principle and multivalued wave function. For topological anyons in a harmonic potential, the energy, degeneracy, and level spacing all depend on $\nu$ \cite{Myrheim1999, Khare2005}. In the ground state, the energy eigenvalues corresponding to the two branches of the wave function only coincide in the fermionic limit, making $\nu = 1$ the only degenerate ground state and giving rise to the observed zero-temperature limit of the entropy. However, as temperature increases, the hard-core nature of topological anyons biases their thermodynamic behavior more towards the fermionic limit. In contrast, the statistical anyon energy spectrum is a much simpler weighted average over the bosonic and fermionic spectrums, leading to thermodynamic behavior that is evenly spaced between the bosonic and fermionic limits for $p_{\mathrm{B}} = 1/2$. This has notable ramifications for thermodynamic applications of anyons, as statistical anyons will retain their ``intermediate" behavior at higher temperature regimes. 

\subsection{Thermodynamic Equivalence}

In this section we have seen that statistical and topological anyons display different thermodynamic behavior. This brings up the question: Is it possible to mimic the richer behavior of topological anyons using the mathematically and experimentally simpler framework of statistical anyons? 

To determine the relation between the topological anyon parameter $\nu$, and the statistical anyon parameter $p_{\mathrm{B}}$ we can set their partition functions equal to each other and solve for $p_{\mathrm{B}}$. This yields,
\begin{equation}
\label{eq:equiv}
p_{\mathrm{B}} = \frac{\ln \left[ \cosh ((\nu -1) \beta  \hbar \omega )\right]}{\ln \left[\cosh (\beta \hbar \omega )\right]}.
\end{equation}
We see that in order to capture the more complicated thermodynamic behavior $p_{\mathrm{B}}$ becomes dependent on the temperature and frequency parameters. If we take the high temperature limit of Eq. \eqref{eq:equiv} we find a simpler, parameter-independent relation, 
\begin{equation}
\label{eq:equivSimp}
p_{\mathrm{B}} = (\nu -1)^2.
\end{equation} 
We note that the approximate relation becomes exact for $\nu =0,1$, as the behavior of the statistical and topological anyons must converge to the same bosonic and fermionic limits. Notably, previous comparisons between Haldane's GES parameter and the topological anyon phase using the second Virial coefficient have also determined a quadratic polynomial relation \cite{Murthy1994}.   

\begin{figure*}
	\centering
	\subfigure[]{
		\includegraphics[width=.22\textwidth]{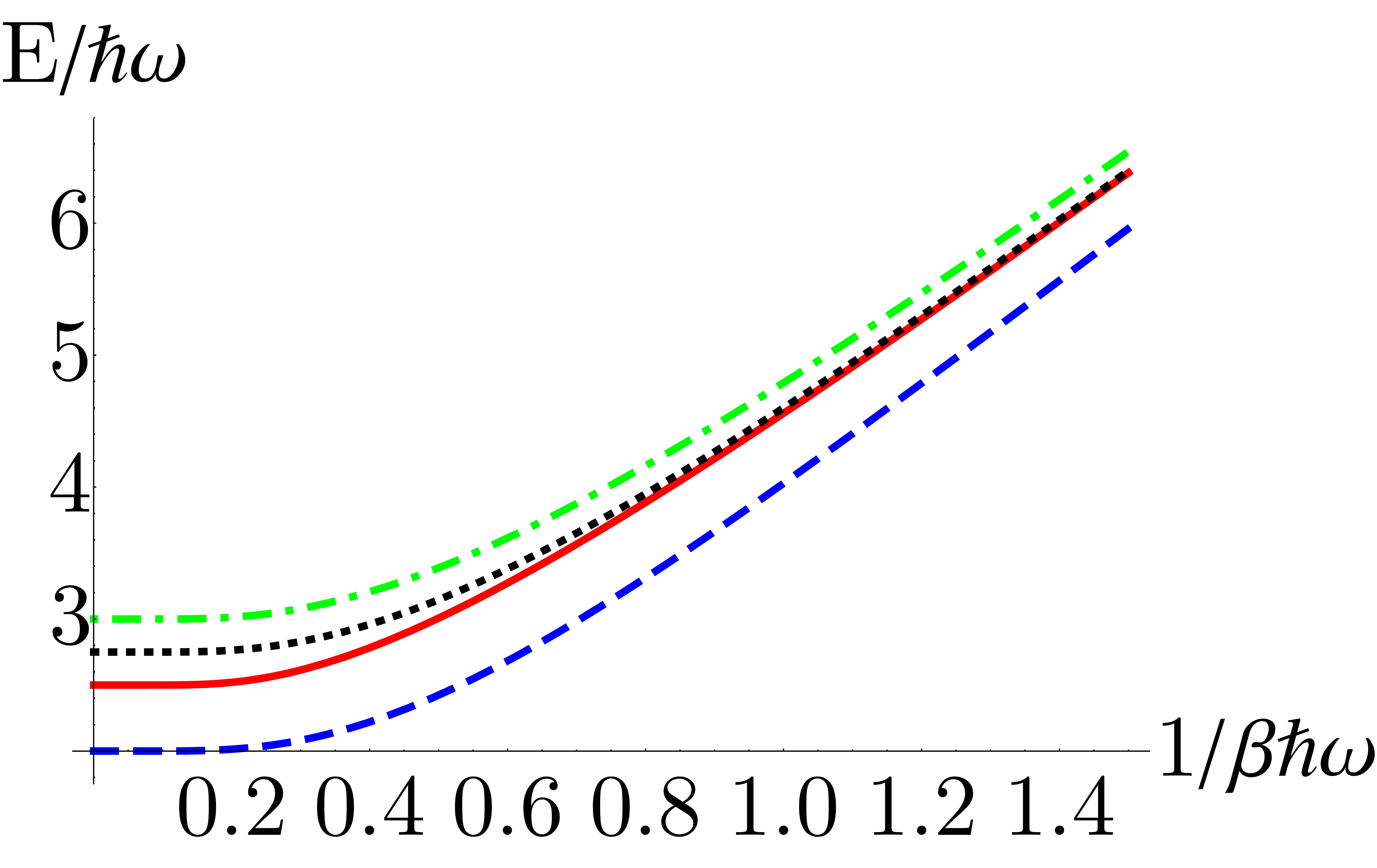}
	}
	\subfigure[]{
		\includegraphics[width=.22\textwidth]{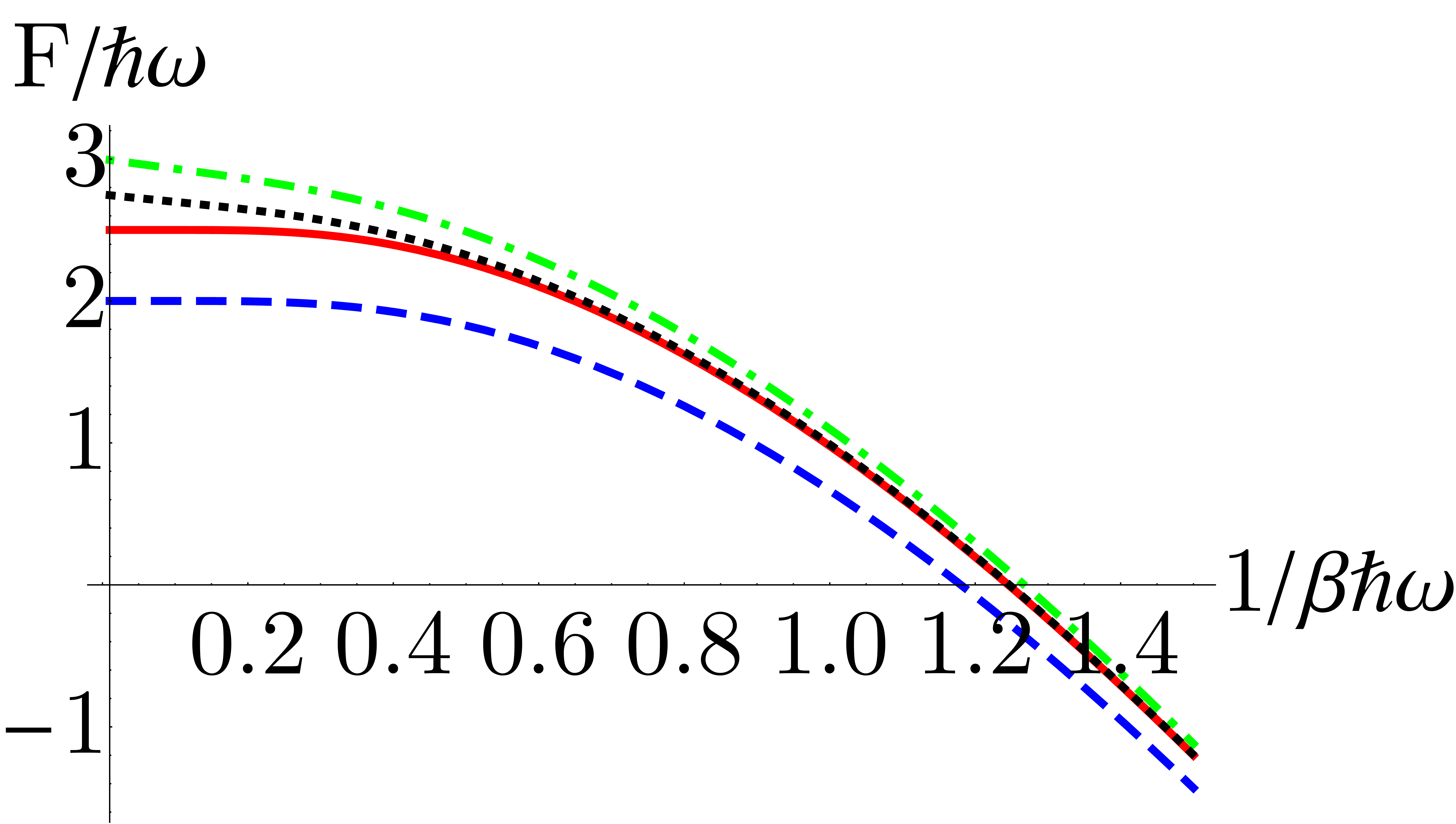}
	}
	\subfigure[]{
		\includegraphics[width=.22\textwidth]{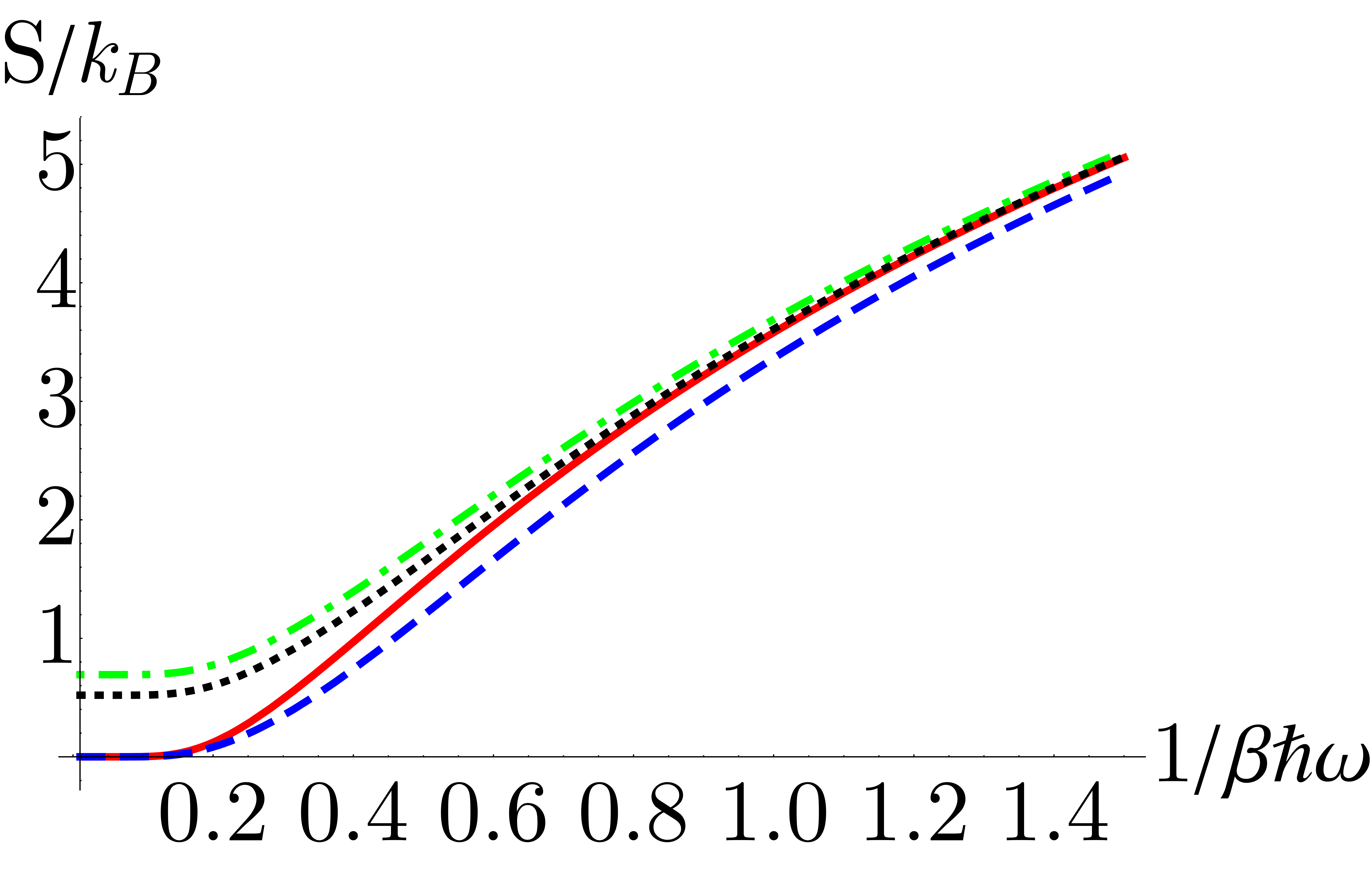}
	}
	\subfigure[]{
		\includegraphics[width=.22\textwidth]{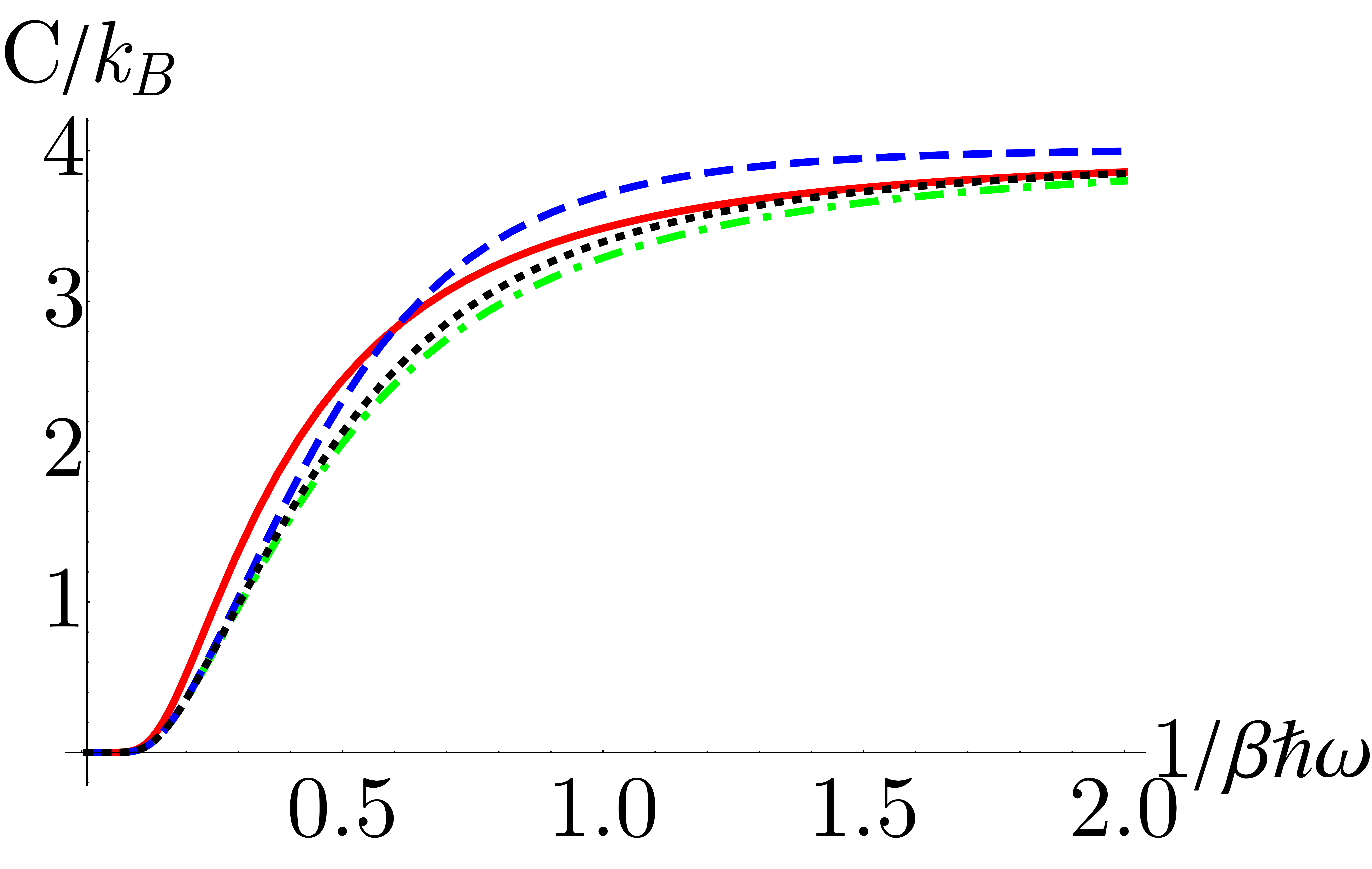}
	}
	\caption{\label{fig:thermo_quant_approx} Comparison of (a) internal energy, (b) free energy, (c) entropy, and (d) heat capacity of two topological anyons with phase parameter $\nu$, against two statistical anyons using phase parameter $p_{\mathrm{B}} = (\nu -1)^2$. The lines correspond to $\nu = 0, p_{\mathrm{B}} =1$ (blue, dashed), $\nu = 1, p_{\mathrm{B}} = 0$ (green, dot-dashed), $\nu = 1/2$ (red, solid), and $p_{\mathrm{B}} = 1/4$ (black, dotted).}
\end{figure*}

Plotting the internal energy, free energy, entropy and heat capacity for topological anyons and statistical anyons using Eq. \eqref{eq:equivSimp} in Fig. \ref{fig:thermo_quant_approx} we see that the statistical and topological anyon behavior rapidly converges. We note that the parameter-independent approximation does not capture the low-temperature behavior of the topological anyon heat capacity. To fully imitate this behavior, we must use Eq. \eqref{eq:equiv}.

The ability to mimic the thermodynamic properties of topological anyons using statistical anyons has important ramifications from an experimental standpoint. The difficulty of detecting and manipulating topological anyons in two-dimensional materials makes probing their thermodynamics exceedingly challenging. Statistical anyons provide an straightforward alternative, both as a model to test thermodynamic control of topological anyons or as a replacement in applications that would rely on their thermodynamics properties.         

\section{Endoreversible Anyonic Engine}
\label{sec:4}
Having established the equilibrium thermodynamic behavior of both statistical and topological anyons, we now continue our exploration of their behavior in the context of heat engines. Studying the thermodynamic properties of a system using the framework of cyclic heat engines has a rich tradition as old as thermodynamics itself \cite{Deffner2019}. In equilibrium thermodynamics the optimal efficiency of any heat engine cycle is bounded by the Carnot efficiency, regardless of the properties of the working medium \cite{Callen}. However, this efficiency is obtained in the limit of infinitely slow, quasistatic strokes, resulting in zero power output. A figure of merit of more practical use, the \textit{efficiency at maximum power} (EMP), was introduced by Curzon and Ahlborn using the framework of \textit{endoreversible thermodynamics} \cite{Curzon1975, Rubin1979, Hoffmann1997}. Curzon and Ahlborn found the EMP of a endoreversible Carnot engine to be,
\begin{equation}
\eta_{\mathrm{CA}} = 1 - \sqrt{\frac{T_c}{T_h}}
\end{equation}  
where $T_c$ ($T_h$) is the cold (hot) reservoir temperature \cite{Curzon1975}. 

In endoreversible thermodynamics the system is assumed to be in a state of \textit{local equilibrium} at all times, but with dynamics that occur quickly enough that \textit{global equilibrium} with the environment is not achieved. This results in a process that is locally reversible, but globally irreversible \cite{Hoffmann1997}. It has been shown that the performance of a quantum Otto engine is dependent on the stroke protocol \cite{Feldmann1996, Rezek2006, Feldmann2012, Campo2014, Beau2016, Zheng2016, Abah2017, Abah2018, Chen2019, Chen20192, Bonanca2019, Abah2019, Funo2019, Lee2020} and the nature of the working medium \cite{Uzdin2014, Pena2014, Zhang2014, Zheng2015, Jaramillo2016, Pena2017, Huang2017, Deffner2018, Li2018, Kloc2019, YungerHalpern2019, Pena2019, Myers2020, Watanabe2020, Pena2020}, with the EMP in particular being determined by the form of the fundamental relation of the working medium \cite{Smith2020}. In Ref. \cite{Deffner2018} it was shown that the EMP of an endoreversible quantum Otto engine with a single particle working medium can exceed the Curzon-Ahlborn efficiency. It is of interest then to examine the role of quantum statistics in the operation of such an engine.

In the following analysis we closely follow the method established in Ref. \cite{Deffner2018}. Let us consider a working medium of two anyons in a harmonic potential, evolving under the Hamiltonian given in Eq. \eqref{eq:1DHamil} with a time-dependent frequency. The Otto cycle consists of four strokes summarized graphically in Fig.~\ref{fig:cycle}: 

\begin{figure}
	\includegraphics[width=.48\textwidth]{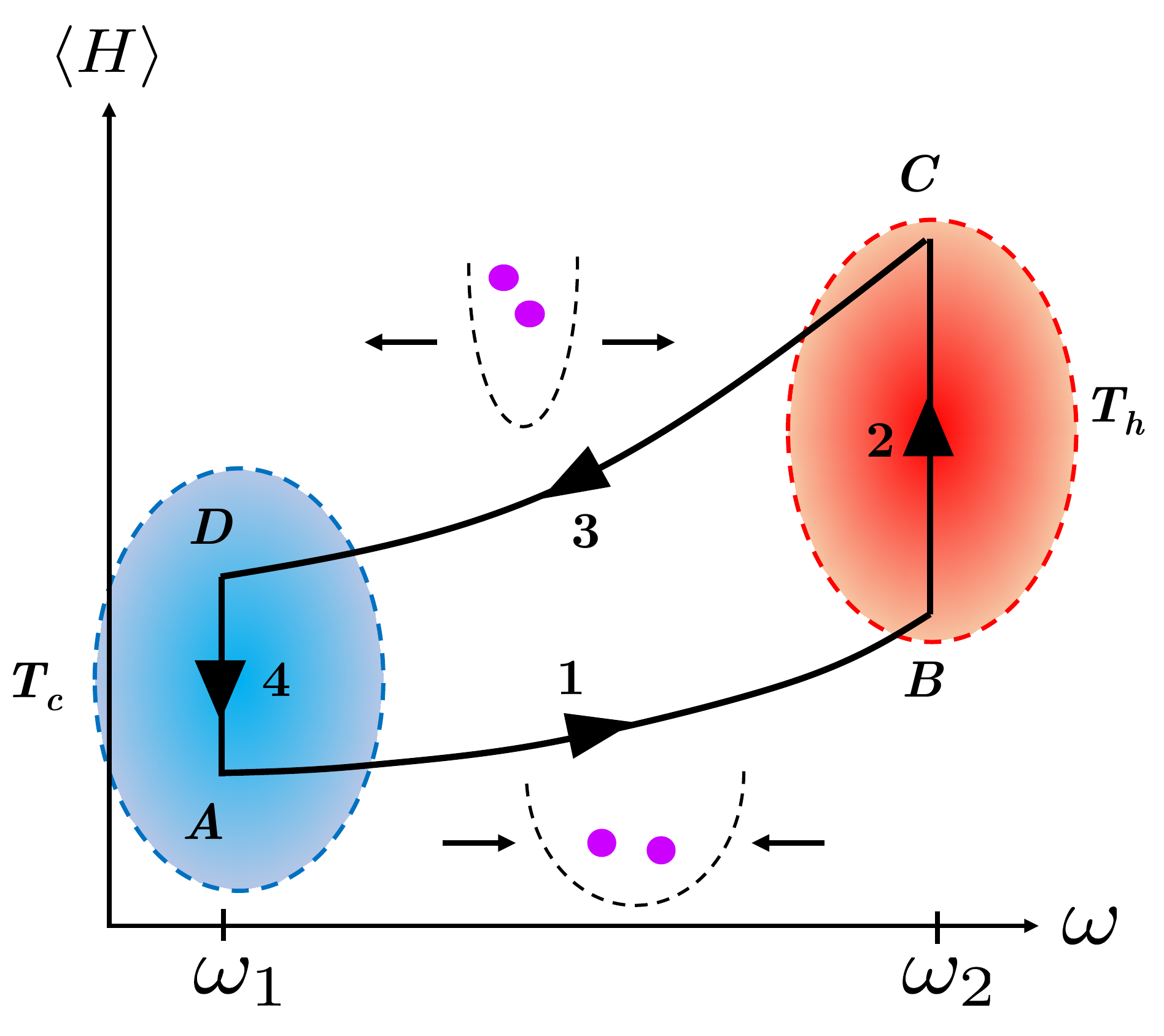}
	\caption{\label{fig:cycle} Energy-frequency diagram of a quantum Otto cycle for a harmonic trapping potential with a working medium of two anyons.}
\end{figure}

(1) \textit{Isentropic compression} \\
During this stroke the working medium remains in a state of constant entropy, exchanging no heat with the environment. Using the first law $\Delta E = Q + W$ we can identify the change in internal energy completely with work,
\begin{equation}
\label{eq:Wcomp}
W_{\mathrm{comp}} = E(T_B, \omega_2) - E(T_A, \omega_1).
\end{equation}  

(2) \textit{Isochoric Heating} \\ 
During this stroke the externally-controlled work parameter (the trap frequency in the case of the harmonic engine) is held constant, resulting in zero work. By the first law we can then identify the change in internal energy completely with heat,      
\begin{equation}
\label{eq:Qh}
Q_{h} = E(T_C, \omega_2) - E(T_B, \omega_2).
\end{equation}
Recalling the conditions of endoreversibility we note that the working medium does not fully thermalize with the hot reservoir during this stroke, giving us the condition $T_B \le T_C \leq T_h$ \cite{Curzon1975}. The change in temperature during the stroke depends on the properties of the working medium can be determined using Fourier's law \cite{Callen},    
\begin{equation}
\label{eq:fh}
\frac{dT}{dt} = -\alpha_h (T(t)-T_h),
\end{equation}
where $\alpha_h$ is a constant determined by the heat capacity and thermal conductivity of the working medium.

(3) \textit{Isentropic expansion} \\
In exactly the same manner as the compression stroke, we can identify the change in internal energy during the expansion with work,
\begin{equation}
\label{eq:Wexp}
W_{\mathrm{exp}} = E(T_D, \omega_1) - E(T_C, \omega_2).
\end{equation} 

(4) \textit{Isochoric Cooling} \\
As in the heating stroke, we identify the change in internal energy during this stroke with heat,
\begin{equation}
Q_{c} = E(T_A, \omega_1) - E(T_D, \omega_1).
\end{equation}
The temperature change can again be determined from Fourier's law,
\begin{equation}
\label{eq:fc}
\frac{dT}{dt} = -\alpha_c (T(t)-T_c),
\end{equation}
where $T_D > T_A \geq T_c$.

The efficiency of the engine is given by the ratio of the total work and the heat exchanged with the hot reservoir,
\begin{equation}
\label{eq:eff}
\eta = -\frac{W_{\mathrm{comp}}+W_{\mathrm{exp}}}{Q_h},
\end{equation}
and the power output by the ratio of the total work to the cycle duration,
\begin{equation}
P = -\frac{W_{\mathrm{comp}}+W_{\mathrm{exp}}}{\gamma (\tau_h + \tau_c)}.
\end{equation} 
Note that only the durations of the heating and cooling strokes are accounted for explicitly, with $\gamma$ serving as a multiplicative factor that implicitly incorporates the duration of the isentropic strokes \cite{Deffner2018}.

\subsection{1D Statistical Anyons}
Combining the internal energy and entropy from Eq. \eqref{eq:1DThermoFunc} with Eqs. \eqref{eq:Wcomp}, \eqref{eq:Qh}, and \eqref{eq:Wexp} and plugging it all into Eq. \eqref{eq:eff} yields a complicated expression that can be considerably simplified. First we note that from the isentropic strokes we have the conditions,
\begin{equation}
\label{eq:conditions}
S(T_A, \omega_1) = S(T_B, \omega_2) \ \ \mathrm{and} \ \ S(T_C, \omega_2) = S(T_D, \omega_1).
\end{equation}       
Using Eq. \eqref{eq:1DThermoFunc} it is straightforward to verify that the conditions in Eq. \eqref{eq:conditions} are satisfied by,
\begin{equation}
\label{eq:freqTemp}
T_A \omega_2 = T_B \omega_1 \ \ \mathrm{and} \ \ T_C \omega_1 = T_D \omega_2.  
\end{equation}
Furthermore, Eq. \eqref{eq:fh} and Eq. \eqref{eq:fc} can be fully solved to yield, 
\begin{equation}
\label{eq:fsolved}
\begin{split}
T_C - T_h  = (T_B - T_h) e^{- \alpha_h \tau_h}, \\
T_A - T_c  = (T_D - T_c) e^{- \alpha_c \tau_c},
\end{split}
\end{equation}
where $\tau_h$ ($\tau_c$) is the duration of the heating (cooling) stroke. Combining Eq. \eqref{eq:eff} with Eq. \eqref{eq:freqTemp} and Eq. \eqref{eq:fsolved} yields a much simplified form for the efficiency,
\begin{equation}
\eta = 1 - \kappa,
\end{equation}
where $\kappa \equiv \omega_1/\omega_2$ is the \textit{compression ratio}. We note that this efficiency is identical to classical, reversible Otto efficiency, as well as the single particle quantum Otto efficiency found in Ref. \cite{Deffner2018} and is completely independent of the quantum statistics of the working medium.

To find the EMP we next need to compute the power, given by Eq. \eqref{eq:power}. Eliminating free parameters using the same simplification process as we did for the efficiency we arrive at the much more cumbersome expression, 
\begin{equation}
\label{eq:power}
\begin{split}
P = \frac{(1-\kappa ) \omega _2 \hbar}{\gamma  \left(\tau _h+\tau _c\right)}[3 \coth(\Gamma)-3 \coth(\Lambda) \\ +\, \text{csch}\,(\Gamma)-\text{csch}\,(\Lambda)],
\end{split}
\end{equation}
where,
\begin{equation}
\begin{split} 
\label{eq:gl}
\Gamma = \frac{\kappa  \omega _2 \hbar  \left(e^{\alpha _c \tau _c+\alpha _h \tau _h}-1\right)}{ k_{\mathrm{B}} \left[\kappa  T_h \left(e^{\alpha _h \tau_h}-1\right)e^{\alpha _c \tau _c}+T_c \left(e^{\alpha _c \tau_c}-1\right)\right]}, \\ 
\Lambda = \frac{\kappa  \omega _2 \hbar  \left(e^{\alpha _c \tau _c+\alpha _h \tau _h}-1\right)}{ k_{\mathrm{B}} \left[\kappa  T_h \left(e^{\alpha_h \tau_h}-1\right)+T_c \left(e^{\alpha _c \tau _c}-1\right) e^{\alpha _h \tau_h}\right]}.
\end{split}
\end{equation}

The first thing that we can note about this expression is that the endoreversible power output does not depend on the statistics of the working medium for one-dimensional statistical anyons. This is consistent with the results of Ref. \cite{Myers2020}, which showed differences in the performance of a one-dimensional harmonic quantum Otto engine arising from the bosonic or fermionic nature of the working medium are a feature of nonequilibrium performance. In the case of endoreversible operation, the only effect of the generalized exclusion principle is to shift the value of the work expended during the compression stroke by $\hbar p_{\mathrm{B}} (\omega_1 - \omega_2)$ and the work extracted during the expansion stroke by $\hbar p_{\mathrm{B}} (\omega_2 - \omega_1)$. Thus when $W_{\mathrm{comp}}$ and $W_{\mathrm{exp}}$ are summed to determine the total work, these contributions exactly cancel each other out. 

We observe that the power vanishes for the case of $\Gamma = \Lambda$. Examining Eq. \eqref{eq:gl} we see that these expressions become equivalent in the limit $\kappa \rightarrow T_c/T_h$. This matches with our physical intuition, as this limit corresponds to quasistatic operation at Carnot efficiency. This provides us with the parameter range $\Gamma < \Lambda$ where the power output is positive and the cycle operates as an engine. This is typically referred to as the \textit{positive work condition}.     

\begin{figure}
	\includegraphics[width=.48\textwidth]{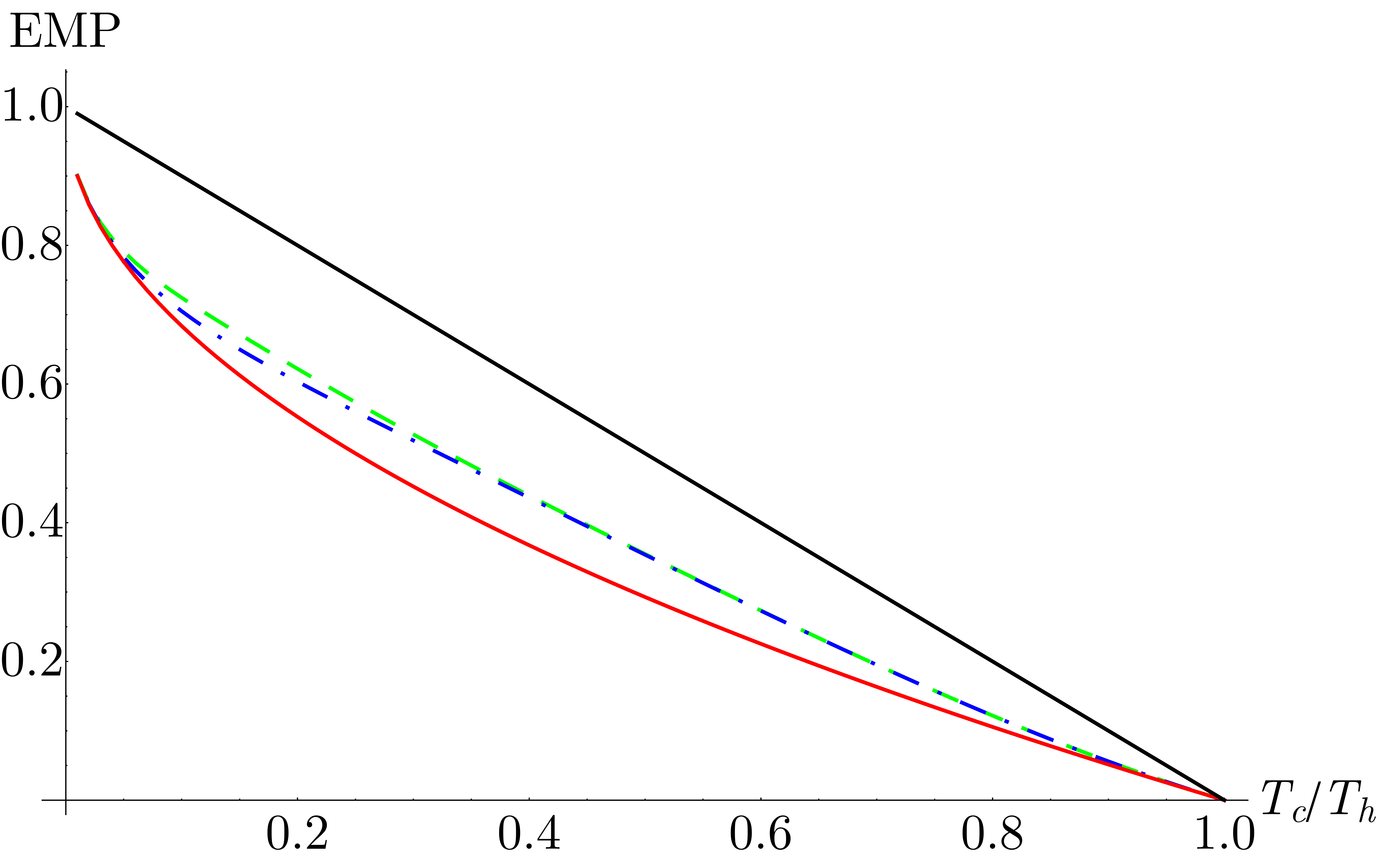}
	\caption{\label{fig:EMP} EMP as a function of the ratio of bath temperatures for two distinguishable quantum particles (dot-dashed, blue) and two indistinguishable statistical anyons (dashed, green) in one dimension. The Curzon-Ahlborn efficiency (bottom solid, red) and the Carnot efficiency (top solid, black) are given in comparison. Operation is in the quantum regime corresponding to $\hbar \omega_2/k_{\mathrm{B}} T_c = 10$. Parameters are $\alpha_c= \alpha_h=\gamma = 1$, and $\tau_c=\tau_h=0.5$.}
\end{figure}

To find the EMP we maximize Eq. \eqref{eq:power} numerically with respect to the compression ratio. The EMP as a function of the ratio of bath temperatures for one-dimensional statistical anyons is shown in Fig.~\ref{fig:EMP}, along with the EMP of distinguishable particles, the Curzon-Ahlborn efficiency, and the Carnot efficiency for comparison. Notably we see that, while the EMP does not depend on the anyonic phase, it does depend on whether or not the particles of the working medium are \textit{distinguishable}. In this case, distinguishable refers to particles that remain sufficiently spatially separated such that the overlap of their wave functions is negligible, negating any behavior that would arise from the exchange forces. Experimentally, we can consider an engine consisting of two distinguishable particles as equivalent to the joint output of two separate single particle engines situated across the lab from each other. We see that, at low temperature ratios, the EMP of the indistinguishable, anyonic working medium outperforms that of two distinguishable quantum particles. We also note that for both indistinguishable and distinguishable quantum working mediums the EMP is greater than the Curzon-Ahlborn efficiency, confirming the results found in \cite{Deffner2018}.  

\subsection{2D Statistical Anyons}

\begin{figure*}
	\includegraphics[width=.6\textwidth]{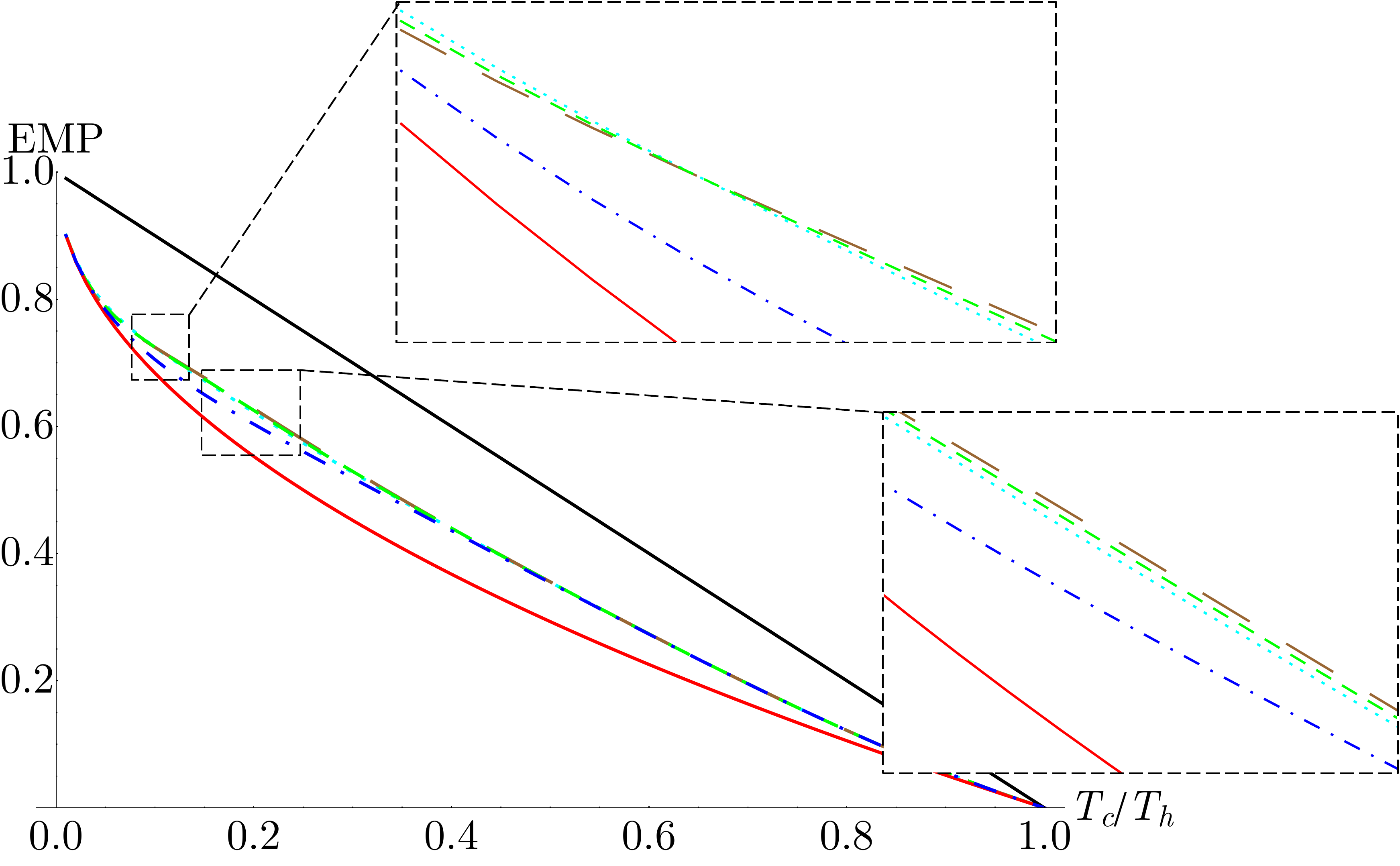}
	\caption{\label{fig:EMP2DSA} EMP as a function of the ratio of bath temperatures for two statistical anyons with $p_{\mathrm{B}} = 1$ (long dashed, brown), $p_{\mathrm{B}} = 1/2$ (short dashed, green), $p_{\mathrm{B}} = 0$ (dotted, cyan), and two distinguishable quantum particles (dot-dashed, blue) in two dimensions. The Curzon-Ahlborn efficiency (solid, red) is given in comparison. The bottom inset highlights the range of bath temperature ratios at which bosonic working mediums display the greatest EMP and the top inset highlights the critical point and transition to the region where fermionic working mediums begin to outperform bosonic ones. Operation is in the quantum regime corresponding to $\hbar \omega_2/k_{\mathrm{B}} T_c = 10$. Parameters are $\alpha_c= \alpha_h=\gamma = 1$, and $\tau_c=\tau_h=0.5$.}
\end{figure*}

It is straightforward to extend the previous analysis to two dimensions. As noted in Section \ref{sec:4}, in two dimensions the entropy is no longer independent of the anyonic phase, in the form of $p_{\mathrm{B}}$. Similarly, in the two-dimensional endoreversible engine the $p_{\mathrm{B}}$ dependence in the work done on expansion and expended during compression no longer cancel each other out, resulting in an anyonic phase dependent power output, 
\begin{equation}
\label{eq:powerSA}
\begin{split}
P = \frac{(1-\kappa) \omega _2 \hbar}{\gamma  \left(\tau _h+\tau _c\right)}\Big[2 \coth(\Gamma)-2 \coth(\Lambda) + \coth(\Gamma/2) \\ 
-\coth(\Lambda/2)+p_{\mathrm{B}} \tanh(\Lambda) -p_{\mathrm{B}} \tanh(\Gamma)\Big],
\end{split}
\end{equation} 
where $\Lambda$ and $\Gamma$ are the same as given in Eq. \eqref{eq:gl}. This dependence manifests in the EMP at small values of the bath temperature ratio, shown in Fig. \ref{fig:EMP2DSA}. We see that, as in one-dimension, all indistinguishable working mediums show greater EMP than a working medium of two distinguishable quantum particles. We see further that the anyonic phase that gives maximum performance depends on the bath temperature ratio. For temperature ratios between around 0.1 and 0.25 we see that bosonic symmetry ($p_{\mathrm{B}} =1$) gives the greatest enhancement to EMP over distinguishable particles and fermionic symmetry $(p_{\mathrm{B}} = 0)$ the least. However, around $T_c/T_h = 0.1$ there is a crossing, after which the fermionic symmetry gives the greatest enhancement to the EMP and bosonic symmetry the least. Intermediate values of $p_{\mathrm{B}}$ interpolate smoothly between these limits. Interestingly, this transition indicates the existence of a critical point at which the EMP becomes equivalent for all values of $p_{\mathrm{B}}$.

The origin of this behavior can be traced to the fact that, unlike in the one-dimensional case, the energy shift arising from the generalized exclusion principle is temperature dependent. In one dimension, while the magnitude of each energy eigenvalue depends on whether the particles are bosons or fermions, the \textit{degeneracy} does not. In two dimensions, both the magnitude and degeneracy of each energy state differ between bosons and fermions, leading to the temperature-dependent shift in the internal energy. Due to this temperature dependence, in general the contributions to the work from $p_{\mathrm{B}}$ on the compression and expansion strokes no longer cancel out. The critical point then corresponds to the unique ratio of bath temperatures such that these contributions become exactly equal, leading to $p_{\mathrm{B}}$-independent performance.           

\subsection{2D Topological Anyons}

\begin{figure}
	\includegraphics[width=.48\textwidth]{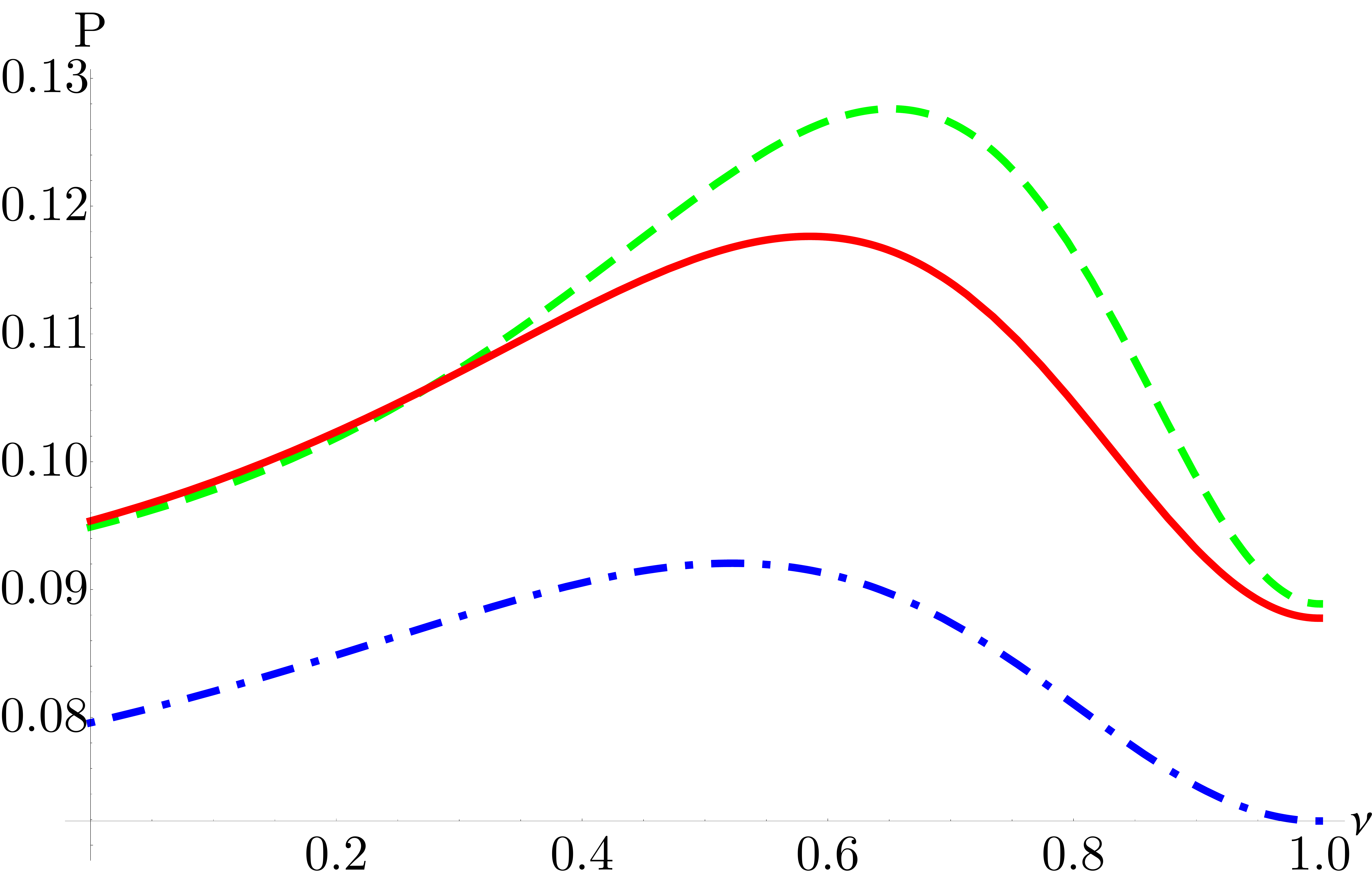}
	\caption{\label{fig:PTA} Power as a function of the anyonic phase for two topological anyons at $T_c/T_h = 0.1$ (dashed, green), $T_c/T_h = 0.2$ (solid, red), and $T_c/T_h = 0.3$ (dot-dashed, blue). Parameters are $\alpha_c= \alpha_h=\gamma =\tau_c=\tau_h=1$, and $\kappa=0.5$, along with $\hbar = k_{\mathrm{B}}=1$. }
\end{figure}

Using the partition function in Eq. \eqref{eq:partitionTrue} we can carry out the endoreversible analysis for a harmonic quantum Otto engine with a working medium of topological anyons. This results in an expression for the power that is very similar to the two-dimensional statistical anyon power, but with an additional factor dependent on the anyonic phase within the phase-dependent hyperbolic trigonometric terms, 
\begin{widetext} 
	\begin{equation}
	\label{eq:powerTrue}
	P = \frac{(1-\kappa) \omega _2 \hbar}{\gamma  \left(\tau _h+\tau_c\right)}\Big\{2 \coth(\Gamma)-2 \coth(\Lambda)+ \coth(\Gamma/2) -\coth(\Lambda/2)+(1-\nu)\tanh\big((1-\nu)\Lambda\big)  -(1-\nu)\tanh\big((1-\nu)\Gamma\big)\Big\}.
	\end{equation}
\end{widetext}
This results in richer behavior, with the power no longer always being maximized by either the bosonic or fermionic limit. Fig. \ref{fig:PTA} shows the power as a function of the anyonic phase for several different bath temperature ratios. We see that as the bath temperature ratio increases, the anyonic phase that maximizes the power output shifts from more fermionic to more bosonic.

\begin{figure*}
	\includegraphics[width=.6\textwidth]{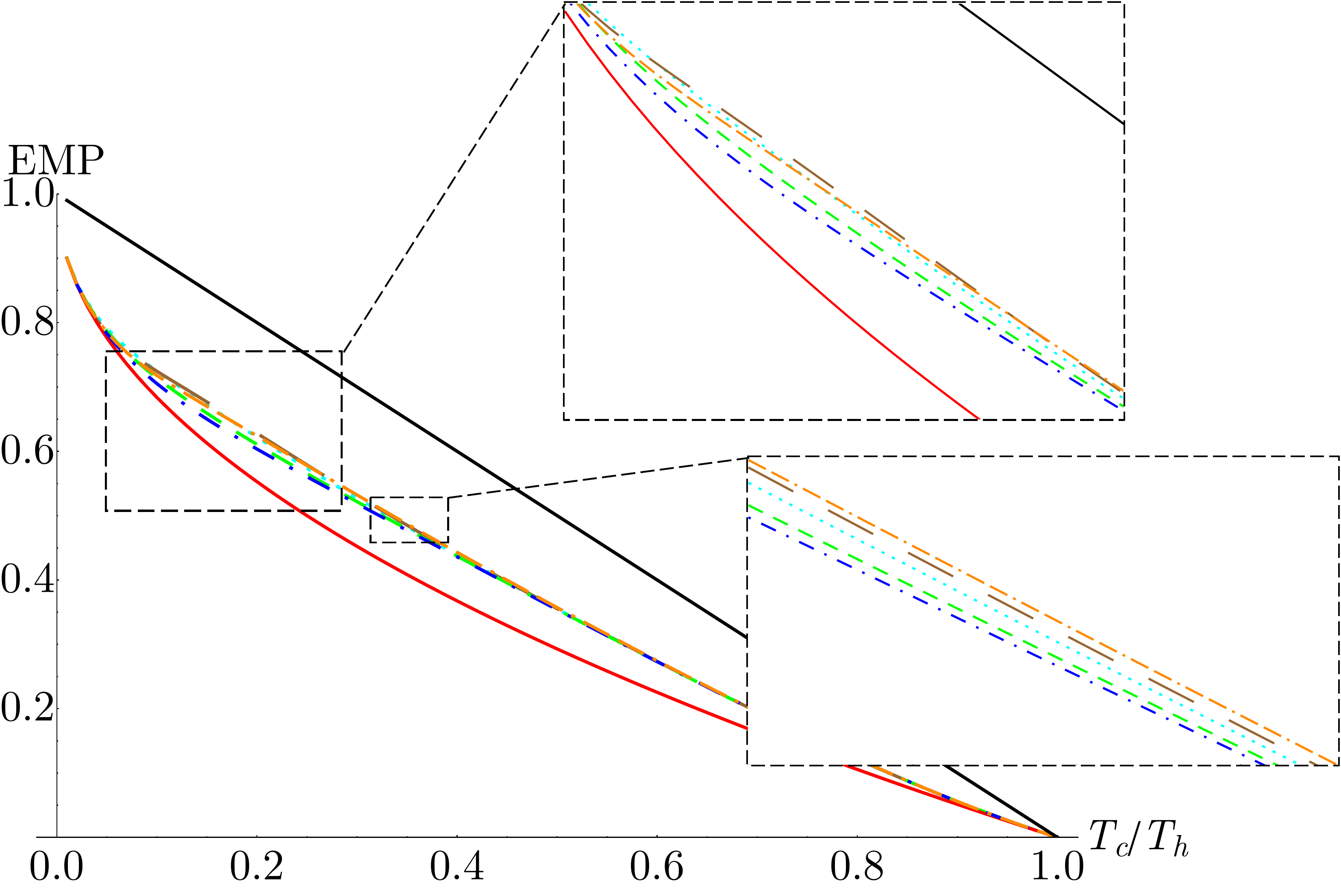}
	\caption{\label{fig:EMPTrue} EMP as a function of the ratio of bath temperatures for two topological anyons with $\nu = 0$ (long dashed, brown), $\nu = 1/2$ (short dashed, green), $\nu = 0.25$ (dot-dash-dashed, orange), $\nu = 1$ (dotted, cyan), and two distinguishable quantum particles (dot-dashed, blue) in two dimensions. The Curzon-Ahlborn efficiency (bottom solid, red) and Carnot efficiency (top solid, black) are given in comparison. The bottom inset highlights a parameter region where an intermediate anyonic phase ($\nu = 0.25$) displays the greatest EMP and the top inset highlights a region of low temperature ratios that exhibits multiple crossings. Operation is in the quantum regime corresponding to $\hbar \omega_2/k_{\mathrm{B}} T_c = 10$. Parameters are $\alpha_c= \alpha_h=\gamma = 1$, and $\tau_c=\tau_h=0.5$.}
\end{figure*}

\begin{figure}
	\includegraphics[width=.48\textwidth]{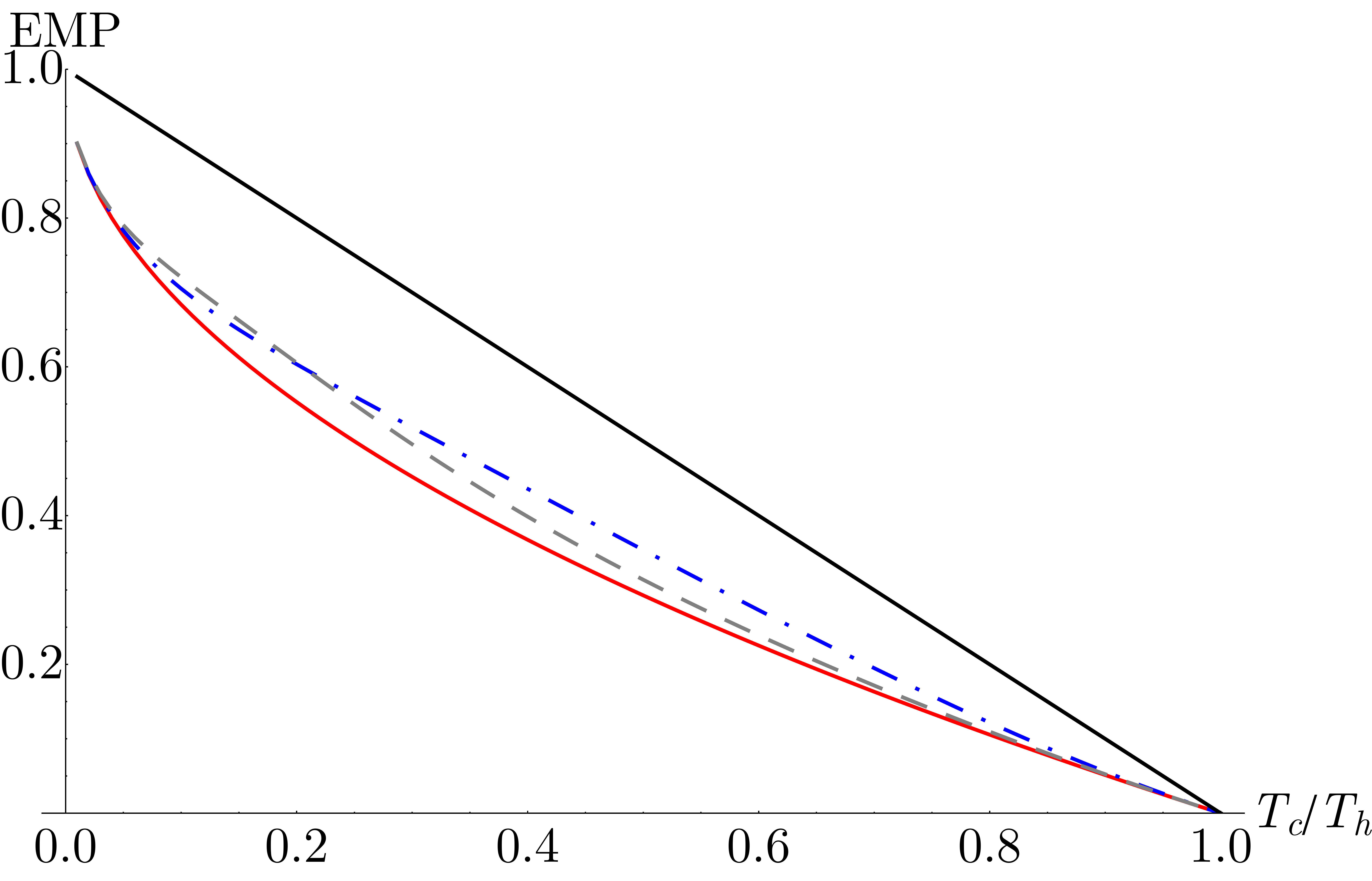}
	\caption{\label{fig:EMP08} EMP as a function of the ratio of bath temperatures for two topological anyons with $\nu = 0.8$ (dashed, gray) and two distinguishable quantum particles (dot-dashed, blue). The Curzon-Ahlborn efficiency (lower solid, red) and Carnot efficiency (upper solid, black) are given in comparison. Operation is in the quantum regime corresponding to $\hbar \omega_2/k_{\mathrm{B}} T_c = 10$. Parameters are $\alpha_c= \alpha_h=\gamma = 1$, and $\tau_c=\tau_h=0.5$.}
\end{figure}

Again maximizing with respect to the frequency, we find an EMP with complex dependence on the anyonic phase, shown in Fig. \ref{fig:EMPTrue}. Consistent with the pure power, we see that the anyonic EMP is no longer bounded by the bosonic and fermionic limits of the anyonic phase. The phase that provides the maximum EMP is highly temperature dependent, and we see intermediate values that provide both better and worse EMP than either bosons or fermions. Furthermore, the anyonic EMP can even fall below that of the distinguishable particles. One such example for the case of $\nu = 0.8$ is shown in Fig. \ref{fig:EMP08}. 

The origins of this more complicated behavior become clear when comparing Eq. \eqref{eq:powerSA} and Eq. \eqref{eq:powerTrue}. We see that, as in the case of the two-dimensional statistical anyons, the energy shifts arising from the statistics no longer cancel out (except in the quasistatic limit of $\Gamma = \Lambda$), giving rise to the two $\nu$-dependent hyperbolic tangent terms in Eq. \eqref{eq:powerTrue}. However, the more complicated energy spectrum arising from the hard-core restriction and multivalued wave function results in an additional $\nu$-dependence in the argument of the trigonometric functions. This highly non-linear dependence is responsible for the large variations in performance for intermediate values of $\nu$ seen in Figs. \ref{fig:EMPTrue} and \ref{fig:EMP08}.

From the standpoint of pure performance, the existence of temperature regimes where intermediate values of $\nu$ exceed the bosonic EMP demonstrates an advantage for topological anyons over statistical anyons as a working medium. However, the complex dependence of the power on $\nu$ also has the consequence that small variations in the anyonic phase can lead to vastly different performance, as seen for the case of $\nu = 0.8$. From a quantum metrology standpoint, the fact that performance is linked directly to the anyonic phase for both statistical and topological anyons indicates that thermal machines may be a useful tool for detecting signatures of anyonic statistics.     

\section{Anyonic Engine: Beyond Endoreversibility} 
\label{sec:5}
Under the conditions of endoreversiblity we have seen that the engine performance depends on the anyonic phase for both statistical and topological anyons in two dimensions, but not for statistical anyons in one dimension. We now move beyond the assumption of local equilibrium, and consider a finite-time quantum Otto cycle with fully nonequilibrium isentropic strokes. To focus on the effects on engine performance arising solely from the quantum statistics, we will make the standard assumption that the thermalization time is short enough and the isochoric strokes long enough that the working medium is in a state of thermal equilibrium with the hot (cold) bath at point C (A) in the cycle, removing the need to explicitly model the interaction with the heat baths \cite{Kosloff1984, Rezek2006, Abah2012, Campo2014, Beau2016, Abah2017, Myers2020, Watanabe2020}. 

For the full nonequilibrium treatment we restrict our analysis to the case of one-dimensional statistical anyons, as of the three working mediums we have explored so far this was the only one that has not yet shown effects arising from the nature of the quantum statistics. In order to calculate the efficiency, power, and EMP of the engine, we must determine the internal energies at points A, B, C, and D in the cycle. These can be found from the density operator in the typical fashion,
\begin{equation}
\label{eq:exH}
\langle H \rangle = \text{tr}\left\{ \rho H \right\}. 
\end{equation}
 
For $N$ independent particle pairs, the total density operator is given by the product of the individual density operators for each pair,
\begin{equation}
\label{eq:denOp}
\rho_N = \displaystyle\prod_{j=1}^{N_{\mathrm{B}}} \rho_{\mathrm{B}}^{(j)} \prod_{k=N_{\mathrm{B}} + 1}^{N} \rho_{\mathrm{F}}^{(k)},
\end{equation}
where $\rho_{\mathrm{B}}$ and $\rho_{\mathrm{F}}$ are the density operators of particle pairs with bosonic and fermionic symmetry, respectively. Combining Eqs. \eqref{eq:exH} and \eqref{eq:denOp} we have,
\begin{align}
\langle H_N \rangle &= \sum_{j=1}^{N_{\mathrm{B}}} \text{tr}\left\{\rho_{\mathrm{B}}^{(j)} H_j \right\} + \sum_{k=N_{\mathrm{B}} +1}^{N} \text{tr}\left\{\rho_{\mathrm{F}}^{(k)} H_k \right\} \\
&= N_{\mathrm{B}} \langle H_{\mathrm{B}} \rangle + (N-N_{\mathrm{B}})\langle H_{\mathrm{F}} \rangle. \nonumber
\end{align} 
Using $N_{\mathrm{B}} = N p_{\mathrm{B}}$ we arrive at the expression for the  internal energy of a single pair of statistical anyons,
\begin{equation}
\label{eq:SAH}
\langle H_{\mathrm{SA}} \rangle = p_{\mathrm{B}} \langle H_{\mathrm{B}} \rangle + (1-p_{\mathrm{B}})\langle H_{\mathrm{F}} \rangle
\end{equation}

In Ref. \cite{Myers2020} both the thermal state and time-evolved density operators are derived for two bosons and fermions in a harmonic potential (for completeness, the full expressions are provided in Appendix \ref{Appendix C}). The corresponding internal energies are, 

\begin{align}
\label{eq:internals}
\la H \ra_A &= \frac{\hbar \omega_1}{2}\,\left( 3 \mathrm{coth}(\beta_1 \hbar \omega_1) + \mathrm{csch}(\beta_1 \hbar \omega_1) \mp 1 \right), \nonumber \\
\la H \ra_B &= \frac{\hbar \omega_2}{2}\, Q^*_{12}\, \left( 3 \mathrm{coth}(\beta_1 \hbar \omega_1) + \mathrm{csch}(\beta_1 \hbar \omega_1) \mp 1 \right), \nonumber \\
\la H \ra_C &= \frac{\hbar \omega_2}{2}\,\left( 3 \mathrm{coth}(\beta_2 \hbar \omega_2) + \mathrm{csch}(\beta_2 \hbar \omega_2) \mp 1 \right), \\
\la H \ra_D &= \frac{\hbar \omega_1}{2}\, Q^*_{21}\, \left( 3 \mathrm{coth}(\beta_2 \hbar \omega_2) + \mathrm{csch}(\beta_2 \hbar \omega_2) \mp 1 \right). \nonumber
\end{align}

Here the $(-)$ corresponds to bosons and the $(+)$ to fermions. $Q^*_{12}$ and $Q^*_{21}$ are protocol-dependent dimensionless parameters that measure the degree of adiabaticity of the isentropic strokes \cite{Husimi1953}. Using Eqs. \eqref{eq:SAH} and \eqref{eq:internals} we can determine the engine efficiency and power output. The full expressions are cumbersome and given in Appendix \ref{Appendix D}. We note that the engine behavior we find here is equivalent to that found in Ref. \cite{Jaramillo2016} for an Otto engine with a working medium of a Calogero–Sutherland gas. However, our underlying construction is very different, with the anyonic nature of the working medium arising from a simple statistical average over bosons and fermions rather than an additional inter-particle interaction term. 

To continue our analysis we must pick a specific protocol for the compression and expansion strokes. For simplicity we choose the ``sudden switch" protocol, which corresponds to an instantaneous quench from the initial to final frequency. For the sudden switch the adiabaticity parameters are,
\begin{equation}
Q^*_{12} = Q^*_{21} = \frac{1 + \kappa^2}{2 \kappa},
\end{equation}
where $\kappa = \omega_1/\omega_2$. 

\begin{figure*}
	\centering
	\subfigure[]{
		\includegraphics[width=.48\textwidth]{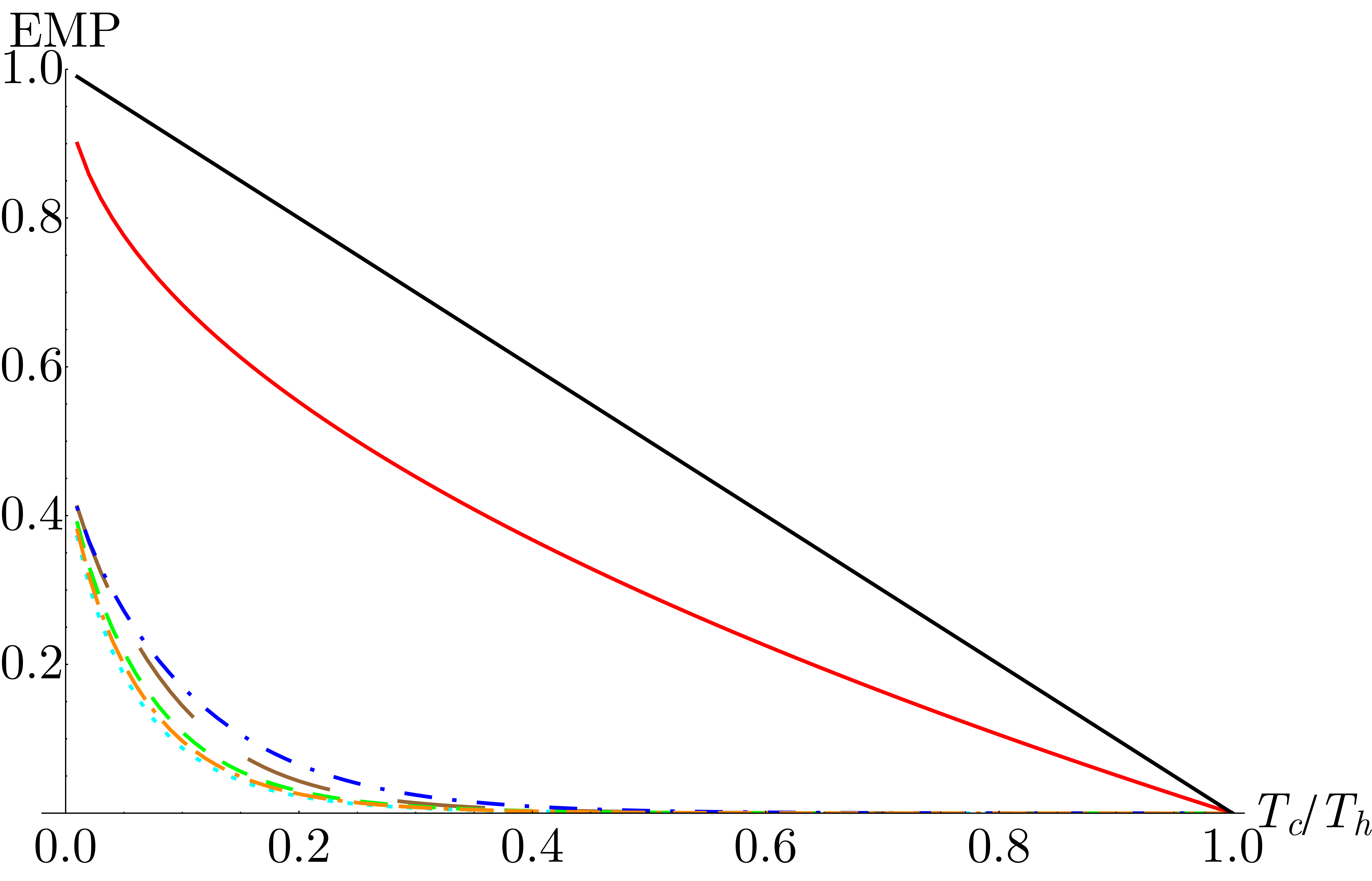}
	}
	\subfigure[]{
		\includegraphics[width=.48\textwidth]{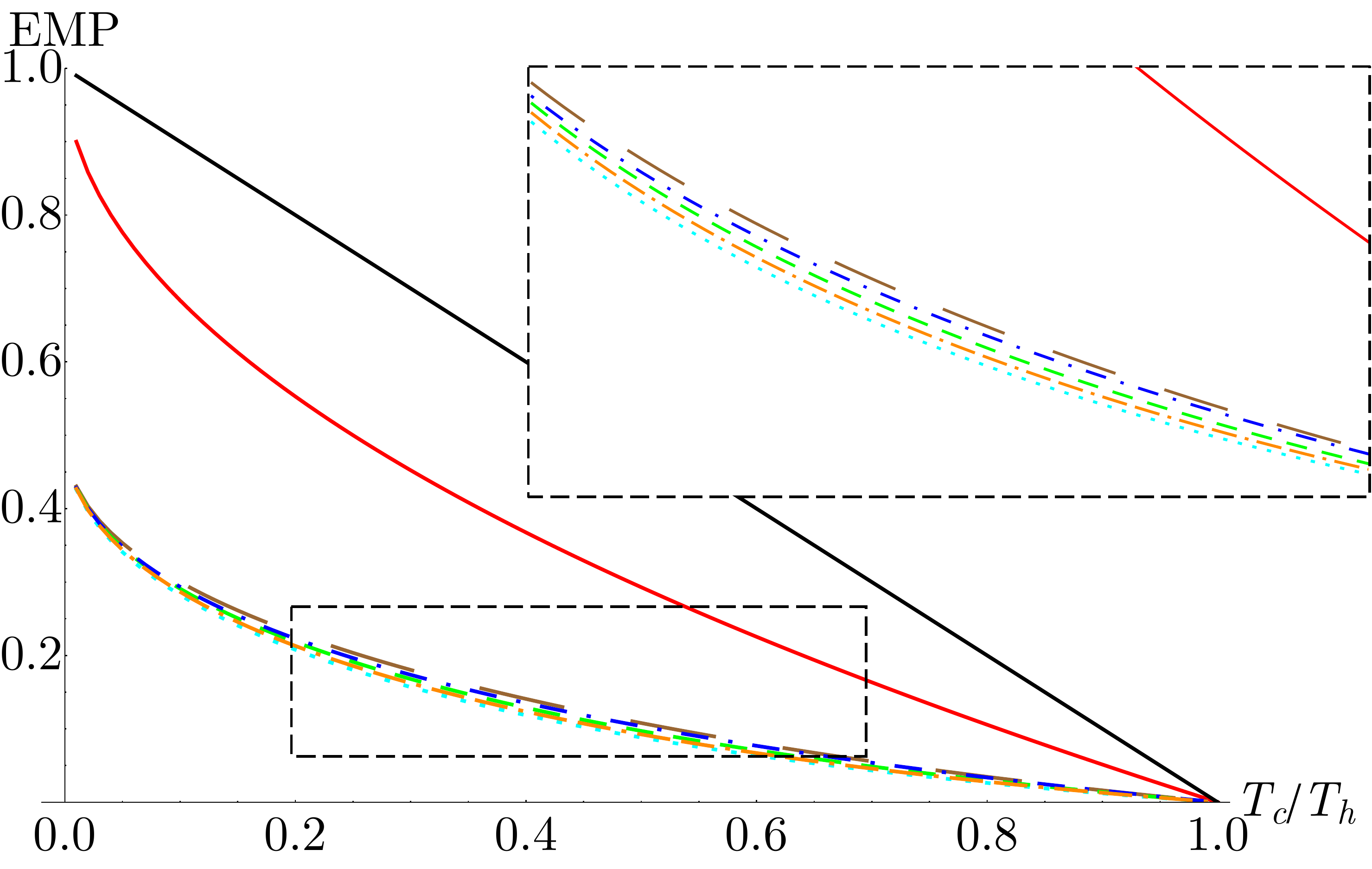}
	}
	\caption{\label{fig:EMPnonSA} Nonequilibrium EMP as a function of the ratio of bath temperatures for two statistical anyons in one dimension with $p_{\mathrm{B}} = 1$ (long dashed, brown), $p_{\mathrm{B}} = 1/2$ (short dashed, green), $p_{\mathrm{B}} = 0.25$ (dot-dash-dashed, orange), $p_{\mathrm{B}} = 0$ (dotted, cyan line), and two distinguishable quantum particles (dot-dashed, blue). The Curzon-Ahlborn efficiency (lower solid, red) and Carnot efficiency (top solid, black) are given in comparison. Plot (a) shows operation in the  regime corresponding to $\hbar \omega_2/k_{\mathrm{B}} T_c = 10$ and (b) in the regime corresponding to $\hbar \omega_2/k_{\mathrm{B}} T_c = 1$. We have set $\tau = 1$.}
\end{figure*}

Following the same method of performance analysis that we used in the endoreversible case, we numerically maximize the power with respect to the frequency ratio in order to determine the EMP. The EMP as a function of the bath temperature ratios is shown in Fig. \ref{fig:EMPnonSA}. We immediately see that in the case of nonequilibrium operation the performance is no longer independent of the anyonic phase. We note fermionic symmetry gives the worst EMP and bosonic symmetry the best, with intermediate values of $p_{\mathrm{B}}$ falling between. In the parameter regimes explored we see no transitions between anyonic phases that provides the optimal EMP, unlike the two-dimensional endoreversible engines.

In general we see that the EMP of the nonequilibrium engine is significantly worse than in the endoreversible case, no longer outperforming the Curzon-Ahlborn efficiency. This is unsurprising, as the sudden switch protocol we have employed is far from adiabatic, resulting in significantly lower power due to the loss of energy to nonadiabatic excitations. We also compare the EMP as we transition from the deep quantum regime characterized by $\hbar \omega_2/k_{\mathrm{B}} T_c = 10$ to a more classical regime characterized by $\hbar \omega_2/k_{\mathrm{B}} T_c = 1$. We see that in the deep quantum regime the indistinguishable particles give worse performance than distinguishable quantum particles. As we transition toward the more classical regime the gaps in EMP between different values of the anyonic phase shrink, but we also see that the indistinguishable particles begin to outperform the distinguishable ones. Fig. \ref{fig:EMPnonSA}b examines the engine performance in the same parameter regime studied in Ref. \cite{Myers2020}, where we see the same bosonic advantage emerge.         

\section{Optimization of Anyonic Engines: Shortcuts to Adiabaticity} 
\label{sec:6}

As we saw in the previous section, nonadiabatic driving significantly hinders engine performance through the loss of energy to nonadiabatic excitations - an effect typically referred to as ``quantum friction." To achieve completely frictionless strokes, the cycle driving must be fully adiabatic, requiring infinite time and leading to zero power output. This trade-off can be circumvented through the use of ``shortcuts to adiabaticity" (STA). STA refer to a set of techniques that can produce the same final state of a system in finite time that it would have achieved under adiabatic driving \cite{Torrontegui2013}. There are numerous established techniques to achieve this, including counterdiabatic driving \cite{Demirplak2003, Berry2009, Deffner2014, Deng2018}, dynamical invariants \cite{Chen2010}, inversion of scaling laws \cite{Campo2012}, the fast forward approach \cite{Masuda2009, Masuda2011, Torrontegui2012, Masuda2014, Deffner2015}, optimal protocols \cite{Stefanatos2013, Acconcia2015, Campbell2015, Rosales2020}, and time-rescaling \cite{Bernardo2020, Roychowdhury2021}. For a recent review on the topic of shortcuts see Ref. \cite{Guery2019}. In this section we will examine implementing optimal protocol and fast forward STA for harmonic anyonic systems in order to determine if the anyonic phase plays a role in shortcut design.        

\subsection{Optimal Protocol Shortcut}

We will first examine the optimal protocol shortcut using the phenomenological framework of linear response theory \cite{Acconcia2015,Deffner2020}. This shortcut is based on separating the total nonequilibrium work into two contributions, the quasistatic work and the excess work,
\begin{equation}
\label{eq:workSum}
\langle W \rangle = \langle W_{\mathrm{qs}} \rangle + \langle W_{\mathrm{ex}} \rangle.
\end{equation}
Here $\langle W_{\mathrm{qs}} \rangle$ corresponds to the work carried out were the process to be fully quasistatic, and $\langle W_{\mathrm{ex}} \rangle$ the work lost due to nonequilibrium excitations. It has been shown that there exist optimal protocols for which $\langle W_{\mathrm{ex}} \rangle$ vanishes, leading to quasistatic performance in finite time. This is a true STA, as for protocols where excess work vanishes no nonadiabatic transitions between eigenstates occur \cite{Acconcia2015}.

Let us consider a process in which we begin with a quantum system in thermal equilibrium with a reservoir at inverse temperature $\beta$. The system is then decoupled from the reservoir and driven by Hamiltonian $H(t) \equiv H(\lambda_t)$ where $\lambda_t$ is a time-dependent external control parameter, $\lambda_t \equiv \lambda_0 + \delta \lambda g(t)$. This process corresponds exactly to the isentropic strokes of the harmonic Otto engine, with $\omega_t^2$ as the external control parameter. 

If the external driving can be considered a weak perturbation, we can derive an expression for the excess work entirely from the equilibrium thermodynamic properties of the system using the tools of linear response theory. In this framework the excess work is given by \cite{Acconcia2015},
\begin{equation}
\label{eq:WexLR}
\langle W_{\mathrm{ex}} \rangle = -(\delta \lambda)^2 \int_{t_0}^{t_f} dt \frac{\pd}{\pd t} g(t) \int_{0}^{t-t_0} ds \mathcal{R}(s) \frac{\pd}{\pd s} g(t-s),  
\end{equation}
where $\mathcal{R}(t)$ is the relaxation function. Note that typically the relaxation function is denoted by $\Psi(t)$, as in Refs. \cite{Acconcia2015,Deffner2020}. Here we use $\mathcal{R}(t)$ to avoid confusion with the wave function. The relaxation function is determined by the quantum response function,
\begin{equation}
\phi(t) = - \frac{\pd}{\pd t} \mathcal{R}(t),
\end{equation}
which is in turn found from the equilibrium state,   
\begin{equation}
\label{eq:resFunc}
\phi(t) = \frac{1}{i \hbar} \text{tr}\left\{\rho_0 [A_0,A_t]\right\}.
\end{equation}
Here $A_t$ is the generalized force,
\begin{equation}
A = \frac{\pd}{\pd \lambda} H(\lambda).
\end{equation}

Let us first determine the excess work for two bosons and two fermions in one dimension. With these results we will be able to construct the excess work for one-dimensional statistical anyons. From the Hamiltonian in Eq. \eqref{eq:1DHamil} we determine the time-dependent generalized force,
\begin{equation}
A_t = \frac{1}{2}m\left[x_1^2(t)+x_2^2(t)\right].
\end{equation}
With Heisenberg's equation of motion and some elementary commutator algebra we determine $x(t)$, and from there the commutator $\left[A_0,A_t\right]$. Plugging this commutator into Eq. \eqref{eq:resFunc} we can then take the trace. To simplify the necessary integrals we first convert the thermal density operator to its Wigner distribution representation. The full expression for the Wigner distribution for bosons and fermions is given in Appendix \ref{Appendix C}. In the Wigner distribution representation $x$ and $p$ are converted from operators to simple commuting variables. The resulting response functions are, 
\begin{equation}
\phi(t) = \frac{\hbar}{4 \omega^2}\left[3 \coth(\beta \hbar \omega) + \text{csch}(\beta \hbar \omega) \mp 1\right]\sin(2 t \omega)
\end{equation}
where the $(-)$ corresponds to bosons and the $(+)$ to fermions. From the response functions we can determine the relaxation functions with a trivial integral,
\begin{equation}
\mathcal{R}(t) = \frac{\hbar}{8 \omega^2}\left[3 \coth(\beta \hbar \omega) + \text{csch}(\beta \hbar \omega) \mp 1\right]\cos(2 t \omega).
\end{equation}

We now have all the pieces we need to calculate the excess work from Eq. \eqref{eq:WexLR}. For simplicity we take $t_0 = 0$ and $t_f = \tau$. We pick a simple linear protocol $g(t) = t/\tau$, as it has been shown that there exist zeros of the excess work for a single particle in harmonic potential \cite{Acconcia2015}. We find the excess work for two bosons or fermions to be,
\begin{equation}
\begin{split}
\label{eq:Wex}
\langle W_{\mathrm{ex}} \rangle = &\frac{\hbar (\delta\omega)^2}{16 \omega_0^5 \tau^2}\sin^2(\omega_0 \tau) 
\\&\times\Big[3 \coth(\beta \hbar \omega_0) +\text{csch}(\beta \hbar \omega_0) \mp 1\Big].
\end{split}
\end{equation}
In Eqs. \eqref{eq:denOp} -- \eqref{eq:SAH} we showed that, as the $N$ particle density operator is simply the product of the density operators of each individual particle pair, the statistical anyon internal energy is simply given by the weighted sum of the boson and fermion internal energies. The exact same mathematical process can be applied here in the calculation of the statistical anyon Wigner distribution, response function, relaxation function, and excess work. Thus we have,
\begin{equation}
\langle W_{\mathrm{ex}}^{\mathrm{SA}} \rangle = p_{\mathrm{B}} \langle W_{\mathrm{ex}}^{\mathrm{B}} \rangle+(1-p_{\mathrm{B}})\langle W_{\mathrm{ex}}^{\mathrm{F}} \rangle.
\end{equation}

\begin{figure}
	\includegraphics[width=.48\textwidth]{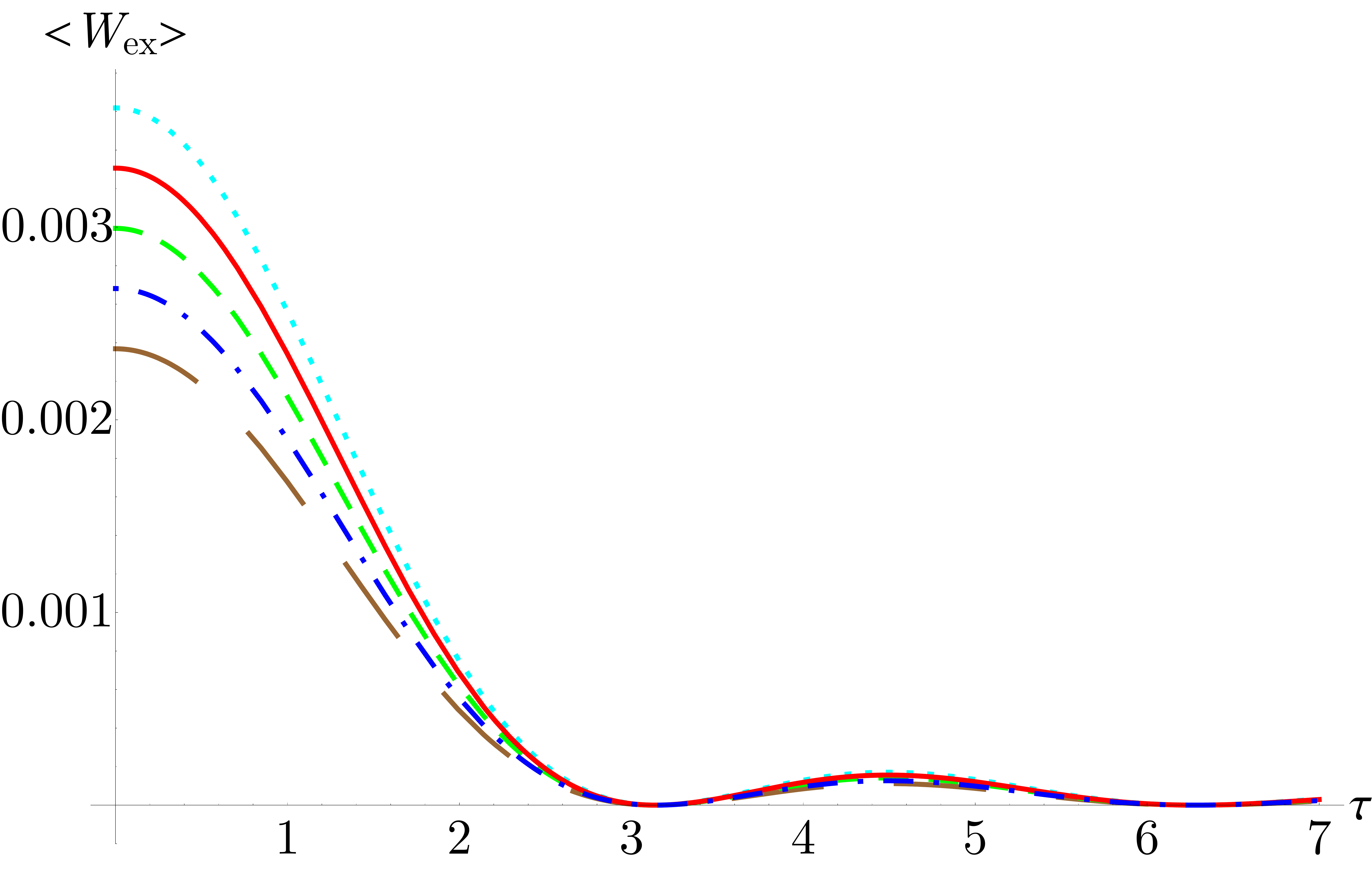}
	\caption{\label{fig:WexSA} Excess work under linear driving as a function of the driving time for two statistical anyons with $p_{\mathrm{B}} = 1$ (long dashed, brown), $p_{\mathrm{B}} = 0$ (dotted, cyan), $p_{\mathrm{B}} = 1/2$ (dashed, green), $p_{\mathrm{B}} = 3/4$ (dot-dashed, blue), and $p_{\mathrm{B}} = 1/4$ (solid, red) using a linear protocol. Parameters are $\omega_0= \beta=\hbar = 1$, and $\delta\omega=0.1$.}
\end{figure}

The statistical anyon excess work is plotted as a function of $\tau$ in Fig. \ref{fig:WexSA}. We see first that excess work varies with the anyonic phase, with fermions having the greatest excess work and bosons the least. We note that its zeros, however, are independent of the statistics. This is clear when examining Eq. \eqref{eq:Wex}. The statistics only come into play in the form of a $\mp 1$ within a multiplicative factor, which has no bearing on the zeros of the function. We see that for a linear protocol the one-dimensional statistical anyon excess work will vanish for all values of the anyonic phase when $\tau = n\pi/\omega$, where $n \in \mathbb{Z}$.     

\begin{figure}
	\includegraphics[width=.48\textwidth]{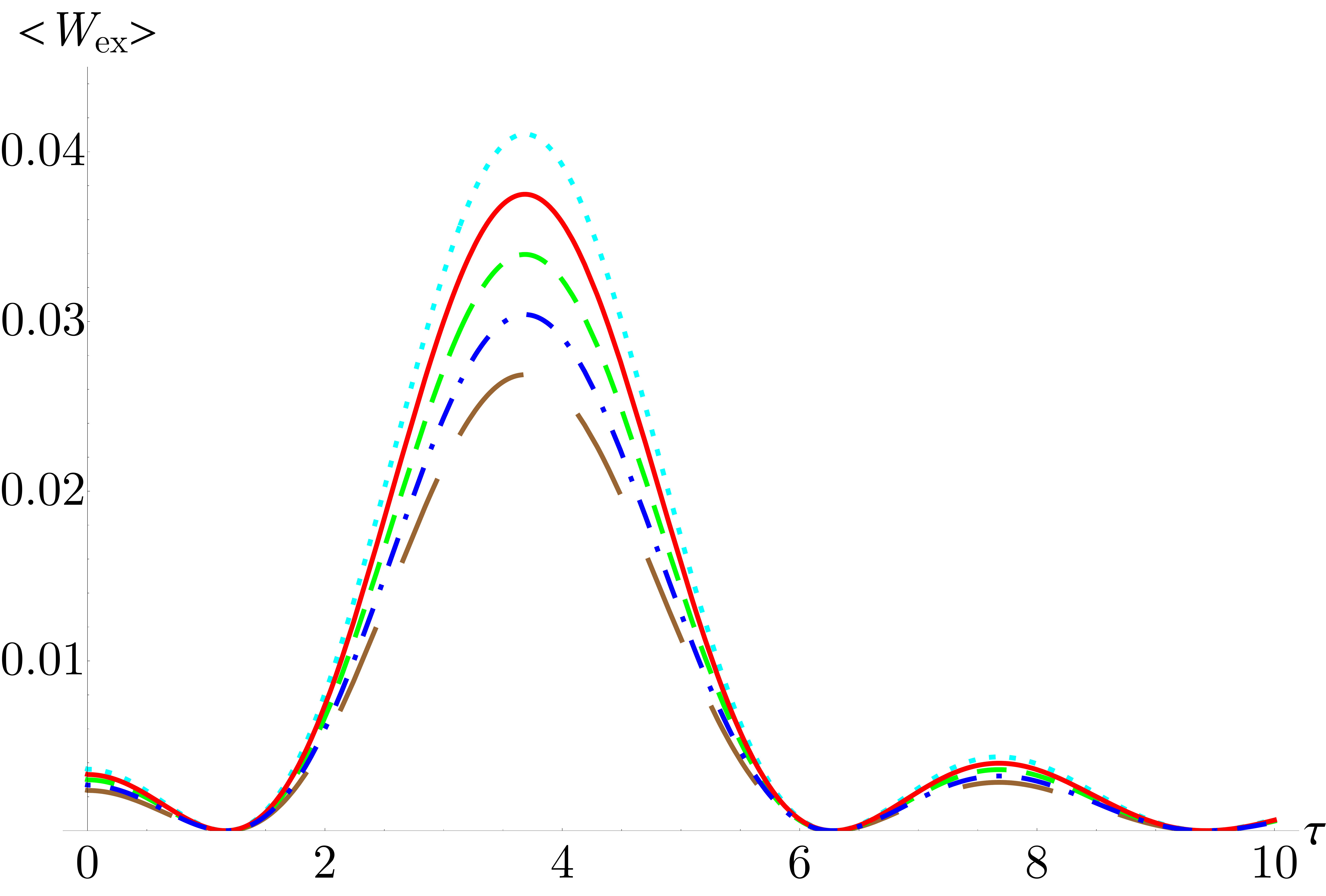}
	\caption{\label{fig:WexSAdeg} Excess work under shortcut protocol driving as a function of the driving time for two statistical anyons with $p_{\mathrm{B}} = 1$ (long dashed, brown), $p_{\mathrm{B}} = 0$ (dotted, cyan), $p_{\mathrm{B}} = 1/2$ (dashed, green), $p_{\mathrm{B}} = 3/4$ (dot-dashed, blue), and $p_{\mathrm{B}} = 1/4$ (solid, red) using the shortcut protocol. Parameters are $\omega_0= \beta=\hbar = 1$, $\alpha =1$, $\kappa =2$ and $\delta\omega=0.1$.}
\end{figure}

In Ref. \cite{Acconcia2015} it was also shown that there exists a family of degenerate shortcut protocols that produce zeros of the excess work for a single particle in a parametric harmonic potential. This family take the form of,
\begin{equation}
g(t) = \frac{t}{\tau}+\alpha \sin(\kappa \pi t/\tau)
\end{equation}
where $\kappa \in \mathbb{Z}$ and $\alpha \in \mathbb{R}$. The excess work from this protocol is shown for one dimensional statistical anyons in Fig. \ref{fig:WexSAdeg}. We arrive at the same conclusion as in the case of the linear protocol. While the excess work itself depends on the statistics, the optimal shortcut protocol does not. Examining Eq. \eqref{eq:WexLR} we can extend this conclusion to any arbitrary protocol for the parametric harmonic potential. Since the factor in the expression for the excess work by which the bosons and fermions differ is independent of time, it will be identical for any $g(t)$.                  

\subsection{Fast Forward Shortcut}

Having shown that the optimal protocol for achieving a shortcut to adiabaticity for one-dimensional statistical anyons in a harmonic Otto engine is independent of the quantum statistics, we next turn to a different shortcut implementation, the fast-forward method. Unlike the optimal protocol, in which the shortcut is determined entirely from the form of the thermal state and system Hamiltonian, the fast forward shortcut is constructed using the instantaneous eigenstates. Since these eigenstates depend explicitly on the anyonic phase, we expect this dependence to carry through to the shortcut. To apply this method we introduce an auxiliary potential to the Schr\"{o}dinger equation,
\begin{equation}
\label{eq:modSE}
i \hbar \frac{\pd}{\pd t}\Psi(\mathbf{x},t) = -\frac{\hbar^2}{2m}\nabla^2\Psi(\mathbf{x},t) + (V+V_{\mathrm{aux}})\Psi(\mathbf{x},t),
\end{equation}
where the form of $V_{\mathrm{aux}}$ ensures the final state of the system after a finite time evolution is identical to that achieved after an adiabatic evolution of the unperturbed system. Let us consider the following ansatz for our time-dependent wave function,
\begin{equation}
\label{eq:Ansatz}
\Psi(\mathbf{x},t) = \psi(\mathbf{x},R_t)e^{i f(\mathbf{x},t)}e^{-\frac{i}{\hbar}\int_{0}^{t} ds \, \epsilon(R_s)},
\end{equation}
where $\psi(\mathbf{x},R_t)$ is the instantaneous eigenstate of the unperturbed Schr\"{o}dinger equation with eigenenergy $\epsilon(R_t)$,
\begin{equation}
\epsilon(R_t) \psi(\mathbf{x},R_t) = -\frac{\hbar^2}{2m}\nabla^2\psi(\mathbf{x},R_t) + V\psi(\mathbf{x},R_t).
\end{equation}
Our goal is now to find the phase, $f(\mathbf{x},t)$, and auxiliary potential $V_{\mathrm{aux}}$ such that the above ansatz and equation are true. Imposing the condition that $f(\mathbf{x},0) = f(\mathbf{x},\tau)$ (where $\tau$ is the duration of the driving) ensures that the final state is identical to that of the adiabatically driven unperturbed equation, and our shortcut is achieved.

To simplify the following analysis we express $\psi(\mathbf{x},R_t)$ in polar representation,
\begin{equation}
\label{eq:polar}
\psi(\mathbf{x},R_t) = \mu(\mathbf{x},R_t) e^{i \gamma(\mathbf{x},R_t)}.
\end{equation} 
Combining Eqs. \eqref{eq:polar}, \eqref{eq:Ansatz}, and \eqref{eq:modSE}, simplifying, and separating the real and imaginary components we find,
\begin{equation}
\label{eq:AuxSoln}
V_{\mathrm{aux}} = -\hbar \frac{\pd \gamma}{\pd t} - \hbar \frac{\pd f}{\pd t} -\frac{\hbar^2}{m} \nabla f \nabla \gamma -\frac{\hbar^2}{m} (\nabla f)^2,
\end{equation}
and,
\begin{equation}
\label{eq:PhaseSoln}
2m \frac{\pd \mu}{\pd t} + 2 \hbar \nabla f \nabla \mu + \hbar \mu \nabla^2 f = 0,
\end{equation} 
where we have written $\mu(\mathbf{x},R_t) = \mu$, $\gamma(\mathbf{x},R_t) = \gamma$, and $f(\mathbf{x},t) = f$ to simplify notation.

The shortcut method has a major caveat in that the auxiliary potential often becomes singular at the nodes of the instantaneous eigenstates, limiting its applicability to the ground state \cite{Patra2017, Jarzynski2017}. However, for the Lewis-Leach family of Hamiltonians, of which the harmonic oscillator is a member \cite{Lewis1982}, it has been shown that the fast forward potential is independent of the energy level \cite{Torrontegui2012, Patra2017}.  

\subsubsection{Statistical Anyons}

As in the optimal protocol analysis let us first consider the fast forward shortcut for the case of two bosons and two fermions, from which we will be able to construct the statistical anyon shortcut. It is straightforward to extend Eqs. \eqref{eq:AuxSoln} and \eqref{eq:PhaseSoln} to a two particle system,
\begin{equation}
\label{eq:AuxSoln2D}
\begin{split}
V_{\mathrm{aux}} = &-\hbar \frac{\pd \gamma}{\pd t} - \hbar \frac{\pd f}{\pd t} -\frac{\hbar^2}{m} (\nabla_1 f \nabla_1 \gamma + \nabla_2 f \nabla_2 \gamma) \\ &-\frac{\hbar^2}{m} \left[ (\nabla_1 f)^2+(\nabla_2 f)^2\right],
\end{split}
\end{equation}
\begin{equation}
\label{eq:PhaseSoln2D}
2m \frac{\pd \mu}{\pd t} + 2 \hbar (\nabla_1 f \nabla_1 \mu +\nabla_2 f \nabla_2 
\mu)+ \hbar \mu (\nabla_1^2 f + \nabla_2^2 f) = 0.
\end{equation}
As we seek shortcuts to optimize performance of our harmonic Otto engine let us again consider a one-dimensional harmonic potential. In this case we have $R(t) = \omega(t)$. We can express $\mu(x_1,x_2,\omega_t)$ for both bosons and fermions in terms of the single particle harmonic oscillator eigenstates in the typical fashion,
\begin{align}
\label{eq:1Dbeta}
\mu(x_1,x_2,\omega_t) = \frac{1}{\sqrt{2(1+\delta_{n_1,n_2})}} \Big[ \mu_{n_1} (x_1) \mu_{n_2} (x_2) \\ \pm \mu_{n_1} (x_2) \mu_{n_2} (x_1) \Big] \nonumber,
\end{align}
where,
\begin{equation}
\mu_n(x)=\frac{1}{\sqrt{2^n n!}} \bigg(\frac{m \omega}{\pi \hbar} \bigg)^{1/4} e^{- \frac{m \omega x^2}{2 \hbar}} H_n \bigg( \sqrt{\frac{m \omega}{\hbar}}x\bigg).
\end{equation}

For a single particle in a harmonic potential Eq. \eqref{eq:PhaseSoln} is solved by \cite{Masuda2009, Torrontegui2012}, 
\begin{equation}
f_1(x,\omega_t) = -\frac{m \dot{\omega_t}}{4 \pi \omega_t} x^2.
\end{equation}
With this in mind, we take our ansatz for the two particle solution to be,
\begin{equation}
\label{eq:1DPhase}
f_2(x_1,x_2,\omega_t) = -\frac{m \dot{\omega_t}}{4 \pi \omega_t} \left(x_1^2+x_2^2\right). 
\end{equation} 
We can see that for any protocol with $\dot{\omega_0} = \dot{\omega_{\tau}}$ the shortcut condition will be fulfilled, as $f(\mathbf{x},0) = f(\mathbf{x},\tau)$. A simple example of a protocol that fulfills this condition is the linear protocol, $\omega(t) = \omega_0 + \alpha t$, where $\alpha$ is a constant.

With Eqs. \eqref{eq:1DPhase} and \eqref{eq:1Dbeta} we can directly verify that our ansatz is correct by plugging in to Eq. \eqref{eq:PhaseSoln2D}. Since $\gamma(x_1,x_2,\omega_t) = 0$ for both bosons and fermions, we can immediately see that not only the phase, but also the auxiliary potential will be identical for both bosons and fermions. Using Eq. \eqref{eq:AuxSoln2D} we determine the explicit form of the auxiliary potential to be,
\begin{equation}
V_{\mathrm{aux}} = -\frac{3 m \dot{\omega_t}}{8 \omega^2}\left(x_1^2+x_2^2\right).
\end{equation}

Following the same method we can extend this analysis to two dimensions. For this case we take the ansatz,
\begin{equation}
\label{eq:2Dphase}
f_2(x_1,x_2,y_1,y_2\omega_t) = -\frac{m \dot{\omega_t}}{4 \pi \omega_t} \left(x_1^2+x_2^2+y_1^2+y_2^2\right),
\end{equation}
which we verify is a solution to Eq. \eqref{eq:PhaseSoln2D}. As in the one-dimensional case, $\gamma(x_1,x_2,y_1,y_2,\omega_t) = 0$ so the auxiliary potential will again be identical for both bosons and fermions.

Since the statistical anyon state is constructed from a statistical average over the boson and fermion states, we know that if there is no difference in the shortcut for bosons and fermions, there will be none for statistical anyons. We conclude that in both one and two dimensions the implementation of a fast forward STA is independent of the quantum statistics. This result is counter to our original hypothesis that the difference in phase of the instantaneous eigenstates should carry through to the design of the auxiliary potential. From the form of Eqs. \eqref{eq:1DPhase} and \eqref{eq:2Dphase} we can see that, physically, the two particle shortcut for the harmonic oscillator potential is accomplished by driving each particle individually.          

\subsubsection{Topological Anyons}

Our previous analysis has shown that the behavior of topological anyons can have a richer dependence on the anyonic phase. Furthermore, topological anyon eigenstates are significantly more complex, and unlike bosonic or fermionic states can not generally be separated into a superposition of the single particle states \cite{Khare2005}. As the fast-forward shortcut depends on the stationary eigenstate of the system it is of interest to explore whether or not a shortcut for topological anyons will show dependence on the anyonic phase.

Let us again consider the situation of two anyons in a harmonic potential. The time-independent Schr\"{o}dinger equation for this problem can be solved by separating the Hamiltonian in Eq. \eqref{eq:1DHamil} into the center of mass and relative components \cite{Lerda1992, Myrheim1999, Khare2005},
\begin{equation}
H = \hbar \omega \left(-\frac{\pd^2}{\pd Z \pd Z^*}-4\frac{\pd^2}{\pd z \pd z^*}+\lvert Z \rvert^2 + \frac{\lvert z \rvert^2}{4}\right)
\end{equation}                     
where we have used the complex coordinates $z_j = \sqrt{mw/\hbar}(x_j + i y_j)$ with $j \in {1,2}$. Here $Z$ is the center of mass coordinate, given by $Z=(z_1+z_2)/2$ and $z$ the relative coordinate, given by $z=z_1-z_2$. In two dimensions the anyonic exchange symmetry is satisfied by two separate families of eigenstates. The harmonic oscillator ground states for each are \cite{Myrheim1999},
\begin{equation}
\label{eq:TopGS}
\begin{split}
&\mu_g^{(I)} = A_{\nu}\omega z^{\nu}e^{-\lvert Z \rvert^2 - \frac{\lvert z \rvert^2}{4}}, \\
&\mu_g^{(II)} = A_{\nu}\omega(z^*)^{2-\nu}e^{-\lvert Z \rvert^2 - \frac{\lvert z \rvert^2}{4}},
\end{split}
\end{equation} 
where $A_{\nu}$ is an anyonic phase-dependent normalization factor. We can construct any desired excited state from these ground states by applying the appropriate raising operators \cite{Myrheim1999}.

In order to determine the fast forward shortcut for the topological anyon system it will be convenient to convert Eqs. \eqref{eq:AuxSoln2D} and \eqref{eq:PhaseSoln2D} into the center of mass and relative coordinate systems,
\begin{equation}
\label{eq:TopAuxPot}
\begin{split}
V_{\mathrm{aux}} = &-\frac{1}{\omega}\frac{\pd \gamma}{\pd t}-\frac{1}{\omega}\frac{\pd f}{\pd t}-\frac{\pd f}{\pd Z}\frac{\pd \gamma}{\pd Z^*}-\frac{\pd \gamma}{\pd Z}\frac{\pd f}{\pd Z^*}-\frac{\pd f}{\pd Z}\frac{\pd f}{\pd Z^*} \\ &-4\frac{\pd f}{\pd z}\frac{\pd \gamma}{\pd z^*}-4\frac{\pd \gamma}{\pd z}\frac{\pd f}{\pd z^*}-4\frac{\pd f}{\pd z}\frac{\pd f}{\pd z^*},
\end{split}
\end{equation}
\begin{equation}
\label{eq:TopPhaseSoln}
\begin{split}
\frac{\pd \mu}{\pd t}& + \omega \Big[ \frac{\pd f}{\pd Z}\frac{\pd \mu}{\pd Z^*}+\frac{\pd \mu}{\pd Z}\frac{\pd f}{\pd Z^*}+\mu \frac{\pd^2 f}{\pd Z \pd Z^*} \\&+ 4\frac{\pd f}{\pd z}\frac{\pd \mu}{\pd z^*}+4\frac{\pd \mu}{\pd z}\frac{\pd f}{\pd z^*}+ 4 \beta \frac{\pd f}{\pd z \pd z^*} \Big]= 0.
\end{split}
\end{equation} 
Motivated by our results for the statistical anyons, we choose the same ansatz for $f$ (converted into the relative and center of mass coordinate system),
\begin{equation}
\label{eq:TopPhase}
f(Z,z,\omega_t) = - \frac{\dot{\omega_t}}{4 \omega^2}\left(\frac{1}{2}\lvert z \rvert^2 + 2 \lvert Z \rvert^2\right).
\end{equation}
Plugging Eq. \eqref{eq:TopPhase} and Eq. \eqref{eq:TopGS} into Eq. \eqref{eq:TopPhaseSoln} we see that our ansatz is indeed a solution to the differential equation. As before, we have $\gamma(Z,z,\omega_t) = 0$ in the center of mass and relative coordinate system. Using Eq. \eqref{eq:TopAuxPot} we can construct the auxiliary potential,
\begin{equation}
V_{\mathrm{aux}} = \frac{\dot{\omega_t}^3}{2 \omega^4}\left(\frac{1}{2}\lvert z \rvert^2 + 2 \lvert Z \rvert^2\right).
\end{equation}
We see that, as in the case of the statistical anyons, the fast forward STA for topological anyons is independent of the anyonic phase. This indicates that, counter to our original intuition, the harmonic oscillator fast forward shortcut is truly independent of any exchange behavior, not just the bosonic and fermionic limits.  

While this result was derived for the ground state we know that, since the harmonic oscillator Hamiltonian is a member of the Lewis-Leach family \cite{Lewis1982}, the above shortcut will also hold for any excited state. This provides a physical motivation for why the shortcut is independent of the anyonic nature of the particles. Both the modified hard-core nature of topological anyons and the generalized exclusion statistics of statistical anyons affect how the particles will be distributed among the available states. Since the same harmonic oscillator shortcut holds for all states, it makes sense that it will be independent of the anyonic phase.            

\section{Concluding Remarks}       
\label{sec:7}

\subsection{Summary}

In this work we have presented a dimension-independent formulation of anyons, dubbed ``statistical anyons," in which anyonic properties arise from averaging over the behavior of a system consisting of a statistical mixture of particles with antisymmetric and symmetric exchange properties. Motivated by the HOM effect, we outlined a quantum optics implementation of statistical anyons. We showed that the statistical anyons are physically equivalent to Haldane's generalized exclusion statistics anyons, broadening the applicability of GES to any system of indistinguishable particles for which such a mixture can be constructed.   

We determined the thermodynamic properties of statistical anyons in one and two dimensions, and compared them to the thermodynamic properties of topological anyons. We found that in two dimensions the internal energy, free energy, entropy, and heat capacity all display a dependence on the anyonic phase, but it is not as rich a dependence as the topological anyons. However, we determined that, with a parameter-dependent choice of the anyonic phase, statistical anyons can exactly imitate the thermodynamic behavior of topological anyons. 

With the thermodynamic properties established, we considered a harmonic quantum Otto engine with a working medium of statistical anyons. We found that in two dimensions endoreversible engine performance depends on the anyonic phase for both statistical anyons and topological anyons. We found that for the nonequilibrium regime, even in one-dimension, the engine EMP for statistical anyons depended on the anyonic phase.

Lastly, we examined the role of the anyonic phase in two STA, the optimal protocol shortcut and the fast forward shortcut. We found both shortcut methods are independent of the anyonic nature of the particles. In the case of the fast forward method, this independence can be considered to arise from the fact that the shortcut does not depend on the energy level for a harmonic potential. 

\subsection{Impacts and Future Directions}

Previous work has been primarily focused on determining the thermodynamic properties of anyons through the calculation of the partition function or Virial coefficients of an anyonic gas \cite{Bhaduri1991, Lerda1992, Sen1992, Sporre1993, Murthy1994, Chen1995, Giacconi1996,Isakov1996, Myrheim1999, Khare2005,Rovenchak2014}, anyonic phase transitions \cite{Nasu2015}, and the distribution of the anyon gas \cite{Wu, Joyce1996, Anghel2013}. In this work we have taken a different approach by examining the thermodynamic properties of anyons in the context of heat engines. This method is motivated by the history of thermodynamics, as a field developed around the optimization of thermal machines, and is an approach commonly used in modern quantum thermodynamics \cite{Deffner2019}. To the authors' knowledge, this approach has only been applied in two works, both based around Calogero–Sutherland GES anyons \cite{Jaramillo2016, Beau2016}.  

While functionally equivalent to GES, statistical anyons present a new paradigm both theoretically and experimentally that extends GES to a range of new settings, including Bose-Fermi mixtures \cite{Fang2011, Dehkharghani2017, Decamp2017, Fukuhara2009, Onofrio2016, Sowinski2019}, optimechanical systems \cite{Holmes2020}, and surface plasmonics \cite{ChenY2018}. While they lack the non-abelian properties necessary for implementing topological quantum computation, they do provide an experimentally tractable method of examining the thermodynamic behavior of abelian topological anyons. This opens up the door to searching for general thermodynamic signatures that may provide alternative methods of detecting and controlling both abelain and non-ableian anyons. 

Using statistical anyons we have connected topics across various subfields of physics, including the Hong-Ou-Mandel effect from quantum optics, heat engines from quantum thermodynamics, GES from quantum statistical mechanics, fractional exchange statistics from topological states of matter, and shortcuts to adiabaticity from quantum control. The establishment of the statistical anyon framework opens up a multitude of new possible research directions and questions for the study of anyons. There remains much to be explored about the thermodynamics of statistical anyons, including Gibbs mixing for statistical anyons \cite{Yadin2020} and the behavior of autonomous quantum engines with statistical anyon working mediums. We have shown in this work that the performance of a cyclic quantum engine is sensitive to the anyonic phase, and we would expect the same to be true for autonomous engines. Implementing the latter in an optical or plasmonic setting has metrological implications as an alternative device for detecting signatures of anyonic behavior.    

It would be of interest to compare statistical anyons to other methods of generating anyonic behavior not covered in this work, such as N00N states subject to Bloch oscillations \cite{Lebugle2015}, quasi-holes in Bose-Einstein condensates \cite{Paredes2001}, or particles possessing ambiguous statistics \cite{Medvedev1997}. Statistical anyons may also have the potential to provide simpler implementations or experimental analogues for other applications of anyons, including Haldane insulators \cite{Lange2017, Sicks2020} or anyon beams \cite{Majhi2019}. Anyonic statistics has also been applied to more exotic systems, such as black hole gasses \cite{Strominger1993, Medvedev1997}, for which statistical anyons may provide a tractable theoretical tool. Finally, a particularly interesting possibility for statistical anyons is the introduction of interparticle interactions. We have shown that non-interacting statistical anyons are capable of recreating GES, which historically has been limited to interacting systems. By combining these paradigms, we add significant complexity to statistical behavior of the particles. It is intriguing to wonder if this additional complexity may be leveraged to expand the state space of possible anyonic phases, effectively recreating the braiding statistics of topological anyons. We leave this multitude of questions open for exploration in future works.       

\begin{acknowledgments}

S.D. acknowledge support from the U.S. National Science Foundation under Grant No. DMR-2010127.  

\end{acknowledgments}
\hfill \break
\appendix

\onecolumngrid

\section{Statistical Anyon Harmonic Oscillator Partition Function}
\label{Appendix A}

In this appendix we derive an expression for the partition function of two statistical anyons in a harmonic potential. We begin with the definition of the partition function for $N$ independent pairs of particles in the position basis,
\begin{equation}
Z_{\mathrm{A}}^N = \tr{ e^{-\beta H}} = \displaystyle\prod_{j=1}^{N} \int dx_j \int dy_j \bra{x_j y_j}e^{-\beta H_j}\ket{x_j y_j},
\end{equation}
where,
\begin{equation}
H_j = \frac{p_{x_j}^2+p_{y_j}^2}{2m}+\frac{1}{2} m \omega^2 (x_j^2+y_j^2).
\end{equation} 
Inserting the identity in the energy basis $I = \sum_{n_1^{(j)},n_2^{(j)} = 0}^{\infty} \ket{n_1^{(j)} n_2^{(j)}} \bra{n_1^{(j)} n_2^{(j)}}$ twice leads to,
\begin{equation}
Z_{\mathrm{A}}^N = \displaystyle\prod_{j=1}^{N} \int dx_j \int dy_j \sum_{n_1^{(j)},n_2^{(j)} = 0}^{\infty} \sum_{m_1^{(j)},m_2^{(j)} = 0}^{\infty} \Psi_{\mathrm{A}}^*(x_j,y_j) \Psi_{\mathrm{A}}(x_j,y_j) e^{-\beta \hbar \omega (m_1^{(j)}+m_2^{(j)}+1)}\braket{n_1^{(j)} n_2^{(j)}}{m_1^{(j)} m_2^{(j)}}.
\end{equation} 
Here $\Psi_{\mathrm{A}}(x_j,y_j)$ is given by Eq. \eqref{eq:anyon} where,
\begin{equation}
\psi_n(x)=\frac{1}{\sqrt{2^n n!}} \bigg(\frac{m \omega}{\pi \hbar} \bigg)^{1/4} e^{- \frac{m \omega x^2}{2 \hbar}} H_n \bigg( \sqrt{\frac{m \omega}{\hbar}}x\bigg).
\end{equation}  

After evaluating the inner product, the integration over each $x_j$ and $y_j$ can be carried out with application of the Hermite polynomial orthogonality to yield, 
\begin{equation}
Z_{\mathrm{A}}^N = \displaystyle\prod_{j=1}^{N} \sum_{n_1^{(j)},n_2^{(j)} = 0}^{\infty} \frac{1}{4} e^{-\beta \hbar \omega (n_1^{(j)}+n_2^{(j)}+1)} (2+e^{-i \pi \Theta\left(j - N_{\mathrm{B}} +1 \right)} \delta_{n_1^{(j)},n_2^{(j)}}+e^{i \pi \Theta\left(j - N_{\mathrm{B}} +1 \right)} \delta_{n_1^{(j)},n_2^{(j)}}).   
\end{equation}
where $\delta_{n_1,n_2}$ is the Kronecker delta. In order to evaluate the step functions we separate the product into one from 1 to $N_{\mathrm{B}}$ and a second from $N_{\mathrm{B}} +1$ to $N$, 
\begin{equation}
Z_{\mathrm{A}}^N = \displaystyle\prod_{j=1}^{N_{\mathrm{B}}} \sum_{n_1^{(j)},n_2^{(j)} = 0}^{\infty} \frac{1}{2} e^{-\beta \hbar \omega (n_1^{(j)}+n_2^{(j)}+1)} (1+ \delta_{n_1^{(j)},n_2^{(j)}}) \displaystyle\prod_{j=N_{\mathrm{B}}+1}^{N} \sum_{n_1^{(j)},n_2^{(j)} = 0}^{\infty} \frac{1}{2} e^{-\beta \hbar \omega (n_1^{(j)}+n_2^{(j)}+1)} (1- \delta_{n_1^{(j)},n_2^{(j)}}).   
\end{equation}
Here we can immediately recognize the individual partition functions for two bosons and two fermions in harmonic potential,
\begin{equation}
Z_{\mathrm{B}} = \sum_{n_1,n_2 = 0}^{\infty} \frac{1}{2} (1+ \delta_{n_1,n_2})e^{-\beta \hbar \omega (n_1+n_2+1)}, \quad Z_{\mathrm{F}} = \sum_{n_1,n_2 = 0}^{\infty} \frac{1}{2} (1- \delta_{n_1,n_2})e^{-\beta \hbar \omega (n_1+n_2+1)}. 
\end{equation}
Thus we can re-write our anyonic partition function in the simple form,
\begin{equation}
Z_{\mathrm{A}}^N = \displaystyle\prod_{j=1}^{N_{\mathrm{B}}} Z_{\mathrm{B}}^{(j)} \displaystyle\prod_{k=1}^{N_{\mathrm{F}}} Z_{\mathrm{F}}^{(k)} = Z_B^{N_{\mathrm{B}}} Z_{\mathrm{F}}^{N_{\mathrm{F}}} = (Z_{\mathrm{B}}^{p_{\mathrm{B}}} Z_{\mathrm{F}}^{p_{\mathrm{F}}})^N.
\end{equation}
Here we have used the fact that for large $N$ we can express the number of bosonic and fermionic particle pairs as $N_{\mathrm{B}} = N p_{\mathrm{B}}$ and $N_{\mathrm{F}} = N p_{\mathrm{F}} = N(1-p_{\mathrm{B}})$ where $p_{\mathrm{B}}$ ($p_{\mathrm{F}}$) is the probability of a particle pair having bosonic (fermionic) symmetry.   

The boson and fermion partition functions can be written in closed-form as,  
\begin{equation}
Z_{\mathrm{B}} = \frac{1}{8} \, \text{csch}^2\left(\frac{\beta  \omega  \hbar }{2}\right)+\frac{1}{4} \, \text{csch}(\beta  \omega  \hbar ), \quad Z_{\mathrm{F}} = \frac{1}{8} \, \text{csch}^2\left(\frac{\beta  \omega  \hbar }{2}\right)-\frac{1}{4} \, \text{csch}(\beta  \omega  \hbar ).
\end{equation}
With this, we arrive at a closed-form expression for the partition function of a single pair of statistical anyons,
\begin{equation}
Z_{\mathrm{A}} = \left[\frac{1}{8} \, \text{csch}^2\left(\frac{\beta  \omega  \hbar }{2}\right)+\frac{1}{4} \, \text{csch}(\beta  \omega  \hbar ) \right]^{p_{\mathrm{B}}}\left[\frac{1}{8} \, \text{csch}^2\left(\frac{\beta  \omega  \hbar }{2}\right)-\frac{1}{4} \, \text{csch}(\beta  \omega  \hbar ) \right]^{1-p_{\mathrm{B}}}.
\end{equation} 

\section{Equilibrium Thermodynamic Quantities}
\label{Appendix B}
In this appendix we give the full expressions for the internal energies, free energies, entropies, and heat capacities of statistical anyons and topological anyons in one- and two-dimensional harmonic potentials.
\subsection{2D Statistical Anyons}
Plugging the respective partition functions for two bosons and two fermions in a harmonic potential into Eq. \eqref{eq:partition}, and substituting that into Eq. \eqref{eq:thermofunc} we arrive at the following expressions for the thermodynamic properties of two-dimensional statistical anyons:
\begin{equation}
E = \hbar \omega \left[2 \coth(\frac{1}{2}\beta \hbar \omega)+\tanh(\frac{1}{2} \beta \hbar \omega)-p_{\mathrm{B}}\tanh(\beta \hbar \omega )\right]
\end{equation}
\begin{equation}
F = -\frac{1}{\beta} \ln\left\{\frac{1}{8} \left[\cosh(\beta \hbar \omega)\right]^{p_{\mathrm{B}}} \text{csch}^2(\frac{1}{2}\beta \hbar \omega)\text{csch}^2(\beta \hbar \omega)\right\}
\end{equation}
\begin{equation}
S = k_{\mathrm{B}} \beta \hbar \omega \left[2 \coth(\frac{1}{2}\beta \hbar \omega)+\tanh(\frac{1}{2} \beta \hbar \omega)-p_{\mathrm{B}} \tanh(\beta \hbar \omega)\right]+k_{\mathrm{B}} \ln\left\{\frac{1}{8} \left[\cosh(\beta \hbar \omega)\right]^{p_{\mathrm{B}}} \text{csch}^2(\frac{1}{2}\beta \hbar \omega)\text{csch}^2(\beta \hbar \omega)\right\}
\end{equation}
\begin{equation}
C = \frac{1}{2} k_{\mathrm{B}} \beta^2 \hbar^2 \omega^2 \left[ 2 \,\text{csch}^2(\frac{1}{2} \beta \hbar \omega) - \text{sech}^2\,(\frac{1}{2} \beta \hbar \omega)+2 p_{\mathrm{B}} \, \text{sech}^2\,(\beta \hbar \omega) \right]
\end{equation}

\subsection{2D Topological Anyons}
Plugging Eq. \eqref{eq:partitionTrue} into Eq. \eqref{eq:thermofunc} we determine the following expressions for the thermodynamic properties of topological anyons: 
\begin{equation}
E = \hbar \omega \left[\coth(\frac{1}{2}\beta \hbar \omega)+2 \coth(\beta \hbar \omega)+(\nu -1)\tanh(\beta \hbar \omega (1-\nu))\right]
\end{equation}
\begin{equation}
F = \frac{1}{\beta} \ln\left[8 \, \text{sech}(\beta \hbar \omega (\nu-1))\sinh^2(\frac{1}{2}\beta \hbar \omega)\sinh^2(\beta \hbar \omega) \right]
\end{equation}
\begin{equation}
S = k_{\mathrm{B}} \beta \hbar \omega \left[\coth(\frac{1}{2}\beta \hbar \omega)+2 \coth(\beta \hbar \omega)+(\nu - 1)\tanh(\beta \hbar \omega (1-\nu)) \right]+k_{\mathrm{B}} \ln\left[\frac{1}{8} \cosh(\beta \hbar \omega (\nu-1))\text{csch}^2 \, (\frac{1}{2}\beta \hbar \omega)\text{csch}^2 \, (\beta \hbar \omega) \right]
\end{equation}
\begin{equation}
C = \frac{1}{2} k_{\mathrm{B}} \beta^2 \hbar^2 \omega^2 \left[ \text{csch}^2(\frac{1}{2} \beta \hbar \omega) + 4 \text{csch}^2\,(\beta \hbar \omega)+2(\nu-1)^2 \text{sech}^2\,(\beta \hbar \omega (\nu-1)) \right]
\end{equation}

\section{1D Boson and Fermion Density Operators and Wigner Distributions}
\label{Appendix C}
In this appendix we give the full expressions for the thermal state and time evolved density operators for a nonequilibrium harmonic quantum Otto engine with a working medium of 1D bosons and fermions. We also include the thermal state Wigner distribution applied in the determination of the optimal protocol shortcut. In each expression the top sign of each plus/minus denotes the boson expression and the the bottom sign the fermion expression.

The thermal state density operator in position representation is given by, 
\begin{equation}
\label{eq:thermal}
\begin{split}
\rho_0(x_1,x_2, y_1,y_2) &= \frac{1}{Z} \frac{m \omega}{2 \pi \hbar \sinh{(\beta \hbar \omega)}}
\bigg[ e^{- \frac{m \omega}{4 \hbar}\{[(x_1+y_1)^2+(x_2+y_2)^2]\tanh{(\beta \hbar \omega/2)}+[(x_1-y_1)^2+(x_2-y_2)^2]\coth{(\beta \hbar \omega/2)}\}} \\
&\pm e^{- \frac{m \omega}{4 \hbar}\{[(x_2+y_1)^2+(x_1+y_2)^2]\tanh{(\beta \hbar \omega/2)}+[(x_2-y_1)^2+(x_1-y_2)^2]\coth{(\beta \hbar \omega/2)}\}}\bigg].
\end{split}
\end{equation}

Solving the time dependent Schr\"{o}dinger equation by means of the appropriate evolution operator for the isentropic strokes of the engine yields the time-evolved density operator, 
\begin{equation}
\begin{split}
&\rho_t(x_1,x_2,y_1,y_2)= \frac{m \omega}{2 \pi \hbar (Y_t^2+X_t^2 \omega^2)} \left(e^{\mp \beta \hbar \omega}-1\right) \\
&\quad\times\bigg\{ e^{\frac{m}{2 \hbar (Y_t^2+X_t^2 \omega^2)}\left[i (x_1^2+x_2^2-y_1^2-y_2^2)(Y_t \dot{Y}_t+X_t\dot{X}_t\omega^2)-\omega (x_1^2+x_2^2+y_1^2+y_2^2) \mathrm{coth}(\beta \hbar \omega)
	+2 \omega (x_1 y_1 +x_2 y_2) \mathrm{csch}(\beta \hbar \omega) \right]} \\
&\quad\pm e^{\frac{m}{2 \hbar (Y_t^2+X_t^2 \omega^2)}\left[i (x_1^2+x_2^2-y_1^2-y_2^2)(Y_t \dot{Y}_t+X_t\dot{X}_t\omega^2)
	- \omega(x_1^2+x_2^2+y_1^2+y_2^2) \mathrm{coth}(\beta \hbar \omega)+2 \omega (x_2 y_1 +x_1 y_2) \mathrm{csch}(\beta \hbar \omega) \right]} \bigg\}.
\end{split}
\end{equation}
Here $X_t$ and $Y_t$ are solutions to the equation of motion of the classical time-dependent harmonic oscillator,
\begin{equation}
\ddot{X}_t + \omega^2(t) X_t = 0.
\end{equation} 

The two particle Wigner distribution is found by carrying out the integral,
\begin{equation}
W(x_1,p_1,x_2,p_2) = \frac{1}{4 \pi^2 \hbar^2} \int du_1 \int du_2\,
\rho\left(x_1+\frac{u_1}{2},x_2+\frac{u_2}{2},x_1-\frac{u_1}{2},x_2-\frac{u_2}{2}\right)
e^{-\frac{i p_1 u_1}{\hbar}}e^{-\frac{i p_2 u_2}{\hbar}}.
\end{equation}
For the thermal state the Wigner distribution is,
\begin{equation}
\begin{split}
&W_0(x_1,p_1,x_2,p_2) = \frac{\mathrm{sech}^2\left(\beta \hbar \omega/2 \right)}{\pi^2 \hbar^2 (\mathrm{csch}^2\left( \beta \hbar \omega/2 \right) \pm 2 \mathrm{csch}\left(\beta \hbar \omega \right))}\\
&\quad\times\left( e^{-\frac{(p_1^2 +p_2^2 +m^2(x_1^2+x_2^2)\omega^2)\mathrm{tanh}\left(\beta \hbar \omega/2 \right)}{m \omega \hbar}}\pm 2 e^{\frac{-(p_1^2+p_2^2 +m^2(x_1^2+x_2^2)\omega^2)\mathrm{coth}(\beta \hbar \omega)+2(p_1 p_2 +m^2 \omega^2 x_1 x_2)\mathrm{csch}(\beta \hbar \omega)}{m \omega \hbar} }\right)\,.
\end{split} 
\end{equation}

\section{Nonequilibrium Engine Characterizations: 1D Statistical Anyons}
\label{Appendix D}
In this appendix we give the full expressions for the power and efficiency of a nonequilibrium harmonic quantum Otto engine with a working medium of 1D statistical anyons. 

\begin{align}
P &= \frac{\hbar}{2 \tau}  \Big\{\left(2 p_{\mathrm{B}}-1\right) \left[\left(Q_{21}-1\right) \omega_1+\left(Q_{12}-1\right) \omega_2\right]+3 \left(\omega_2-Q_{21} \omega_1\right) \coth \left(\beta_2 \omega_2 \hbar \right) \\
&+3 \left(\omega_1-Q_{12} \omega_2\right) \coth \left(\beta_1 \omega_1 \hbar \right)+\left(\omega_2-Q_{21} \omega_1\right) \text{csch}\left(\beta_2 \omega_2 \hbar \right)+\left(\omega_1-Q_{12} \omega_2\right) \text{csch}\left(\beta_1 \omega_1 \hbar \right)\Big\} \nonumber
\end{align}

\begin{equation}
\eta = 1+ \frac{\omega_1}{\omega_2}\left[ \frac{3 \coth \left(\beta_1 \omega_1 \hbar \right)+\text{csch}\left(\beta_1 \omega_1 \hbar \right)+2 \left(Q_{21}-1\right) p_{\mathrm{B}}-Q_{21} \left(3 \coth \left(\beta_2 \omega_2 \hbar \right) +\text{csch}\left(\beta_2 \omega_2 \hbar \right)+1\right)+1}{3 \coth \left(\beta_2 \omega_2 \hbar \right)+\text{csch}\left(\beta_2 \omega_2 \hbar \right)+2
	\left(Q_{12}-1\right) p_{\mathrm{B}}-Q_{12} \left(3 \coth \left(\beta_1 \omega_1 \hbar \right) +\text{csch}\left(\beta_1 \omega_1 \hbar \right) +1\right)+1} \right]
\end{equation}
 
\twocolumngrid

\bibliography{refs}
		
\end{document}